\pdfoutput=1
\documentclass[12pt,letterpaper]{article}
\usepackage{jheppub}
\usepackage[mathscr]{eucal}
\usepackage{bm}
\usepackage{booktabs}
\usepackage{afterpage}

\bibpunct{[}{]}{,}{n}{}{;}

 \def\Nfour{$\mathcal{N}\,{=}\,4$}
 
 \def\Nc{N_\text{c}}

 \def\tr{\text{tr}\,}
 \def\half{\tfrac{1}{2}}

 \def\rh{r_{\rm h}}
 \def\rtilde{\tilde r}
 \def\uh{u_{\rm h}}
 \def\rhbar{\bar r_{\rm h}}

 \def\Re{\mathrm{Re}}
 \def\Im{\mathrm{Im}}
 
 \def\rave{r_0}
 \def\Th{T_{\rm h}}
 \def\teq{t_{\rm eq}}
 \def\rhomax{\rho_{\rm max}}

 \def\B{{\cal B}}
 \def\E{{\cal E}}

 \def\epsB{\varepsilon_{\cal B}}
 \def\epsL{\varepsilon_{L}}
 \def\tpeak{t_{\rm peak}}
 \def\suck[#1]#2{\includegraphics[#1]{#2}}        

 \title{Far-from-equilibrium dynamics of a strongly coupled
     non-Abelian plasma with non-zero charge density or
     external magnetic field}
 \author{John~F.~Fuini~III,}
 \author{Laurence~G.~Yaffe}
 \affiliation{Department of Physics, University of Washington, Seattle WA 98195, USA}
 \emailAdd{fuini@u.washington.edu}
 \emailAdd{yaffe@phys.washington.edu}

 \abstract
    {%
    Using holography, we study the evolution of a spatially homogeneous,
    far from equilibrium, strongly coupled \Nfour\ supersymmetric
    Yang-Mills plasma with a non-zero charge density
    or a background magnetic field.
    This gauge theory problem corresponds,
    in the dual gravity description,
    to an initial value problem in Einstein-Maxwell theory
    with homogeneous but anisotropic initial conditions.
    We explore the dependence of the equilibration process
    on different aspects of the initial departure from equilibrium
    and, while controlling for these dependencies, examine how the 
    equilibration dynamics are affected by the presence of a 
    non-vanishing charge density or an external magnetic field.
    The equilibration dynamics are remarkably insensitive to 
    the addition of even large chemical potentials or magnetic fields;
    the equilibration time is set primarily by the form of the initial
    departure from equilibrium.
    For initial deviations from equilibrium which are well localized
    in scale, we formulate a simple model for equilibration times
    which agrees quite well with our results.
    }
 \arxivnumber{1503.07148}
 \begin{document}
 \maketitle


\section{Introduction}

The discovery of gauge/gravity duality (or ``holography'')
has enabled the study of
previously intractable problems involving the dynamics of strongly
coupled gauge theories.%
\footnote
    {%
    See, for examples,
    refs.~\cite{Aharony:1999ti,
    D'Hoker:2002aw,
    Son:2007vk} 
    and references therein.
    }
In the limit of large gauge group rank $\Nc$,
and large `t Hooft coupling $\lambda$,
the strongly coupled quantum dynamics of certain gauge field theories
may be mapped, precisely, into classical gravitational dynamics of
higher dimensional asymptotically anti-de Sitter (AdS) spacetimes~%
\cite{Maldacena:1997re, Witten:1998qj, GKP}.
Numerical studies of the resulting 
gravitational dynamics can shed light on poorly understood
aspects of the quantum dynamics of strongly coupled gauge theories.

Using the simplest example of gauge/gravity duality,
applicable to maximally supersymmetric $SU(\Nc)$ Yang-Mills theory
(\Nfour\ SYM),
this approach has been applied to a succession of problems
of increasing complexity
involving far from equilibrium dynamics.
These include homogeneous isotropization~%
\cite{Chesler:2008hg,Heller:2012km,Heller:2013oxa},
colliding shock waves~%
\cite{Chesler:2010bi,
Casalderrey-Solana:2013aba,
Casalderrey-Solana:2013sxa,
vanderSchee:2013pia,
Bantilan:2014sra,
Chesler:2015wra},
and turbulence in two-dimensional fluids~%
\cite{Adams:2013vsa,Adams:2012pj}.
A detailed presentation of the methods used in most of these works
is available \cite{Chesler:2013lia}.

In this paper, we extend previous work on the dynamics of
homogeneous but anisotropic \Nfour\ SYM plasma~%
\cite{Chesler:2008hg,Heller:2012km,Heller:2013oxa}.
We examine the influence on the equilibration dynamics
of a non-zero global $U(1)$ charge density, or a background
magnetic field.
Inclusion of these effects is motivated by
the physics of relativistic heavy ion collisions
\cite{Shuryak,Shuryak:2004cy,CasalderreySolana:2011us}.
Hydrodynamic modeling of near-central events clearly
indicates that the baryon chemical potential $\mu_B$
in the mid-rapidity region
is significantly smaller than the temperature, but
not by an enormously large factor at RHIC energies.%
\footnote
    {Inferred values of $\mu_B/T$ at
    chemical freeze-out are about 0.15 for RHIC collisions
    at $\sqrt{s_{NN}} = 200$ GeV, 
    and roughly 0.005 for LHC heavy ion collisions
    with $\sqrt{s_{NN}} = 2.8$ TeV
    \cite{Stachel:2013zma,Andronic:2012dm}.
    }
Hence, it is desirable to understand the sensitivity
of the plasma equilibration dynamics to the presence of
a baryon chemical potential and associated
non-zero baryon charge density.
Similarly, it is clear that large, but transient,
electromagnetic fields are generated in heavy ion collisions.
A growing body of work~%
\cite{Skokov:2009qp,Gursoy:2014aka,Agasian:2008tb,Fukushima:2010vw,Kharzeev:2007jp} 
suggests that electromagnetic effects
may play a significant role despite the small value of the
fine structure constant.
Electromagnetic effects on equilibrium QCD properties are also under study
using lattice gauge theory 
\cite{Bali:2014kia, D'Elia:2010nq, Buividovich:2009wi, Abramczyk:2009gb}. 

The large $\Nc$, strongly coupled \Nfour\ SYM plasma we study is,
of course, only a caricature of a real quark-gluon plasma.
But it is a highly instructive caricature which correctly
reproduces many qualitative features of QCD plasma
(such as Debye screening, finite static correlation lengths,
and long distance, low frequency dynamics described by
neutral fluid hydrodynamics).
Moreover,
in the temperature range relevant for heavy ion collisions,
quantitative comparisons of bulk thermodynamics,
screening lengths, shear viscosity, and other observables
show greater similarity between \Nfour\ SYM and QCD
than one might reasonably have expected
\cite{Bak:2007fk,Gubser:2009md}.
Since the composition of a plasma
depends on the chemical potentials, or associated charge
densities, of its constituents,
studying the dependence of the equilibration dynamics on
a conserved charge density provides a simple means to
probe the sensitivity of the dynamics to the precise
composition of the non-Abelian plasma.
This, in small measure, may help one gauge the degree to
which \Nfour\ SYM plasma properties can be extrapolated
to real QCD plasma.
At the very least, strongly coupled \Nfour\ SYM theory provides
a highly instructive toy model in which one may explore,
quantitatively, non-trivial aspects of non-equilibrium gauge field dynamics.%
\footnote
    {%
    Previous work examining thermalization in plasmas with non-zero chemical
    potential (not involving numerical solutions of far from equilibrium
    geometries) includes refs.~%
    \cite{Galante:2012pv,Caceres:2012em,Giordano:2014kya,Caceres:2014pda}.
    }

The remainder of the paper is organized as follows.
Section \ref{sec:setup} 
summarizes necessary background material.
This includes the coupling of an Abelian background gauge field
to a $U(1)$ subgroup of the $SU(4)_R$ global symmetry group
of \Nfour\ SYM.
This $U(1)$ symmetry may be regarded as analogous
to either the baryon number $U(1)_B$ or electromagnetic
$U(1)_{\rm EM}$ flavor symmetries of QCD.
Turning on a background magnetic field implies an enlargement
of the theory under consideration from 
\Nfour\ SYM to \Nfour\ SYM coupled to electromagnetism 
(which we abbreviate as SYM+EM).
The combined theory is no longer scale invariant;
this has important implications which we discuss.
This section describes
the 5D Einstein-Maxwell theory which provides the holographic
description of the states of interest,
presents our coordinate ansatz
(based on a null slicing of the geometry), 
and summarizes relevant portions of the holographic dictionary
relating gravitational and dual field theory quantities.
This section also records the reduced field equations
which emerge from our symmetry specializations, describes
the relevant near-boundary asymptotic behavior, and summarizes
properties of the static equilibrium geometries to which our
time dependent solutions asymptote at late times.

The following section \ref{sec:methods} briefly describes our numerical
methods, which are based on the strategy presented in
ref.~\cite{Chesler:2013lia}.
When studying states with a non-zero charge density
(but no background magnetic field)
appropriate numerical methods for asymptotically AdS Einstein-Maxwell
theory are immediate generalizations of methods which have previously
been found to work well for pure gravity.
However, the inclusion of a background magnetic field induces
a trace anomaly in the dual
quantum field theory which,
in the gravitational description, manifests in the appearance
of logarithmic terms in the near-boundary behavior of fields.
Such non-analytic terms degrade the performance of spectral methods,
on which we rely, and necessitate careful attention to numerical issues.
Section~\ref{sec:methods} also describes the specifics of our chosen
initial data.

Results are presented in section~\ref{sec:results}.
We focus on the evolution of the expectation value of the
stress-energy tensor.
We first discuss the sensitivity of the equilibration dynamics
to features in the initial data and, in particular, examine
the extent to which the evolution shows nonlinear dependence
on the initial departure from equilibrium.
We find that only disturbances in the geometry originating
deep in the bulk, very close to the horizon, generate significant
nonlinearities.
This is broadly consistent with earlier work
\cite{Heller:2012km,Heller:2013oxa}.
However, for a very wide variety of initial disturbances,
including ones which generate extremely large pressure anisotropies,
we find remarkably little nonlinearity in the equilibration dynamics,
often below the part-per-mille level.

We then present comparisons of the equilibration dynamics as a function of
the charge density or background magnetic field.
We focus on comparisons in which the form of the initial departure
form equilibrium and the energy density,
or the equilibrium temperature,
is held fixed while either the charge density or magnetic field is varied.
These comparisons reveal surprisingly little sensitivity to the
charge density, or magnetic field, even at early times when the
departure from equilibrium is large.

We verify the late time approach to the expected equilibrium states,
and extract the leading quasinormal mode (QNM) frequency from
the late time relaxation.
Quasi-normal mode frequencies extracted from our full
nonlinear dynamics are compared, where possible,
with independent calculations of QNM frequencies based
on a linearized analysis around the equilibrium geometry.
This provides a useful check on our numerical accuracy.

We define an approximate equilibration time
based on the relative deviation of the pressure anisotropy
from its equilibrium value,
and examine the dependence of this time on
charge density or external magnetic field. 
Once again, changes in this quantity are largest for initial
disturbances which originate very close to the horizon, but
the overall sensitivity of the equilibration time
to the charge density or magnetic field is remarkably modest.

The final section~\ref{sec:discussion} discusses and attempts
to synthesize the implications of our results.
We present a simple model of equilibration times, for
initial disturbances which are well localized in scale,
which agrees rather well with our numerical results
(but becomes less accurate for disturbances localized
extremely close to the horizon).
We end with a few concluding remarks.%
\footnote
    {%
    As this paper neared completion, we learned of 
    the somewhat related work by A. Buchel, M. Heller, and 
    R. Myers \cite{Buchel:2015saa}. These authors examine quasinormal mode
    frequencies in $\mathcal{N} = 2^*$ SYM and argue that, 
    in this non-conformal deformation of $\mathcal{N} = 4$ SYM, 
    the longest equilibration times are largely set
    by the temperature with little sensitivity to other scales.
    }


\section{Ingredients}\label{sec:setup}

\subsection{\Nfour\ SYM in an external field}\label{sec:N=4}

We study maximally supersymmetric $SU(\Nc)$ Yang-Mills theory (\Nfour\ SYM)
on four dimensional Minkowski space
when the conserved current for a
$U(1)$ subgroup of the $SU(4)_R$ global symmetry group
either (a) has a non-vanishing charge density,
or (b) is coupled to a background Abelian gauge field
describing a uniform magnetic field.
The embedding of the $U(1)$ symmetry is chosen such that
the $U(1)$ commutes with an $SU(3)$ subgroup
of the $SU(4)_R$ global symmetry.

The coupling to the external field has the usual form%
\footnote
    {%
    We use a mostly-plus Minkowski space metric,
    $\eta_{\mu\nu} \equiv \mathop{\rm diag}({-}1,{+}1,{+}1,{+}1)$.
    }
\begin{equation}
\label{eq:Sint}
    S = S_{\rm SYM} + \int d^4x \> j^\alpha(x) A_\alpha^{\rm ext}(x) \,,
\end{equation}
where $j^\alpha(x)$ is the conserved $U(1)$ current normalized such that
the four Weyl fermions of \Nfour\ SYM have charges $\{+3,-1,-1,-1\}/\sqrt 3$
and the three complex scalars have charge $+2/\sqrt 3$.
The overall factor of $1/\sqrt 3$ in these charge assignments has no physical
significance, but is chosen so that the trace anomaly and electromagnetic
beta function (induced when this current is gauged) have convenient
coefficients, as will be seen below.%
\footnote
    {%
    These charge assignments are $1/\sqrt 3$ times those used in
    ref.~\cite{D'Hoker:2009mm}.
    Overall rescaling of these charge assignments
    has implications for the holographic description which
    are noted below in footnote~\ref{fn:xi}.
    }
The background $U(1)$ gauge field $A_\alpha^{\rm ext}(x)$ 
we take to have the form
\begin{equation}
    A_\alpha^{\rm ext}(x) \equiv
    \mu \, \delta_\alpha^0
    + \half {\B}\, (x^1 \delta_\alpha^2-x^2 \delta_\alpha^1) \,,
\label{eq:external A}
\end{equation}
with $\mu$ the chemical potential which,
in equilibrium, will be conjugate to the charge density $j^0$,
and $\B$ the
amplitude of a constant magnetic field pointing in the $x^3$ direction.
Although it should be straightforward
to study dynamics when both the charge density $j^0$
and magnetic field $\B$ are non-zero, in this paper we focus for
simplicity on the cases of either a non-zero charge density
with vanishing magnetic field, $j^0 \ne 0$ and $\B = 0$,
or non-zero magnetic field
with vanishing charge density, $j^0 = 0$ and $\B \ne 0$.

With a non-zero magnetic field $\B$ in the $x^3$ direction,
changes in the background gauge field under a
translation in the $x^1$ or $x^2$ directions,
or a rotation in the $x^1$-$x^2$ plane,
can be compensated by a suitable $U(1)$ gauge transformation.
Hence, the theory retains full spatial translation invariance
as well as rotation invariance in the $x^1$-$x^2$ plane.

We will be interested in initial states which:~%
({\em i})
    have non-trivial expectation values
    $\langle T^{\alpha\beta}(x)\rangle$ and
    $\langle j^{\alpha}(x)\rangle$
    for the stress-energy tensor
    and $U(1)$ current density, respectively;
({\em ii})
    are invariant under spatial translations as well as
    $O(2)$ rotations in the $x^1{-}x^2$ plane; and
({\em iii})
    are invariant under the $SU(3)_R$ subgroup of the $SU(4)_R$
    global symmetry which commutes with our chosen $U(1)$.

Since all \Nfour\ SYM fields transform in the adjoint representation
of the $SU(\Nc)$ gauge group, the stress-energy and $U(1)$ current
expectation values both scale as $O(\Nc^2)$ in the large $\Nc$ limit.
For later convenience, we define a rescaled energy density $\varepsilon$
and charge density $\rho$,
via
\begin{equation}
    \langle T^{00} \rangle \equiv \kappa \, \varepsilon  \,,\qquad
    \langle j^0 \rangle \equiv \kappa  \, \rho \,,
\end{equation}
with
\begin{equation}
    \kappa \equiv (\Nc^2{-}1) / (2\pi^2) \,.
\end{equation}

\Nfour\ SYM is a conformal field theory with a traceless stress-energy
tensor.
Adding a chemical potential $\mu$ introduces a physical scale, but
does not modify the microscopic dynamics of the theory and hence
does not affect the tracelessness of the stress-energy tensor.
In contrast, introducing an external magnetic field does affect
the microscopic dynamics and, in particular, generates a non-zero
trace anomaly,%
\footnote
    {%
    We define the external gauge field such that no factor of an
    electromagnetic gauge coupling appears in the interaction (\ref{eq:Sint}),
    in our $U(1)$ covariant derivatives, or in the trace anomaly
    (\ref{eq:traceanom}).
    The coefficient of $-\tfrac 14 F_{\mu\nu}^2$ in the trace anomaly 
    (\ref{eq:traceanom}) equals the EM beta function coefficient
    $b_0$, given below in eq.~(\ref{eq:b0}).
    (Note that the sign of the trace anomaly depends on the
    metric convention in use.)
    }
\begin{equation}
    T^\alpha{}_\alpha
    = -\tfrac{1}{4} \, \kappa \, \big(F_{\mu\nu}^{\rm ext}\big)^2
    = -\tfrac{1}{2} \, \kappa \, \B^2 \,.
\label{eq:traceanom}
\end{equation}

The trace anomaly generated by the external magnetic field implies that
the theory is no longer scale invariant.
For example, the ground state energy density, as a function of
magnetic field, need not have the simple form of some pure number
times $\B^2$.
This will be seen explicitly below.
The trace anomaly implies that there must be logarithmic dependence on
a renormalization point.
To interpret this dependence,
it is appropriate to adopt the perspective that
adding an external magnetic field means that the theory under consideration
has been enlarged --- it is now \Nfour\ SYM coupled to $U(1)$ electromagnetism
(SYM+EM).
The complete action of the theory is the SYM action, minimally coupled
to the $U(1)$ gauge field, plus the Maxwell action for $U(1)$ gauge field,

\begin{equation}
    S_{\rm SYM+EM} = S_{\rm SYM,\,min.\;coupled}
    + S_{\rm EM} \,,
\label{eq:SYM+EM}
\end{equation}
with
\begin{equation}
    S_{\rm EM} \equiv
    -\int d^4x \> \tfrac 1{4 e^2}
    \, F_{\mu\nu}^2 \,.
\label{eq:S_EM}
\end{equation}
The electromagnetic coupling $e^2$
(having been scaled out of covariant derivatives)
appears as an inverse prefactor of the Maxwell action.
We regard the electromagnetic coupling $e^2$ as arbitrarily weak.
Hence, quantum fluctuations in the $U(1)$ gauge field are negligibly small,
allowing us to view the EM gauge field as a classical background field.%
\footnote
    {%
    In an arbitrary background $SU(4)$ gauge field,
    the divergence of the $SU(4)_R$ current acquires an anomalous contribution,
    $\partial^\mu J_\mu^a \propto d^{abc} F_{\mu\nu}^b F^{\mu\nu c}$.
    This anomaly, when specialized to our chosen $U(1)$ subgroup,
    is proportional to the sum of the cubes of our fermion charges
    and is non-zero,
    $
	\sum_\alpha (q^\alpha_{\rm f})^3 = 8/\sqrt 3
    $.
    To make the combined SYM+EM theory well defined,
    one could add to the theory additional fermions, 
    charged under the $U(1)$ but with no SYM interactions,
    which would cancel this $U(1)$ anomaly.
    As we are not concerned with quantum fluctuations in
    the $U(1)$ gauge field, the presence of this $U(1)$ anomaly
    (in the absence of compensating spectators) is irrelevant
    for our purposes.
    \label{fn:anomaly}
    }

However, just as in QED, 
fluctuations in the SYM fields which are electromagnetically
charged will cause the electromagnetic coupling $e^2$ to run with scale.
The associated renormalization group (RG) equation for the inverse coupling
has the usual form,
\begin{equation}
    \mu \frac d{d\mu} \, e^{-2}
    \equiv \beta_{1/e^{2}}(e^{-2}) = -b_0 + O(e^2) \,,
\label{eq:EMRG}
\end{equation}
with the one-loop beta function coefficient%
\footnote
    {%
    A non-renormalization theorem in supersymmetric \Nfour\ SYM
    implies that the short distance behavior of the current-current
    correlation cannot depend on the `t Hooft coupling $\lambda$
    \cite{Freedman:1998tz}.
    This implies that the leading EM beta function coefficient $b_0$ does not
    depend on $\lambda$, and hence may easily be evaluated in the $\lambda \to 0$ limit.
    }
\begin{equation}
    b_0
    \equiv
    \kappa
    \Big[
	\tfrac 1{6} \sum_\alpha (q^\alpha_{\rm f})^2
	+ \tfrac 1{12} \sum_a (q^a_{\rm s})^2 
    \Big]
    =
    \kappa
    \,.
\label{eq:b0}
\end{equation}
Here,
$
    q^\alpha_{\rm f}
    =
    ( 3, -1,-1,-1 )/\sqrt 3
$
and
$
    q^a_{\rm s}
    =
    ( 2, 2, 2 )/\sqrt 3
$
are the charge assignments of the four Weyl fermions and
three complex scalars, respectively.
Integrating this renormalization group equation leads, as usual, to
\begin{equation}
    1/e^2(\mu) = b_0 \ln(\Lambda_{\rm EM}/\mu) + O[\ln(\ln \Lambda_{\rm EM}/\mu)] \,,
\label{eq:EMcoupling}
\end{equation}
with the RG invariant scale $\Lambda_{\rm EM}$ denoting the Landau pole scale
where the (one loop approximation to the) electromagnetic coupling diverges.

The total stress-energy tensor derived from the combined action (\ref{eq:SYM+EM})
will equal the \Nfour\ SYM stress-energy tensor,
augmented with minimal coupling terms to the EM gauge field,
plus the classical Maxwell stress-energy.
An essential point, however, is that while the total stress-energy tensor is
well-defined, partitioning the stress-energy tensor into separate SYM
and EM contributions is inherently ambiguous, as the individual pieces
depend on the renormalization point.
We define
\begin{equation}
    T^{\alpha\beta}_{\rm tot}
    \equiv
    T^{\alpha\beta}_{\rm EM}(\mu) 
    +
    \Delta T^{\alpha\beta}_{\rm SYM}(\mu) \,,
\label{eq:Ttot}
\end{equation}
with%
\footnote
    {%
    Note that $T^{\alpha\beta}_{\rm EM}(\mu)$
    is not the metric variation of some renormalized EM action
    (whose separation from the total action would not be well-defined).
    Rather, eq.~(\ref{eq:TEMren}) is simply defining
    $T^{\alpha\beta}_{\rm EM}(\mu)$ as the classical EM stress-energy
    tensor multiplied by the scale-dependent inverse EM coupling.
    }
\begin{equation}
    T^{\alpha\beta}_{\rm EM}(\mu)
    \equiv
    \frac 1{e^2(\mu)}
    \left[
	F^{\alpha\nu} F^\beta{}_\nu - \tfrac 14 \eta^{\alpha\beta}
	F^{\mu\nu} F_{\mu\nu}
    \right] ,
\label{eq:TEMren}
\end{equation}
and
\begin{equation}
    \Delta T^{\alpha\beta}_{\rm SYM}(\mu)
    \equiv
    T^{\alpha\beta}_{\rm SYM,\, min.\;coupled}(\mu) \,.
\label{eq:TSYMren}
\end{equation}
The partitioning (\ref{eq:Ttot}) of the stress-energy tensor
puts all quantum corrections other than the running of the EM
coupling into the SYM contribution $\Delta T^{\alpha\beta}_{\rm SYM}(\mu)$.
The scale dependence must, of course, cancel between the two terms
because the total stress-energy tensor is a physical quantity.
Therefore, the scale dependence in the SYM contribution to the stress-energy
must simply compensate the known running of the inverse electromagnetic
coupling (\ref{eq:EMRG}) in the Maxwell stress-energy tensor (\ref{eq:TEMren}),
\begin{equation}
    \mu \frac d{d\mu} \,
    \Delta T^{\alpha\beta}_{\rm SYM}(\mu)
    =
    -\mu \frac d{d\mu} \,
    T^{\alpha\beta}_{\rm EM}(\mu)
    =
    b_0
    \left[
	F^{\alpha\nu} F^\beta{}_\nu - \tfrac 14 \eta^{\alpha\beta}
	F^{\mu\nu} F_{\mu\nu}
    \right] .
\label{eq:TSYMRG}
\end{equation}

Specializing to zero temperature states in a constant static
magnetic field $\B$,
the scale dependence (\ref{eq:TSYMRG})
plus dimensional analysis
implies that the SYM contribution to the ground state energy
density is a non-analytic function of magnetic field,
\begin{equation}
    \varepsilon(\mu)
    = c_0 \, \B^2 - \tfrac 14 \B^2 \ln (|\B|/\mu^2)
    = \tfrac 14 \B^2 \ln \!\big[\B^*(\mu)/|\B| \big] ,
\label{eq:magenergy}
\end{equation}
with $c_0$ some pure number.
(Here and henceforth, when considering physics in a
non-zero magnetic field
$
    \varepsilon(\mu) \equiv \Delta T^{00}_{\rm SYM}(\mu)/\kappa
$
denotes the SYM portion of the rescaled energy density.)
In the second form of eq.~(\ref{eq:magenergy}),
the analytic term has been absorbed by defining a
scale dependent ``fiducial'' magnetic field amplitude,
\begin{equation}
    \B^*(\mu) \equiv \mu^2 e^{4 c_0} \,.
\label{eq:B*}
\end{equation}
Note that the ground state energy acquires a simple quadratic
form when the renormalization point is chosen to scale with
the magnetic field,
$
    \varepsilon({|\B|^{1/2}}) = c_0 \, \B^2
$.
Our numerically determined value for the coefficient $c_0$
is given below in eq.~(\ref{eq:c_0}).

When considering low temperature physics in a background magnetic field,
$T^2 \ll |\B|$,
it is natural to choose a renormalization point $\mu = O(|\B|^{1/2})$,
as this is the relevant scale which cuts off long range fluctuations
in the charged SYM fields.
We will employ two choices for the renormalization point.
One choice is $\mu = 1/L$, with $L$ the AdS curvature scale (discussed below);
this choice is computationally convenient but not physically significant.
We will also report and discuss results with $\mu =|\B|^{1/2}$.
For later convenience, we define abbreviations for the (rescaled)
energy density evaluated at these two renormalization points,
\begin{equation}
    \epsL \equiv \varepsilon(1/L) \,,\qquad
    \epsB \equiv \varepsilon(|\B|^{1/2}) \,.
\end{equation}

\subsection{Holographic description}

The holographic description of SYM states, within our sector of interest,
in the limit of large $\Nc$ and large `t Hooft coupling $\lambda$,
is given by classical Einstein-Maxwell theory on 5-dimensional spacetimes
which are asymptotically AdS$_5$~\cite{Chamblin:1999tk}.
The 5D bulk action is
\begin{equation}
    S_5 \equiv \frac 1{16\pi G_5} \int d^5x \> \sqrt {-G} \,
    \left( R -2\Lambda -  L^2 \, F_{MN} F^{MN} \right) ,
\label{eq:bulk action}
\end{equation}
with $G_5 \equiv \frac \pi2 L^3/\Nc^2$
the 5D Newton gravitational constant,
$\Lambda \equiv -6/L^2$ the cosmological constant,
and $L$ the AdS curvature scale.%
\footnote
    {%
    The coefficient of the Maxwell action may,
    of course, be set to an arbitrary value by suitably
    rescaling the bulk gauge field $A_M$.
    However, as the on-shell variation of the
    gravitational action with respect to the boundary value
    of the gauge field defines the associated current,
    such rescaling changes the normalization of the
    $U(1)$ current in the holographic description.
    It will be seen below that the coefficient of the
    Maxwell term in our action (\ref{eq:bulk action}) is correctly
    chosen so that the $U(1)$ current normalization is
    consistent with our previous charge assignments.
    If charge assignments are chosen, for example,
    to be larger by a factor of $\sqrt 3$,
    then either the Maxwell term in the action 
    (\ref{eq:bulk action}) must be multiplied by a factor of 3,
    or else one must regard the boundary value of the bulk gauge
    field as equaling $\sqrt 3$ times the QFT gauge field
    (and the charge density in the bulk theory as equal to the QFT
    charge density divided by $\sqrt 3$),
    as was done in ref.~\cite{D'Hoker:2009mm}.
    \label{fn:xi}
    }
Setting to zero the variation of the action
with respect to the metric gives
the Einstein equation,
\begin{align}
    R_{KL} + (\Lambda - \half R) \, G_{KL} 
    = 2 L^2
    \left(
	F_{KM} F_L{}^M - \tfrac 14 G_{KL} \, F_{MN} F^{MN} 
    \right) ,
\label{eq:eins}
\end{align}
while varying the bulk gauge field
(with $F_{MN} \equiv \nabla_M A_N - \nabla_N A_M$)
gives the usual sourceless Maxwell equation,
$
    \nabla_K F^{KL} = 0
$.

A 5D Chern-Simons term,
$A \wedge F \wedge F$,
could be added to the action (\ref{eq:bulk action})
and would appear with a known coefficient
in a consistent truncation of 10D supergravity.
(See, for example, refs.~\cite{Chamblin:1999tk,D'Hoker:2009mm}.)
However, as stated above, in this paper we consider
solutions with non-zero chemical potential $\mu$ or
non-zero magnetic field $\B$,
but not both $\mu$ and $\B$ non-zero.
For such solutions, the Chern-Simons term makes no
contribution to the dynamics and hence may be neglected.

As usual in holography,
the expectation value $\langle T^{\alpha\beta}(x)\rangle$
of the stress-energy tensor is determined by the subleading
near-boundary behavior of the 5D metric $G_{MN}$.
The leading near-boundary behavior of the bulk gauge field $A_M$
will be fixed by our chosen external $U(1)$ gauge field
(\ref{eq:external A}),
while the
expectation value $\langle j^{\alpha}(x)\rangle$
of the $U(1)$ current density is determined by the subleading
near-boundary behavior of the bulk gauge field.
The precise relations will be shown below.

Following ref.~\cite{Chesler:2013lia},
we choose a coordinate ansatz, based on
generalized Eddington-Finklestein (EF) coordinates,
which is natural for gravitational infall problems.
The metric has the general form
\begin{equation}
    ds^2 = \frac{r^2}{L^2} \> g_{\alpha\beta}(x,r) \, dx^{\alpha}dx^{\beta}
    - 2 \, w_{\alpha}(x) \, dx^{\alpha}dr,
\label{eq:ansatz}
\end{equation}
where $r$ is the bulk radial coordinate and
$x \equiv \{ x^{\alpha} \}$, $\alpha = 0,{\cdots},3$,
denotes the four remaining spacetime coordinates.
The spacetime boundary lies at $r=\infty$;
the $\{ x^\alpha \}$ may be regarded as coordinates
on the spacetime boundary where the dual field theory ``lives''.
Curves of varying $r$, with $x$ held fixed,
are radially infalling null geodesics, affinely parameterized by $r$.
The one-form $\widetilde w \equiv w_\alpha \, dx^\alpha$
(which is assumed to be timelike) depends only on $x$, not on $r$.
These infalling coordinates remain regular across future null horizons.

The form of the ansatz (\ref{eq:ansatz}) remains invariant under
$r$-independent diffeomorphisms,
\begin{equation}
\label{xdiffeo}
    x^{\alpha} \rightarrow \bar{x}^{\alpha} \equiv f^{\alpha}(x) \,,
\end{equation}
as well as radial shifts (with arbitrary $x$ dependence), 
\begin{equation}
\label{rdiffeo}
    r \rightarrow \bar{r} \equiv r + \lambda(x) \,.
\end{equation}
We use the diffeomorphism freedom \eqref{xdiffeo} to transform the timelike
one-form $\widetilde w$ to the standard form $-dx^0$
(or $w_\alpha = -\delta_\alpha^0$).
Our procedure for dealing with the radial shift invariance \eqref{rdiffeo}
is discussed below in subsection~\ref{sec:horizon}.

We are interested in geometries which, at large $r$,
asymptotically approach (the Poincar\'e patch of) AdS$_5$.
This will be the case if
$
    g_{\alpha\beta}(x,r)
$
approaches
$
    \eta_{\alpha\beta}
$
as $r \to \infty$,
with $\eta_{\alpha\beta} \equiv \mathrm{diag}(-1,1,1,1)$
the usual Minkowski metric tensor.
Demanding that the metric and bulk gauge field
satisfy the Einstein-Maxwell equations,
one may derive the near-boundary asymptotic behavior of the fields.
Using radial gauge, $A_r = 0$, for the bulk gauge field,
and a suitable choice of the radial shift
\eqref{rdiffeo}
(which eliminates $O(1/r)$ terms in $g_{\alpha\beta}$),
one finds that for solutions of interest, 
the metric and gauge field have asymptotic expansions
of the form 
\begin{subequations}\label{eq:gAasymp}%
\begin{align}
    g_{\alpha\beta}(x,r) &\sim
    \eta_{\alpha\beta}
    + \left[ g_{\alpha\beta}^{(4)}(x)
	+ h_{\alpha\beta}^{(4)}(x) \, \ln \tfrac rL
    \right]
	(L^2/r)^4
    + O\big[ (L^2/r)^5 \big] \,,
\\[4pt]
    A_\alpha(x,r) &\sim A_\alpha^{\rm ext}(x)
    + A_\alpha^{(2)}(x) \, (L^2/r)^2 + O\big[ (L^2/r)^3 \big] .
\end{align}
\end{subequations}
The coefficient $h^{(4)}_{\alpha\beta}$ of
the logarithmic term in the metric
is only non-zero when there is an external EM field,
\begin{equation}
    h^{(4)}_{\alpha\beta}
    =
	F_{\alpha\nu} F_\beta{}^\nu
	-\tfrac 13 \, \eta_{\alpha\beta} \,
	(F_{\mu\nu} F^{\mu\nu} + F_{0\nu} F_0{}^\nu ) \,.
\label{eq:h_ab}
\end{equation}
For a constant magnetic field in the $x^3$ direction,
$
    \| h^{(4)}_{\alpha\beta} \|
    =
    \tfrac 13 \, \B^2 \> {\rm diag}(+2,\, +1,\, +1,\, -2)
$.
The subleading asymptotic coefficients
$g_{\alpha\beta}^{(4)}(x)$ and $A_\alpha^{(2)}(x)$
cannot be determined solely from a near-boundary analysis of the field equations,
and depend on the form of the solution throughout the bulk.
However, asymptotic analysis does show that
$
    \sum_{i=1}^3 \> g^{(4)}_{ii}
    = -\tfrac 13 \, F_{0\nu} F_0{}^\nu
$.
The subleading metric coefficients $g_{\alpha\beta}^{(4)}(x)$ and
$h_{\alpha\beta}^{(4)}(x)$
encode the expectation value of the SYM stress-energy tensor
\cite{D'Hoker:2009bc,deHaro:2000xn}.
The appropriate holographic relation is
\begin{equation}
    \langle T_{\mu\nu} \rangle
    =  \kappa \left\{
	\widetilde g^{(4)}_{\mu\nu}
	- \eta_{\mu\nu} \, \tr (\widetilde g^{(4)})
	+  [\ln(\mu L) + \mathcal C] \> \widetilde h_{\mu\nu}^{(4)} 
	\right\} ,
\label{eq:<T>}
\end{equation}
where%
\footnote
    {%
    In Fefferman-Graham (FG) coordinates, for which
    $
	ds^2 \equiv (L^2/\rho^2)
	    \big[
		\widetilde g_{\alpha\beta}(\tilde x,\rho) \,
		d\tilde x^\alpha \, d\tilde x^\beta + d\rho^2
	    \big]
    $,
    one has
    $
	\widetilde g_{\alpha\beta}(\tilde x,\rho)
	\sim
	\eta_{\alpha\beta} + 
	    \big[
	    \widetilde g^{(4)}_{\alpha\beta}(\tilde x)
	    + \widetilde h^{(4)}_{\alpha\beta} \, \ln \tfrac L \rho
	    \big]
	    \,\rho^4 
	+ O(\rho^6 \ln \rho)
    $
    as $\rho\to 0$.
    Eq.~\eqref{eq:EFtoFG} gives the relation between the
    subleading asymptotic metric coefficients 
    in our infalling EF coordinates and
    FG coordinates.
    }
\begin{align}
	\widetilde
	g^{(4)}_{\mu\nu}
	&\equiv
	g^{(4)}_{\mu\nu}
	+ \tfrac 14 \, \eta_{\mu\nu}
	    \left( g^{(4)}_{00} + \tfrac{1}{4} \, h^{(4)}_{00} \right) ,
\qquad
	\widetilde h_{\mu\nu}^{(4)}
	\equiv
	h_{\mu\nu}^{(4)} + \tfrac 14\,\eta_{\mu\nu} \, h^{(4)}_{00} \,,
\label{eq:EFtoFG}
\end{align}
$\kappa \equiv {L^3}/(4\pi G_N) = (\Nc^2-1)/(2\pi^2)$,
and $\mathcal C$ is an arbitrary 
renormalization-scheme dependent constant.%
\footnote
    {%
    To perform the required holographic renormalization
    one must add a counterterm depending logarithmically
    on the UV cutoff.
    (See, for example, refs.~\cite{deHaro:2000xn,Henningson:1998gx, Taylor:2000xw}.)
    As always, such a logarithmic counterterm comes with an
    inevitable finite ambiguity.
    }
We adopt a specific value,
\begin{equation}
    \mathcal{C}\equiv -\tfrac 14 \,,
\label{eq:C}
\end{equation}
which will make the subsequent explicit expression \eqref{eq:T00}
for the energy density as simple as possible.

Inserting expression (\ref{eq:h_ab}) into relation (\ref{eq:EFtoFG})
shows that $\widetilde h_{\alpha\beta}^{(4)}$, the coefficient of the
renormalization point dependent part of the holographic SYM stress-energy,
is proportional to the classical EM stress-energy tensor,
\begin{equation}
    \widetilde h^{(4)}_{\alpha\beta}
    =
	F_{\alpha\nu} F_\beta{}^\nu
	-\tfrac 14 \, \eta_{\alpha\beta} \, F_{\mu\nu} F^{\mu\nu} \, ,
\label{eq:h4tilde}
\end{equation}
or
$
    \| \widetilde h^{(4)}_{\alpha\beta} \|
    =
    \tfrac 12 \, \B^2 \> {\rm diag} (+1,\, +1,\, +1,\, -1)
$
for a constant magnetic field in the $x^3$ direction.
Using the above relations, one also finds that
$
    \tr(\widetilde g^{(4)}) = \tfrac 1{12} \, F_{\mu\nu}F^{\mu\nu}
$.
Since $\tilde h^{(4)}$ is traceless, the holographic relation (\ref{eq:<T>})
yields the stress-energy trace
\begin{equation}
    \langle T^\alpha{}_\alpha \rangle
    =
    -3 \kappa \, \tr (\widetilde g^{(4)})
    =
    -\tfrac 14 \, \kappa \, F_{\mu\nu} F^{\mu\nu} \,,
\end{equation}
or
$
    \langle T^\alpha{}_\alpha \rangle
    =
    -\tfrac 12 \, \kappa \, \B^2
$
in a constant magnetic field,
in agreement with
the earlier field theory result (\ref{eq:traceanom}).
Similarly,
the renormalization point dependence of the stress-energy (\ref{eq:<T>})
coincides with the QFT result (\ref{eq:TSYMRG}).%
\footnote
    {%
    As in ref.~\cite{D'Hoker:2009mm},
    one can also use a comparison of holographic and QFT evaluations
    of the $U(1)$ anomaly
    to confirm that the $U(1)$ current normalizations are consistent.
    }

Finally, the subleading asymptotic coefficient
$A_\alpha^{(2)}(x)$ for the bulk gauge field encodes
the $U(1)$ current density.
One finds
\begin{equation}
    \langle j_\nu \rangle
    = 2\, \kappa\, A^{(2)}_\nu \,.
\label{eq:<j>}
\end{equation}

\subsection{Symmetry specialization}

As noted earlier, we are interested in studying solutions of
Einstein-Maxwell theory which are spatially homogeneous.
This implies that all metric functions depend only on $x^0$ and $r$.
The arbitrary function $\lambda$ in the residual
radial shift diffeomorphism \eqref{rdiffeo}
will depend only on $x^0$.
Henceforth, for convenience, we will use $v$ as a synonym for $x^0$;
$v$ is a null time coordinate.
(In other words, $v = \rm const.$ surfaces are null slices of the geometry.)
At the boundary, $v$ coincides with the time $t$
of the dual field theory.

We also impose invariance
under $O(2)$ rotations in the $x^1$-$x^2$ plane.
This implies that only the $g_{00}$, $g_{03}$, $g_{33}$,
and $g_{11}=g_{22}$ components of $g_{\alpha\beta}$ are non-zero.
Our Einstein-Maxwell theory (without a Chern-Simons term)
is also invariant under spatial parity,
or $x^3 \to -x^3$ reflections,
and for simplicity we will also impose parity invariance.
This requires the vanishing of $g_{03}$.

For the bulk gauge field, the choice of radial gauge, $A_r = 0$,
plus our imposed symmetries imply that
\begin{equation}
    A_\alpha(x,r) = A_\alpha^{\rm ext}(x) - \phi(v,r) \, \delta^0_\alpha \,.
\label{eq:bulk A}
\end{equation}

The corresponding bulk field strength,
which is what appears in the field equations,
can have a constant ($x$ and $r$ independent) magnetic field plus a 
radial electric field,
\begin{equation}
    F_{12}(x,r) = \B \,,\quad
    F_{0r}(x,r) = \partial_r \phi(v,r) \equiv -\E(v,r) \,,
\label{eq:field strength}
\end{equation}
with all other components vanishing.

As in ref.~\cite{Chesler:2013lia}, it is convenient to
rename the non-vanishing metric components as
\begin{equation}
    \frac{r^2}{L^2} \, g_{00} \equiv -2 A \,,\quad
    \frac{r^2}{L^2} \, g_{11} =
    \frac{r^2}{L^2} \, g_{22} \equiv \Sigma^2 \, e^{B} ,\quad
    \frac{r^2}{L^2} \, g_{33} \equiv \Sigma^2 \, e^{-2B} ,
\end{equation}
where $A$, $B$, and $\Sigma$ are functions of $v$ and $r$.
The resulting line element is
\begin{equation}
    ds^2 = 
    2 dv \left[ dr - A(v,r) \, dv \right]
    + \Sigma(v,r)^2
	\left[e^{B(v,r)} (dx^2 + dy^2) + e^{-2 B(v,r)} dz^2\right] .
\label{eq:ansatz2}
\end{equation}
Henceforth,
$A$ will always denote the metric function multiplying $dv^2$ 
(times $-1/2$), not the bulk gauge field.
The function $\Sigma$ is the spatial scale factor
(with $\Sigma^3\, dx \, dy \, dz$ the spatial volume element),
while $B$ characterizes the spatial anisotropy
(which should not be confused with
the magnetic field amplitude $\B$).

The radial derivative $\partial_r$ is a directional derivative
along infalling radial null geodesics.
It proves convenient to define a corresponding directional
derivative along outward radial null geodesics,
\begin{equation}
\label{dplus}
    d_+ \equiv \partial_v + A(v,r) \, \partial_r \,.
\end{equation}

The field equations which result from varying the action
(\ref{eq:bulk action}),
inserting the above symmetry specializations,
and re-expressing $v$-derivatives in terms of the $d_+$ modified
time derivative \eqref{dplus},
take a remarkably compact form.
The Einstein equations are:%
\begin{subequations}%
\label{eineqns}%
\begin{align}
\label{sigmaeqn}
     \Sigma'' + \half (B')^2 \, \Sigma & = 0 \,,
\\[2pt]
\label{Aeqn} 
    A''  
	-6(\Sigma'/\Sigma^2) \, d_+\Sigma
	+ \tfrac{3}{2} \, B' d_+B
	&=
	+ \tfrac{5}{3} \B^2 L^2  \, e^{-2 B} \, \Sigma^{-4}
	+ \tfrac{7}{3} \, \E^2 L^2 
	- 2/L^2
	\,,
\\[2pt]
\label{Bdoteqn}
     (d_+ B)'
	 + \tfrac 32 (\Sigma'/\Sigma) \, d_+B
	 + \tfrac 32 B' \, (d_+\Sigma)/\Sigma 
	 & =
	 - \tfrac 23 \B^2 L^2 \, e^{-2 B} \, \Sigma^{-4} 
	 \,,
\\[2pt]
\label{sigmadoteqn} 
    (d_+\Sigma)'/\Sigma
	+ 2(\Sigma'/\Sigma^2) \, d_+\Sigma
	&=
	- \tfrac 13 \B^2 L^2 \, e^{-2 B} \, \Sigma^{-4}
	- \tfrac 13 \E^2 L^2 
	+ 2 /L^2
	\,,
\\[2pt]
\label{sigmaddoteqn} 
    d_+ (d_+\Sigma)
	- A' \, (d_+\Sigma) + \tfrac{1}{2}\Sigma \, (d_+ B)^2
	&= 0 \,,
\end{align}
\end{subequations}
where primes denote radial derivatives, $h' \equiv \partial_r h$.
As discussed in ref.~\cite{Chesler:2013lia},
the anisotropy function $B$ encodes the essential propagating
degrees of freedom.
The functions $\Sigma$ and $A$ may be regarded as
auxiliary fields, determined by solving eqns.~(\ref{sigmaeqn})
and (\ref{Aeqn}) using data on a single time slice.  
Information about the time evolution of $B$ is contained in equation~\eqref{Bdoteqn}.
Equations~\eqref{sigmadoteqn} and \eqref{sigmaddoteqn} may
be viewed as boundary value constraints ---
if they hold at one value of $r$,
then the other equations ensure that these equations hold
at all values of $r$.

Maxwell's equations reduce to the statements that neither
the magnetic field $\B$,
nor the radial electric flux density $\E \, \Sigma^3$,
have any radial or temporal variation.
In other words,
$\B = \rm const.$,
as already indicated in (\ref{eq:field strength}),
and
\begin{equation}
    \E(v,r) = \rho \, L \, \Sigma^{-3}(v,r) \,,
\label{eq:radial E}
\end{equation}
for some constant $\rho$ which,
from eqs.~(\ref{eq:<j>})--(\ref{eq:field strength})
plus (\ref{eq:leadingasymp}) below,
one sees is precisely the $U(1)$ charge density
(rescaled by $\kappa$), 
\begin{equation}
		\label{eq:j0}
    \langle j^0 \rangle = \kappa \, \rho \,.
\end{equation}
The form (\ref{eq:radial E}) of the radial electric field
simply reflects Gauss' law in 4+1 dimensions, combined with
charge conservation and spatial translation invariance,
which imply that $\rho$ cannot have any temporal or spatial variation.

The bulk gauge field $A_M$ does not appear in the
field equations (except via the field strength),
but one may choose to regard $A_M$ as satisfying
the radial gauge condition, $A_r = 0$, plus
the condition that the time component $A_v$
vanish at the horizon.
This fixes the residual $r$-independent gauge freedom
which remains after imposing radial gauge.
With these choices,
the chemical potential $\mu$ is the
boundary value of $A_v$ in the late time
($v \to \infty$) equilibrium limit.
Equivalently (in radial gauge),
the chemical potential $\mu$ equals the
difference between the boundary and horizon
values of $A_v$, in the equilibrium
geometry.
This coincides with the
line integral of the radial electric field
from horizon to boundary,
\begin{equation}
    \mu = \lim_{v \to \infty} A_v \big|_{\rh}^\infty
    = \int_{\rh}^\infty dr \> \E (\infty,r) \,,
\label{eq:mu}
\end{equation}
which gives the work needed to move a unit charge
from the boundary to the horizon.
As usual, the charge density and the chemical potential
are thermodynamically conjugate.
One may consider the chemical potential $\mu$
to be a function of the (rescaled) charge density $\rho$, or vice-versa.
The choice of perspective (``canonical'' vs.\ ``grand canonical'')
has no bearing on the dynamics.

\subsection{Asymptotic analysis}\label{sec:asymptotics}

Asymptotic analysis of these equations is straightforward.
We impose a flat boundary geometry with the requirement that
$
    \lim_{r\to\infty} \, g_{\alpha\beta}(x,r)
    =
    \eta_{\alpha\beta}
$,
implying
\begin{equation}
    \lim_{r\to\infty} (L/r)^2 \, A(v,r) = \half \,,
    \quad
    \lim_{r\to\infty} (L/r) \, \Sigma(v,r) = 1 \,,
    \quad
    \lim_{r\to\infty} B(v,r) = 0 \,,
\label{eq:leadingasymp}
\end{equation}
for our renamed metric functions.
Solutions to Einstein's equations \eqref{eineqns}
with this leading behavior
may be systematically expanded in integer powers of $1/r$
and (for non-zero magnetic field) logarithms of $r$.
One finds:
\begin{subequations}%
\label{metricexpansions}%
\begin{align}
\label{sigmaexpn}
    \Sigma(v,r) &\sim
    L^{-1} [r + \lambda(v)]
    + \mathcal{O}[(L/r)^7\ln^2 \tfrac rL ],
\\[5pt]
\label{Aexpn} 
    A(v,r) &\sim
	\half L^{-2} [r + \lambda(v)]^2
	- \partial_v \lambda(v)
\nonumber\\ &\quad {}
	+ L^4 \left[ a_4 - \tfrac 13 \B^2 \ln \tfrac rL \right] (L/r)^2
\nonumber\\ &\quad{}
    - L^3 \left[ 2 a_4 \, \lambda(v)
    +\tfrac 13 \B^2 \, \lambda(v) \, (1-2\ln \tfrac rL ) \right] (L/r)^3
    + \mathcal{O}[(L/r)^4\ln \tfrac rL ],
\\[5pt]
\label{Bexpn} 
    B(v,r) &\sim
    L^4 \left[ b_4(v) + \tfrac 13 \B^2 \ln \tfrac rL \right] (L/r)^4
\nonumber\\ &\quad {}
    + L^3 \left[
	    L^2 \, \partial_v b_4(v) - 4 b_4(v) \, \lambda(v)
	    + \tfrac 13 \B^2 \, \lambda(v) \, (1-4\ln \tfrac rL )
	\right] (L/r)^5
\nonumber\\ &\quad {}
	+ \mathcal{O}[(L/r)^6\ln \tfrac rL ],
\end{align}
\end{subequations}
The constant $a_4$ and the function $b_4(v)$
cannot be determined just using asymptotic analysis,
and the radial shift $\lambda(v)$ is completely arbitrary.
The coefficient $a_4$
encodes the energy density which, due to homogeneity,
cannot vary in time,
while $b_4(v)$ encodes the anisotropy in the spatial stress.
Using the holographic relation \eqref{eq:<T>} and our
convention \eqref{eq:C} for defining the stress-energy tensor,
one finds
\begin{subequations}\label{eq:T}%
\begin{align}
    \langle T^{00} \rangle 
    &= \kappa \left( -\tfrac{3}{2} a_4
	    + \half \, \B^2 \, \ln \mu L
	    \right) ,
\label{eq:T00}
\\
    \langle T^{11} \rangle =
    \langle T^{22} \rangle
    &= \kappa \left(
		-\tfrac{1}{2} a_4 + b_4
		- \tfrac{1}{4}\B^2
		+ \half \, \B^2 \, \ln \mu L
	    \right) ,
\label{eq:T11}
\\
    \langle T^{33} \rangle
    &= \kappa \left( -\tfrac{1}{2} a_4 - 2 b_4
		- \half \, \B^2 \, \ln \mu L
		\right) .
\label{eq:T33}
\end{align}
\end{subequations}

\subsection	{Scaling relations}\label{sec:scaling}

Consider independent rescaling of the boundary and radial
coordinates,
\begin{equation}
    x \equiv \alpha \, \widetilde x \,,\qquad
    r \equiv \alpha^{-1} \gamma^2 \, \widetilde r \,,
\label{eq:scaling1}
\end{equation}
with $\alpha$ and $\gamma$ arbitrary positive numbers.
If the metric functions $\{A, \Sigma, B\}$ satisfy
the Einstein equations (\ref{eineqns}),
with asymptotic behavior (\ref{eq:leadingasymp}),
then the rescaled metric functions
\begin{subequations}\label{eq:scaling2}%
\begin{align}
    \widetilde B(\widetilde x,\widetilde r) 
    &\equiv
    B(x(\widetilde x),r(\widetilde r))  \,,
\\
    \widetilde\Sigma(\widetilde x, \widetilde r)
    &\equiv
    (\alpha/\gamma) \, \Sigma(x(\widetilde x),r(\widetilde r)) \,,
\\
    \widetilde A(\widetilde x, \widetilde r) \,
    &\equiv
    (\alpha/\gamma)^2
    A(x(\widetilde x),r(\widetilde r)) \,,
\end{align}%
\end{subequations}
also satisfy the Einstein equations (and our asymptotic conditions)
with rescaled parameters%
\begin{align}
    \widetilde L &\equiv \gamma^{-1} \, L \,,\qquad
    \widetilde \B \equiv \alpha^2 \, \B \,,\qquad
    \widetilde \rho \equiv \alpha^3 \, \rho \,.
\label{eq:scaling3}
\end{align}
The subleading asymptotic coefficients $a_4$ and $b_4$ become
\begin{equation}
    \widetilde a_4
    \equiv 
    \alpha^4 \left[ a_4 - \tfrac 13 \, \B^2 \ln (\gamma/\alpha) \right],
    \quad
    \widetilde b_4
    \equiv 
    \alpha^4 \left[ b_4 + \tfrac 13 \, \B^2 \ln (\gamma/\alpha) \right] .
\end{equation}
Using the holographic relation (\ref{eq:T}) for the stress-energy expectation,
one finds that
\begin{equation}
    \widetilde T^{\mu\nu}(\widetilde \mu)
    =
    \alpha^4 \, T^{\mu\nu}(\mu) \,,
\end{equation}
with a rescaled renormalization point
$
    \widetilde \mu \equiv \alpha \, \mu
$.

If $\alpha = \gamma$, then these transformations are just a trivial
rescaling of all quantities according to their dimension.
But transformations with $\alpha \ne \gamma$ are non-trivial
and scale bulk and boundary quantities by different amounts.
In particular, transformations with $\alpha = 1$ but $\gamma \ne 1$
rescale the AdS curvature scale $L$ without affecting the
boundary coordinates or boundary parameters ($\B$, $\rho$, or $\mu$),
showing that the value of $L$ has no physical significance
(in the large $\Nc$, large $\lambda$ limit for which classical
gravity provides the dual description).
This illustrates, explicitly, the independence of the boundary field theory
on the AdS curvature scale $L$.

\subsection	{Apparent horizon} \label{sec:horizon} 

\begin{figure}
\centering
\suck[width=0.38\textwidth]{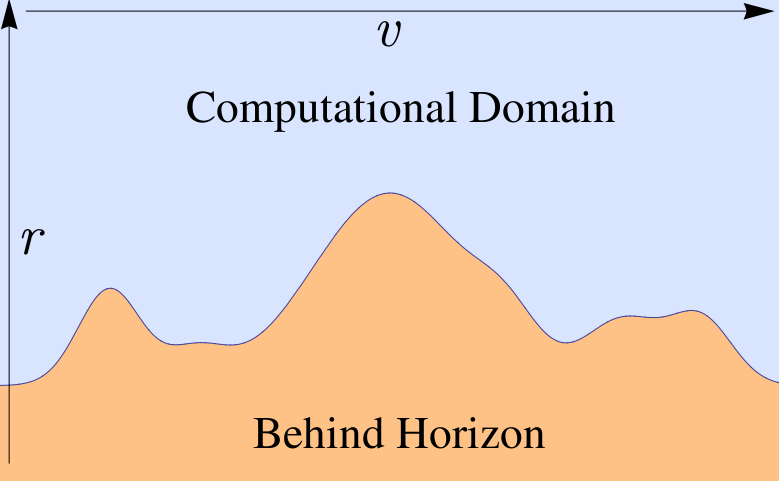}
\quad
\quad
\suck[width=0.38\textwidth]{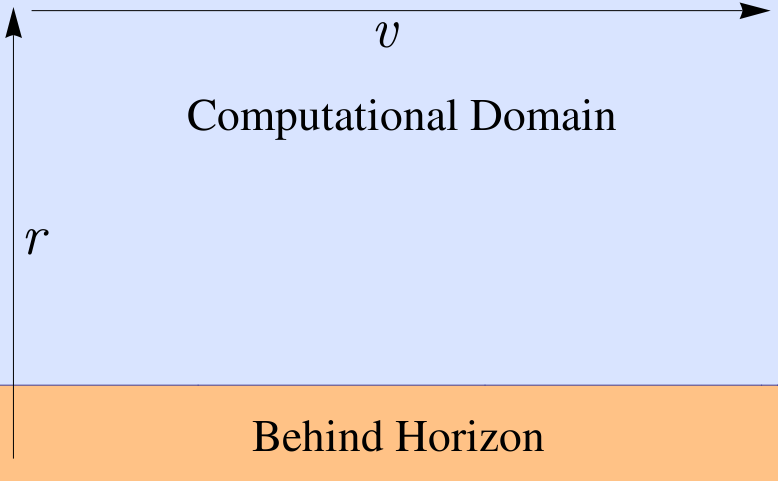}
\caption
    {%
    With a generic choice of the radial shift $\lambda(v)$ (left panel),
    the radial position of the horizon will change with time.
    It may be kept fixed (right panel)
    with a suitable choice of $\lambda(v)$.%
    \label{fig:compdomain}
    }
\end{figure}

With a non-zero homogeneous energy density,
the dual geometries of interest will have an apparent horizon
at some radial position, $r = \rh(v)$ \cite{Chesler:2013lia}.
Since we are investigating non-equilibrium dynamics,
one might expect the horizon position to change significantly
before ultimately settling down as the system equilibrates.
However,
as illustrated in fig.~\ref{fig:compdomain}
it is possible, and very convenient, to use the
residual radial shift diffeomorphism freedom
\eqref{rdiffeo}
to place the apparent horizon at a fixed radial
position,
\begin{equation}
    \rh(v) \equiv \rhbar \,.
\end{equation}

A short exercise \cite{Chesler:2013lia}
shows that the condition for an apparent horizon
to be present at $r = \rhbar$ is that this location
be a zero of the modified time derivative of the
spatial scale factor,
\begin{equation}
    d_+\Sigma \big|_{r=\rhbar} = 0 \,.
\label{eq:horizon}
\end{equation}
This condition serves to fix the radial shift $\lambda(v)$.
It is convenient to regard this condition as a combination
of a constraint on initial data
(implemented by finding the radial shift $\lambda(v_0)$
which is needed to satisfy \eqref{eq:horizon} at 
some initial time $v_0$)
together with the requirement that the horizon
be time-independent, $\partial \rh/\partial v = 0$,
which requires that the time derivative of $d_+\Sigma$
vanish at the apparent horizon.
Evaluating this condition, and using the
Einstein equations (\ref{sigmadoteqn}, \ref{sigmaddoteqn}) to simplify
the result, determines
the value of the metric function $A$ at the horizon,
\begin{equation}
    A\big|_{r=\rhbar} = -\tfrac 14 \, L^2 \, (d_+B)^2 \,.
\label{eq:Ahorizon}
\end{equation}
For the metric to be non-singular on and outside
the apparent horizon, 
the spatial scale factor $\Sigma$ must
be non-vanishing for $r \ge \rhbar$.

\subsection	{Equilibrium solutions} \label{sec:eqsolns}

Given some initial non-equilibrium state of the system,
the dynamical evolution should asymptotically approach
a thermal equilibrium state.
In the gravitational description, this implies that the geometry
should, at late times, approach some static black brane solution.
The specific black brane solution will depend on the values
of the conserved energy and charge densities in the
chosen initial state, and on the value of the background
magnetic field.

\subsubsection*{Schwarzschild}

For initial states with vanishing charge density and magnetic field,
the bulk geometry will equilibrate to the 5D
AdS-Schwarzschild black brane solution.
A standard form of this metric is
\begin{equation}
    ds^2 = - U(\rtilde) \, dt^2 + \frac{d\rtilde^2}{U(\rtilde)}
    + \frac{\rtilde^2}{L^2} \, (dx^i)^2
\label{eq:AdSBH}
\end{equation} 
($i = 1,2,3$), with
\begin{equation}
\label{eq:USBB}
    U(\tilde{r}) \equiv \frac{\rtilde^2}{L^2} - \frac{m \, L^2}{\rtilde^2} \,.
\end{equation}
The radial coordinate $\tilde{r}$ should not be confused with our
Eddington-Finklestein coordinate~$r$.
The zero of $U(\rtilde)$ determines the horizon location,
\begin{equation}
    \rtilde_{\rm h} = m^{1/4} L \,,
\end{equation}
and the horizon temperature
[given by $(2\pi)^{-1}$ times the surface gravity at the horizon]
is proportional to the horizon radius,
\begin{equation}
    \pi \Th = \rtilde_{\rm h} L^{-2} = m^{1/4}/L \,.
\end{equation}
In our infalling EF coordinates \eqref{eq:ansatz2},
this AdS-Schwarzschild solution is described by
\begin{equation}
    \Sigma(r)
    = (r {+} \lambda)/L  \,,
\qquad
    A(r) = \half \Sigma(r)^2 - \half m \, \Sigma(r)^{-2}\,,
\qquad
    B(r) = 0 \,.
\label{eq:AdS-BH}
\end{equation}
Using the holographic relation \eqref{eq:T},
one sees that the parameter $m$ is related
to the (rescaled) equilibrium energy density
$\varepsilon \equiv \langle T^{00} \rangle/\kappa$
via
\begin{equation}
    \varepsilon 
    = \tfrac 34 \, m \, L^{-4}\,.
\label{eq:BHeden}
\end{equation}

\subsubsection*{Reissner-Nordstrom}

If the initial state has a non-zero charge density but vanishing magnetic field,
then the bulk geometry will equilibrate to a 5D Reissner-Nordstrom
(RN) black brane~\cite{Chamblin:1999tk}.
This metric may be written in the form (\ref{eq:AdSBH}), with
\begin{equation}
\label{eq:URN}
    U(\rtilde) \equiv
    \frac{\rtilde^2}{L^2}
    - m \, \frac{L^2}{\rtilde^2}
    + \tfrac 13 \, (\rho L^3)^2 \, \frac{L^4}{\rtilde^4} \,.
\end{equation}
The charge density $\rho$ of the Reissner-Nordstrom brane
is bounded from above by the extremal charge density $\rhomax$,
given by
\begin{equation}
    \label{eq:rhoext}
    (\rhomax L^3)^4 = \tfrac{4}{3} \, m^3 \,.
\end{equation}
The relation (\ref{eq:BHeden}) between the energy density and the
mass parameter $m$ is unchanged.
Hence, the extremal charge density
$\rhomax = \frac 43 \, \varepsilon^{3/4}$.

It is convenient to express $\rho$ in terms
of the fraction $x$ of the extremal charge density,
\begin{equation}
    x \equiv \rho/\rhomax \,.
\end{equation}
The horizon radius $\rtilde_{\rm h}$ is given by the outermost positive
root of $U(\rtilde)$; explicitly,
\begin{equation}
    \rtilde_{\rm h}/L = (\tfrac 13 m)^{1/4}
    \left[
	(-x^2 +i \sqrt{1-x^4})^{1/3} +
	(-x^2 -i \sqrt{1-x^4})^{1/3}
    \right]^{1/2} .
\label{eq:rh}
\end{equation}
The horizon radius (divided by $L$)
varies from $m^{1/4}$ down to $(\frac 13 m)^{1/4}$ as $x$ varies from 0 to 1.
The horizon temperature $\Th$ is given by
\begin{equation}
    \pi \Th L =
	(\rtilde_{\rm h} /L)
	\left[ 1 - (\tfrac 13 m)^{3/2} \, x^2 \, (L/\rtilde_{\rm h})^6 \right] .
\label{eq:RNtemp}
\end{equation}
The horizon temperature decreases with increasing charge density,
and vanishes as the charge density approaches
the extremal value (or $x \to 1$).

\begin{figure}
\centering
\suck[width=0.44 \textwidth]{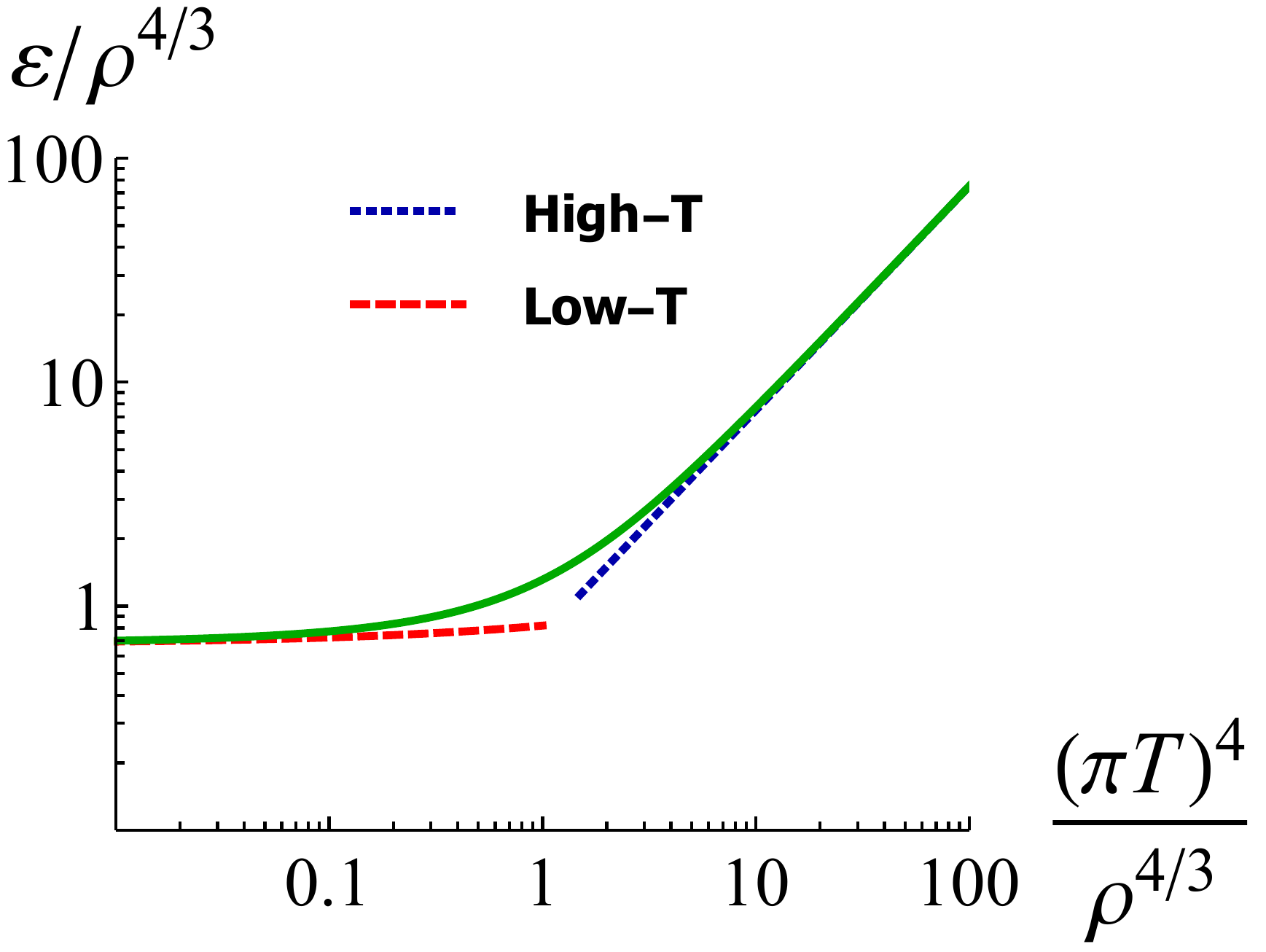}
\caption
    {%
    The one parameter family of non-extremal equilibrium 
    Reissner-Nordstrom charged black brane solutions (solid line)
    in the plane of
    $\varepsilon/|\rho|^{4/3}$ vs.\ $(\pi T)^4/|\rho|^{4/3}$.
    Also shown are the high and low temperature asymptotic forms (dashed lines).
    In the high temperature regime, $\pi T \gg \rho^{1/3}$,
    the curve approaches the Schwarzschild result,
    $\varepsilon = \tfrac 34(\pi T)^4$.
    In the low temperature (or near extremal) regime, $\pi T \ll \rho^{1/3}$,
    the charge density
    $
	\rho \sim
	\rhomax
	\big[
	    1 - \tfrac {1}{8}(\tfrac 34)^{-2/3} (\pi T)^2 \rhomax^{-2/3}
	\big]
    $
    and
    $
	\varepsilon/\rho^{4/3} \sim
	(\tfrac 34)^{4/3} + \tfrac 18 \, (\tfrac 34)^{-1/3} \, (\pi T)^2 \rho^{-2/3} 
    $.
    \label{fig:oneparamRN}
    }
\end{figure}

From the perspective of the dual field theory,
for any given value of the charge density
there is a lower bound on the energy density,
$\varepsilon_{\rm min}(\rho)$,
which must be a monotonically increasing (and convex)
function of $\rho$.
This implies that for any given value of the energy density,
there will be a maximum charge density, corresponding to
a ground state with vanishing temperature.
The equilibrium chemical potential $\mu$,
thermodynamically conjugate to the charge density $\rho$,
is given by 
\begin{equation}
    \mu = \half \, \rho \, (L^2/\rtilde_{\rm h})^2 \,.
\end{equation}
Physically distinct non-extremal solutions may be labeled by the value
of one dimensionless ratio such as
$\varepsilon/|\rho|^{4/3}$
[or $(\pi T)^4/|\rho|^{4/3}$ or $\varepsilon/(\pi T)^4$, etc.].
Figure~\ref{fig:oneparamRN} shows a log-log plot of the curve representing
these solutions in the plane of $\varepsilon/|\rho|^{4/3}$ and $(\pi T)^4/|\rho|^{4/3}$ .

In our infalling EF coordinates, the RN black-brane solution is described by
\begin{equation}
    \Sigma(r)
    = (r {+} \lambda)/L  \,,
\qquad
    A(r)
    = \half \Sigma(r)^2 - \half m \, \Sigma(r)^{-2}
    + \tfrac 16 \rho^2 L^6 \, \Sigma(r)^{-4}\,,
\label{eq:RNeqsolns}
\end{equation}
and $B(r) = 0$.

\subsubsection*{Magnetic branes}

When the magnetic field is non-zero, the bulk geometry
will equilibrate to a stationary magnetic black brane solution.
These solutions are not known analytically, but have been studied
numerically \cite{D'Hoker:2009mm,D'Hoker:2009bc}.
In our infalling coordinates,
the solutions satisfy the static specialization of
eqs.~\eqref{eineqns}.%
\footnote
    {%
    The resulting equations may be written explicitly as:
    \begin{subequations}\label{magbrane}%
    \begin{align}
	 \big( A' \Sigma^3 \big)' \,\Sigma^{-3}
	&=
	+\tfrac 23 \B^2 L^2 e^{-2B} \,\Sigma^{-4}
	+\tfrac 43 \E^2 L^2 
	+ 4 /L^2 \,,
    \\
	 \big( A \Sigma' \Sigma^2 \big)' \,\Sigma^{-3}
	&=
	-\tfrac 13 \B^2 L^2 e^{-2B} \,\Sigma^{-4}
	-\tfrac 13 \E^2 L^2 
	+ 2 /L^2 \,,
    \\
	 \big( A B' \Sigma^3 \big)' \,\Sigma^{-3}
	&=
	-\tfrac 23 \B^2 L^2 e^{-2B} \,\Sigma^{-4} \,.
    \end{align}
    \end{subequations}
    }
The near-boundary behavior of asymptotically AdS$_5$ solutions
is given by the expansions
\eqref{metricexpansions} (but with no time dependence).

The extremal, zero-temperature magnetic brane solution interpolates smoothly
between AdS$_3 \times \mathbb R^2$ near the horizon and AdS$_5$ near the boundary.
In our infalling coordinates, a series in fractional powers of
$\delta r \equiv r {-} \rh$ describes deviations from
the AdS$_3 \times \mathbb R^2$ geometry near the horizon,
\begin{subequations}%
\begin{align}
    A(r) &= \tfrac 32 (\delta r/L)^2
	\left[ 1 + \eta \, \delta r^\gamma
	+ O\big( \eta^2 \, \delta r^{2\gamma} \big) \right],
\\
    \Sigma(r) &= (\B L \, \delta r)^{1/3}
	\left[ 1 - \tfrac 17 (3{+}\gamma) \, \eta \, \delta r^\gamma
	+ O\big( \eta^2 \, \delta r^{2\gamma} \big) \right],
\\
    B(r) &= -\tfrac 16 \ln [27 \, \delta r^4 /(\B^{2} L^{8})]
	- \tfrac 1{14} (13 {+} 2\gamma) \, \eta \, \delta r^\gamma
	+ O\big( \eta^2 \, \delta r^{2\gamma} \big) \,,
\end{align}
\end{subequations}
with $\gamma \equiv -1 + \tfrac 13 \sqrt{57}$.
The constant $\eta$ cannot be determined from a near-horizon
analysis and must be suitably adjusted after integrating
eqs.~\eqref{magbrane} to obtain the desired boundary geometry.
There is a single extremal magnetic brane solution, modulo
the rescaling transformations (\ref{eq:scaling1})-(\ref{eq:scaling3})
(which relate solutions with any non-zero values of the magnetic field $\B$
and curvature scale $L$),
and radial shift diffeomorphisms (\ref{rdiffeo}).

For non-extremal solutions
(with non-zero $\B$ but vanishing $\rho$),
metric functions near the horizon
have power series expansions
in $\delta r \equiv r {-} \rh$ of the form
\begin{subequations}%
\label{eq:maghorexp}%
\begin{align}
    A(r) &= a_0 \, \delta r \, L^{-2}
	    \left[
		1
		- (1 - \tfrac 56 \, \B^2 L^4 s_0^{-4} ) \, a_0^{-1} \delta r
		+ O(\delta r^2)
	    \right],
\\
    \Sigma(r) &= s_0/\beta
	    \left[
		1
		+ (2 - \tfrac 13 \, \B^2 L^4 s_0^{-4}) \, a_0^{-1} \, \delta r 
		+ O(\delta r^2)
	    \right],
\\
    B(r) &=
	    2 \ln \beta
	    -\tfrac 23 \, \B^2 L^4 s_0^{-4} \, a_0^{-1} \, \delta r
		+ O(\delta r^2) \,.
\end{align}
\end{subequations}
The coefficient $a_0$ is proportional to the horizon temperature,
\begin{equation}
    T = a_0 / (2 \pi L^2) \,.
\end{equation}
The other two undetermined constants, $s_0$ and $\beta$,
which control the horizon values of the spatial scale factor
and the anisotropy function,
must be suitably adjusted after integrating eqs.~(\ref{magbrane})
to select solutions which have
the desired near-boundary behavior
(with an isotropic boundary metric).
If $\B^2 \ll T^4$, then the resulting magnetic brane geometry is a
small perturbation away from the Schwarzschild black brane (\ref{eq:AdS-BH}),
while if $\B^2 \gg T^4$ then the geometry may be regarded
as interpolating between the BTZ black brane ($\times \, \mathbb R^2$)
near the horizon and AdS$_5$ near the boundary \cite{D'Hoker:2009mm}.

\begin{figure}
\centering
\suck[width=0.44\textwidth]{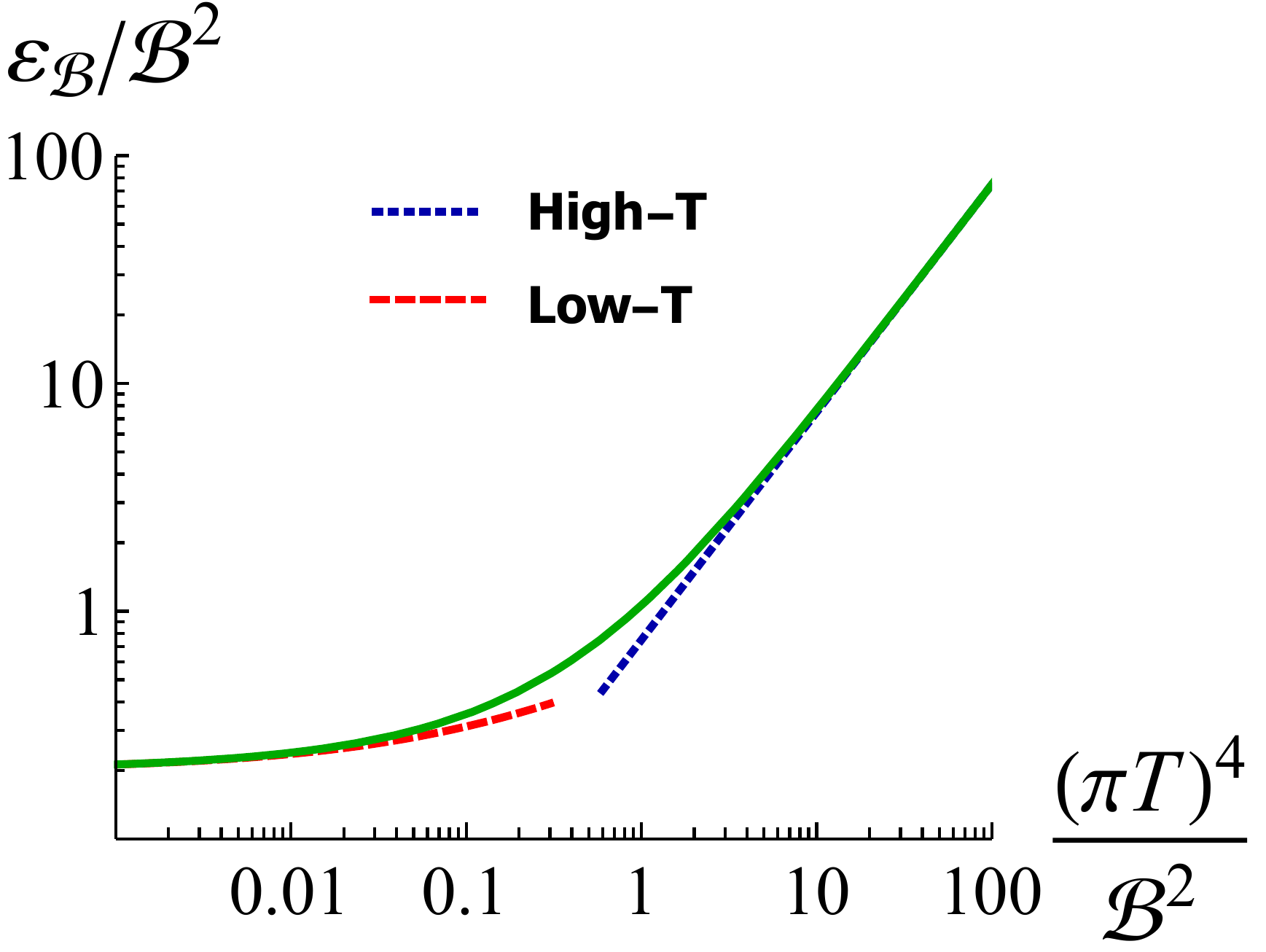}
\hfill
\suck[width=0.44\textwidth]{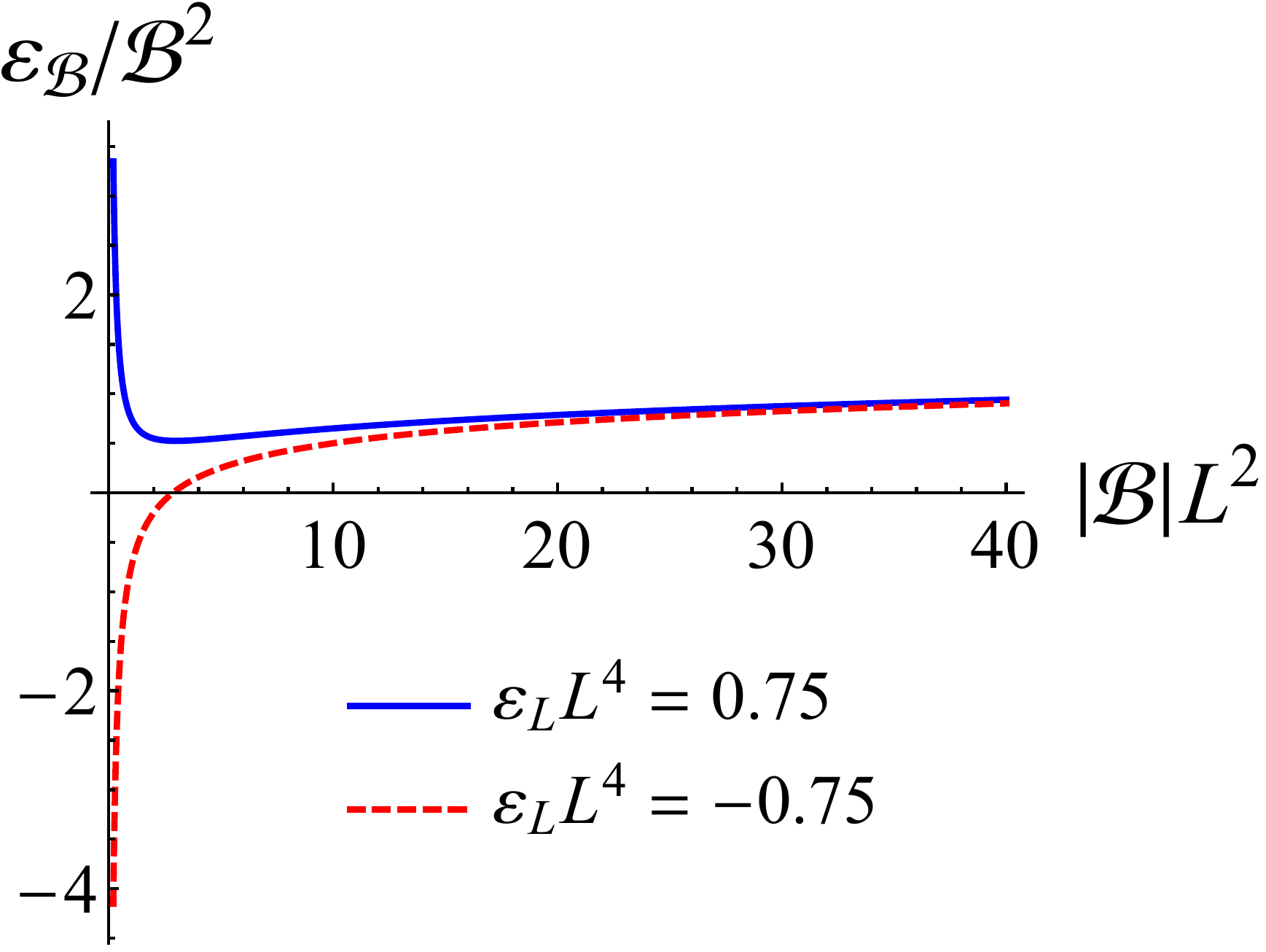}
\caption
    {%
    Left:
    The one parameter family of non-extremal equilibrium 
    magnetic brane solutions in the plane of
    $\epsB/\B^2$ vs.\ $(\pi T)^4/\B^2$.
    Also shown are the high and low temperature asymptotic forms (dashed lines).
    For high temperatures, $\pi T \gg |\B|^{1/2}$,
    the curve approaches the Schwarzschild result
    $\varepsilon = \tfrac 34(\pi T)^4$.
    For low temperatures, $\pi T \ll |\B|^{1/2}$,
    the form
    $
	\epsB/\B^2 \sim c_1 + c_2 \, (\pi T)^2 / |\B|
    $
    provides a good fit to our data for $c_1 = 0.35$ and $c_2 = 0.20$.
    Right: 
    The relation between the intrinsic parameter
    $\epsB/\B^2$ labeling magnetic brane solutions
    and the value of the magnetic field in curvature scale units, $|\B|L^2$,
    for two different fixed values of the curvature scale energy density,
    $\epsL L^4= \pm 0.75 $.
    \label{fig:oneparamMAG}
    }
\end{figure}

There is a one parameter family of non-extremal solutions,
modulo the rescaling transformations (\ref{eq:scaling1})-(\ref{eq:scaling3})
(and radial shift diffeomorphisms).
Physically distinct solutions may be labeled by the value
of the dimensionless ratio $\epsB/\B^2$
[or $(\pi T)^4/\B^2$
or $\epsB/(\pi T)^4$, etc.].
The left panel of
figure~\ref{fig:oneparamMAG} shows a log-log plot of our
numerically determined curve representing these solutions
in the plane of $\epsB/\B^2$ and $(\pi T)^4/\B^2$. 
Extrapolating our lowest temperature numerical results to 
zero temperature, we find estimates of
\begin{equation}
    c_0 \approx 0.18 \,,\qquad
    \B^*(\mu) \approx 2.0 \, \mu^2 \,,
\label{eq:c_0}
\end{equation}
for the coefficient $c_0$
or the equivalent fiducial scale $\B^*$
defined by eqs.~(\ref{eq:magenergy}) and (\ref{eq:B*}).

If one chooses to measure energy density and magnetic field in
units set by the curvature scale $L$, then one may traverse the
one-parameter family of magnetic brane solutions by varying $|\B|L^2$ for
a fixed value of $\epsL L^4$ (or vice versa).
The holographic relation (\ref{eq:T}) shows that these curvature
scale dependent quantities are related to the
intrinsic dimensionless parameter $\epsB/\B^2$
via
\begin{equation}
    \frac{\epsB}{\B^2}
    = \frac{\epsL \, L^4 }{ (|\B| L^2)^2}
    + \tfrac 14 \ln(|\B|L^2) \,.
		\label{eq:MAGepsrel}
\end{equation}
This relation between $\epsB/\B^2$ and $|\B|L^2$,
for two different fixed values of the curvature scale energy density
$\epsL L^4 = \pm 0.75$,
is plotted in the right panel of fig.~\ref{fig:oneparamMAG}.
Note that two different values of $|\B|L^2$ yield the same value
of $\epsB/\B^2$ (and hence describe the same
physical solution) when $\epsL > 0$.

\begin{figure}
\centering
\!\suck[width=0.50\textwidth]{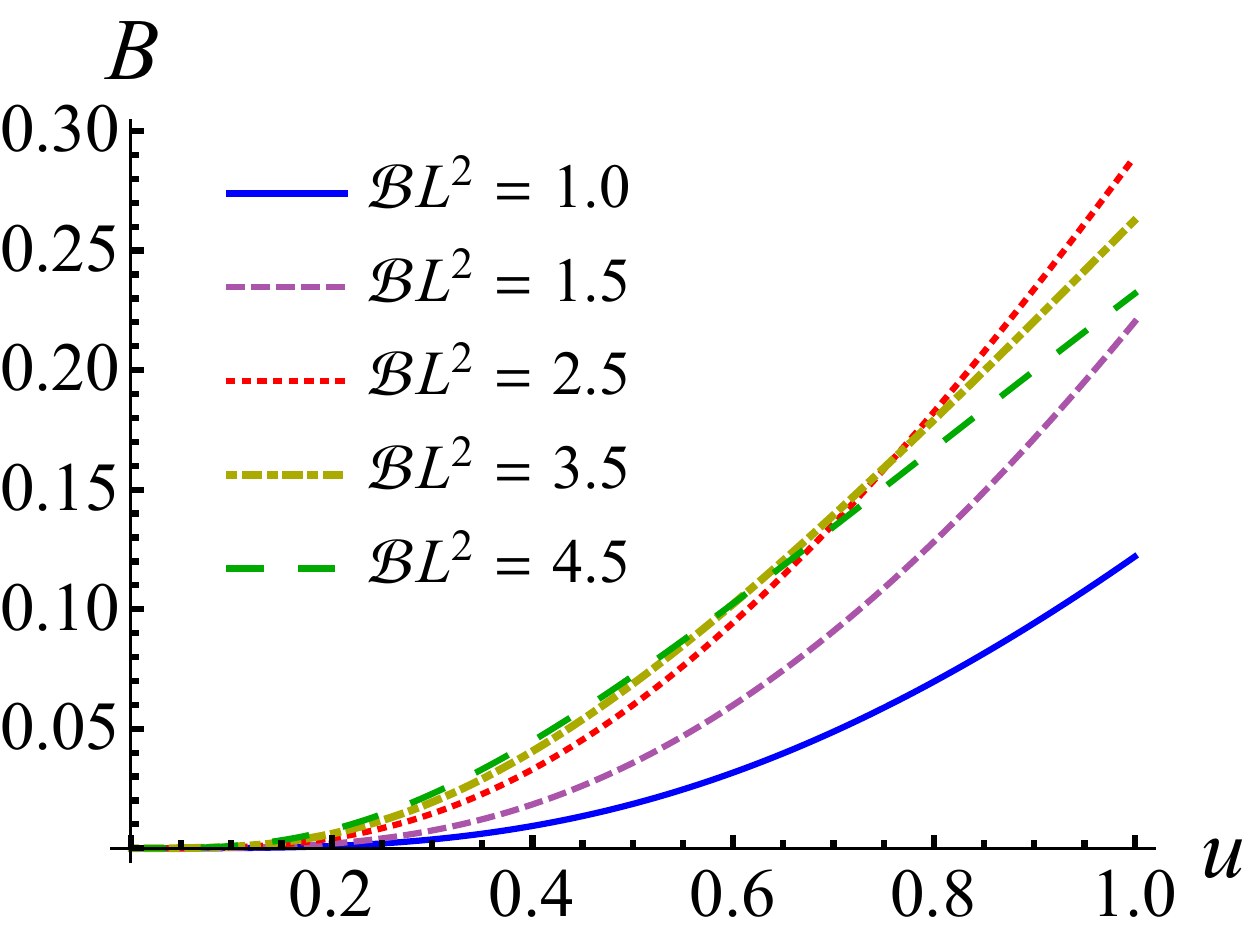}\!
\!\suck[width=0.50\textwidth]{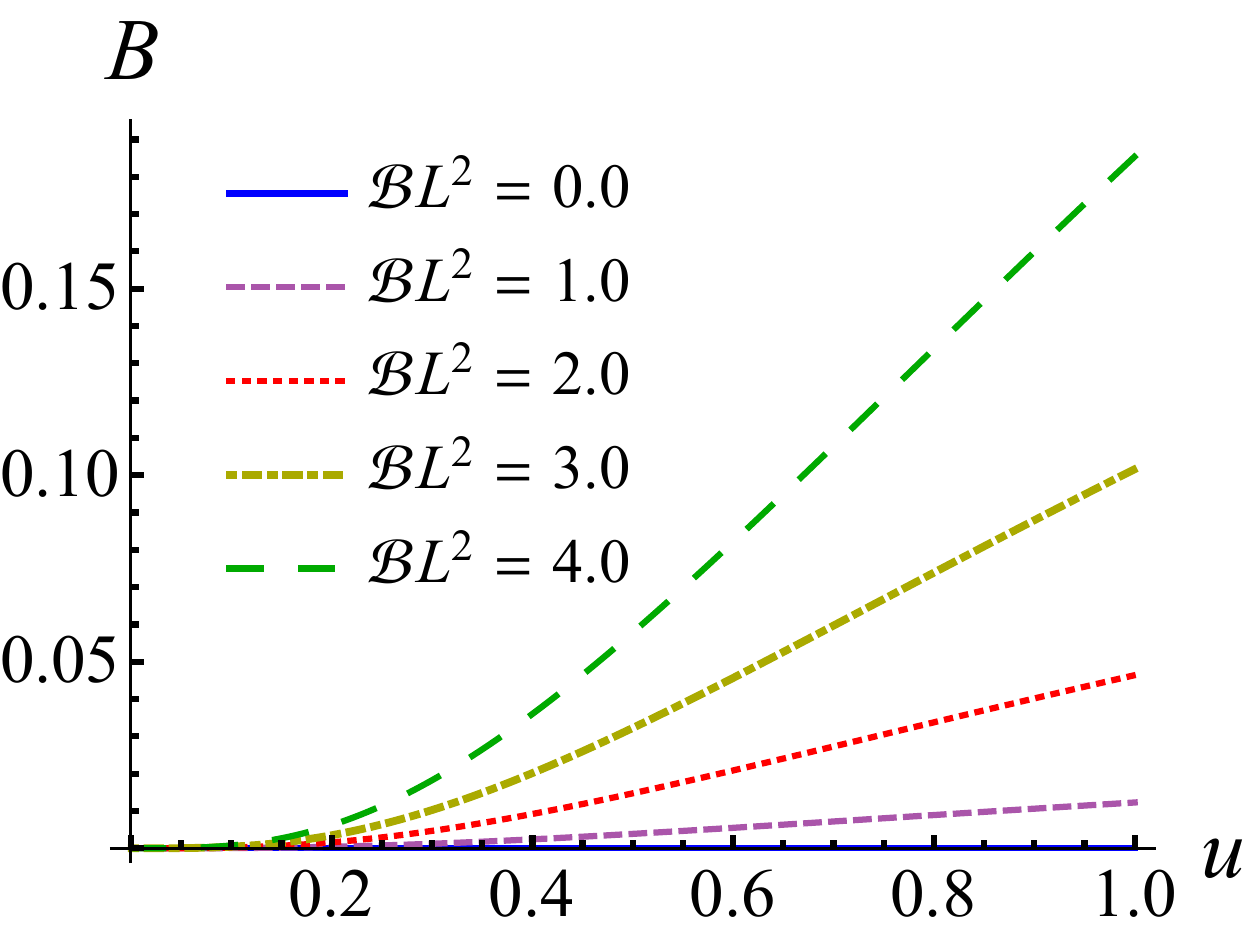}\!
\caption
    {%
    The anisotropy function $B(r)$ as a function of inverse radius $u \equiv 1/r$
    for equilibrium magnetic brane solutions
    with different values of the magnetic field.
    Left panel: Solutions at fixed energy density $\epsL = 0.75\, L^{-4}$ with
    $\B L^2 = 1.0$, 1.5, 2.5, 3.5, and 4.5.
    Right panel: Solutions at fixed energy density $\epsB = 8.0\, L^{-4}$ with
    $\B L^2 = 0.0$, 1.0, 2.0, 3.0, and 4.0.
    In all cases,
    the radial shift $\lambda$ has been
    suitably adjusted to fix the horizon radius at $u = 1$.
    Note that the horizon value of the anisotropy function is not a
    monotonic function of magnetic field at fixed $\epsL$,
    but is monotonic when $\epsB$ is held fixed.
    \label{fig:BinEQ2}
    \label{fig:BinEQ}
    }
\end{figure}

For these non-extremal magnetic brane solutions,
the anisotropy function $B(r)$ increases 
(and the scale factor $\Sigma$ decreases) smoothly
as one moves inward from the boundary toward the horizon.
Figure~\ref{fig:BinEQ} (left) plots the resulting anisotropy function $B(r)$
for several values of the magnetic field when
the energy density at the scale $1/L$ is held fixed,
$\epsL L^4 = \tfrac 34$.
(From eq.~(\ref{eq:T00}), this is the same as fixing
the asymptotic coefficient $a_4 L^4 = -\half$.)
The horizon temperatures for this series of solutions,
in order of increasing magnetic field, are given by
$\pi TL = 0.873$, 0.806, 0.879, 1.103, and 1.347.
From the figure, one may see that
the horizon value of the anisotropy function is not a monotonic
function of magnetic field (for fixed $a_4 L^4$).

The right panel of fig.~\ref{fig:BinEQ2} shows
a similar set of solutions with increasing magnetic field,
but now with the energy density at the scale
$|\B|^{1/2}$ held fixed,
$\epsB = 8 \, L^{-4}$. 
With the physical parameter $\epsB$ held fixed,
the horizon value of the anisotropy function is now
monotonically increasing with magnetic field. 
The temperatures of these solutions (in order of increasing $\B$)
are given by $\pi TL = 1.807$, 1.797, 1.738, 1.620, and 1.433.


\section	{Techniques}\label{sec:methods}

\subsection	{Computational strategy}\label{sec:strategy}

We apply the computational strategy presented in
ref.~\cite{Chesler:2013lia}
to our case of homogeneous isotropization
in Einstein-Maxwell theory.
For convenience,
we choose units in which the AdS curvature scale $L = 1$.

Required initial data, on some $v = v_0$ time slice,
consists of an initial choice for the anisotropy function $B(v_0,r)$
and the radial shift $\lambda(v_0)$, along with
chosen (time independent) values of the energy density
$\varepsilon$, charge density $\rho$,
and magnetic field $\B$.
As noted above, the holographic relation
\eqref{eq:T00} shows that fixing the energy density $\varepsilon$
at the scale $\mu = 1/L$ is equivalent to
fixing the asymptotic coefficient $a_4$.
Our choices for the initial anisotropy function will be detailed
below in subsection \ref{sec:initial}.

Given a set of initial data, the linear second order radial
ordinary differential equation (ODE)
\eqref{sigmaeqn} may be integrated to find the spatial
scale factor $\Sigma(v_0,r)$.
The two leading terms in the asymptotic behavior \eqref{sigmaexpn}
provide the integration constants needed to specify uniquely the
desired solution.
Next, one solves eq.~\eqref{sigmadoteqn}, a linear first order radial
ODE for $d_+\Sigma$.
The near-boundary asymptotic behavior of this function is
$
    d_+\Sigma \sim \half (r{+}\lambda)^2 + a_4 \, r^{-2} +
    \mathcal{O}(r^{-3})
$
and homogeneous solutions to eq.~\eqref{sigmadoteqn} behave
as $r^{-2}$ near the boundary.
Hence, the chosen value of the energy density $\varepsilon$
uniquely specifies the desired solution.
With $B$, $\Sigma$, and $d_+\Sigma$ determined
on the $v_0$ time slice,
one next solves eq.~\eqref{Bdoteqn},
a linear first order radial ODE for $d_+B$.
The desired asymptotic behavior of this function is
$
    d_+B \sim -2 b_4 \, r^{-4} + \mathcal{O}(r^{-5})
$
while
homogeneous solutions to eq.~\eqref{Bdoteqn} behave as
$r^{-3/2}$ near the boundary.
So the needed integration constant corresponds to requiring
the absence of any such homogeneous solution.
Finally, one solves the second order linear ODE \eqref{Aeqn}
to determine $A(v_0,r)$.
Homogeneous solutions are linear or constant functions of $r$.
From the asymptotic behavior \eqref{Aexpn}, one sees that
the value of the radial shift $\lambda(v_0)$ fixes the
coefficient of the homogeneous solution linear in $r$
and provides one of the two needed boundary conditions.
The second boundary condition, needed to fix the constant
homogeneous solution, is provided by the horizon
stationarity condition (\ref{eq:Ahorizon}), which determines the
value of $A$ on the apparent horizon.

Having solved for the modified time derivative $d_+B(v_0,r)$,
and $A(v_0,r)$,
one reconstructs the ordinary time derivative of the anisotropy function via
\begin{equation}
    \partial_v B(v_0,r) =
    d_+ B(v_0,r) - A(v_0,r) \, \partial_r B(v_0,r) \,.
\end{equation}
The time derivative of the radial shift,
$\partial_v \lambda (v_0)$,
is extracted from the asymptotic behavior
\eqref{Aexpn} of $A$
by evaluating the $r \to \infty$ limit of $A - \half (r{+}\lambda)^2$.
These time derivatives of $B$ and $\lambda$ provide the information
needed to advance in time.
Using a standard numerical integration scheme,
one takes a small step forward in time,
advancing $v$ to $v_0 + \Delta v$
for some timestep $\Delta v$.
Repeating this process, one progressively determines the
metric functions on a sequence of equally spaced time slices,
$v = v_k \equiv v_0 + k \, \Delta v$.
On each time slice, the asymptotic coefficient $b_4(v)$,
needed to determined the stress tensor \eqref{eq:T},
is extracted from the large $r$ behavior of the
anisotropy function $B$.
(In the presence of a non-zero magnetic field,
one extracts $b_4$ from the large $r$ limit of a subtracted,
rescaled version of the anisotropy function which removes the leading 
logarithmic piece in eq.~(\ref{Bexpn}).
This is detailed in the next subsection.)

\subsection	{Numerical methods}

We use an inverted radial coordinate $u = 1/r$,
and arbitrarily choose
\begin{equation}
    \rhbar = 1 \,,
\end{equation}
as our apparent horizon location.
This makes our computational domain a fixed radial interval,
$0 \le u \le \uh \equiv 1$.
We use a 4th order Runge-Kutta method
(described in ref.~\cite{Chesler:2013lia})
for time integration.
This requires four integrations of our radial ODEs per time step,
but yields much better accuracy, for a given timestep $\Delta v$,
than a lower order method.

To integrate the radial ODEs (\ref{sigmaeqn}--\ref{sigmadoteqn}),
we have used both traditional short-range finite difference approximations,
and spectral methods~\cite{Boyd:2001}.
In the latter approach,
one implicitly represents the radial dependence of functions
as a (truncated) series of Chebyshev polynomials.
Explicitly, functions are represented by the list of their values on a
discrete, finite collocation grid consisting of the points
\begin{equation}
    u_k = \half\Big[ 1 + \cos \frac{\pi \, k}{M{-}1} \Big] \,,
    \quad k = 0,\cdots,M {-} 1 \,,
\end{equation}
and derivatives are represented by (dense) $M \times M$ matrices
acting on the finite list of function values.
The (truncated) spectral expansion converts each ODE into
a straightforward linear algebra problem.
Boundary conditions are simply encoded into the first or
last rows of the resulting matrix~\cite{Boyd:2001}.

Although there are subtleties (described momentarily) in applying
spectral methods to our problem, we have found the use
of spectral methods to be clearly superior to finite difference approximations,
yielding both faster computation and more accurate results.
Using an $M$ point discretization of the computation domain,
short-range finite difference methods have errors which
scale as an inverse power of $M$ while spectral methods,
in favorable cases, produce errors which fall
exponentially with increasing $M$.

Spectral methods 
presume that one is approximating functions
which are regular and well-behaved on the computational domain.
However,
our radial ODEs have regular singular points at $u = 0$
or $r = \infty$ (due to the $r^2$ growth of the scale factor
near the boundary),
and our functions $\Sigma$, $d_+\Sigma$, and $A$ all
diverge at the $u = 0$ endpoint.
Therefore, we define subtracted functions
in which the known singular near-boundary behavior is removed.
To minimize loss of numerical precision in spectral approximations
of derivatives, it is also helpful to rescale the subtracted
functions so that they do not vanish faster than linearly
as $u \to 0$.
If the magnetic field is zero, so no logarithmic terms are
present in the near-boundary behavior (\ref{metricexpansions}),
then our subtracted functions are analytic in a neighborhood of the
$u\in [0,1]$ radial interval and
spectral methods converge exponentially.
With a non-zero magnetic field, logarithmic terms are
unavoidably present,
showing that $u = 0$ is a branch point of the metric functions.
This degrades convergence of the spectral series,
leading to power-law convergence at a rate which depends on
the behavior of the leading non-analyticity.
Consequently, it is desirable to subtract logarithmic terms to
as high an order as is practicable.
We chose to introduce subtracted/rescaled functions
(denoted with a subscript `s') via:
\begin{subequations}%
\label{eq:subtractions}%
\begin{align}
\label{sigmasub}
    \Sigma(r) &= (r {+} \lambda) + r^{-5} \, \Sigma_s(r)\,,
\\[5pt]
\label{Asub} 
    A(r) &=
	\half (r{+}\lambda)^2
	- \tfrac 13 \B^2 \, \ln r
	\left[ r^{-2} - 2 \lambda \,r^{-3}
		+ 3 \lambda^2\, r^{-4}
		- 4 \lambda^3 \,r^{-5} \right]
	+ A_s(r)\,.
\\[5pt]
\label{sigmadotsub} 
    d_+{\Sigma}(r) &= \tfrac 12 \Sigma(r)^2
	- \tfrac 13 \B^2 \, \ln r
	\left[ r^{-2} - 2 \lambda \,r^{-3}
		+  3 \lambda^2\, r^{-4}
		- 4 \lambda^3 \,r^{-5} \right]
	+ (d_+{\Sigma})_s(r)\,,
\\[5pt]
\label{Bdotsub} 
    d_+{B}(r) &= 
	-\tfrac 23 \B^2 \, \ln r
	\left[ r^{-3} - 3 \lambda \,r^{-4}
	    + 6 \lambda^2\, r^{-5}
	    - 10 \lambda^3 \,r^{-6} \right]
	+ r^{-2} \, (d_+{B})_s(r)\,.
\end{align}
\end{subequations}
All our numerical work 
is performed using these subtracted/rescaled functions;
we directly solve for
$\Sigma_s$, $(d_+{\Sigma})_s$, $(d_+{B})_s$, and $A_s$.%
\footnote
    {%
    Note that $(d_+{\Sigma})_s \neq d_+ (\Sigma_s)$,
    and likewise for $(d_+{B})_s$;
    these are just names for the 
    subtracted/rescaled functions for
    $d_+{\Sigma}$ and $d_+B$, respectively.
    }
The expressions~(\ref{eq:subtractions}) are used to reconstruct the original
functions when needed.
We also use a subtracted/rescaled anisotropy function $B_s(r)$,
introduced via
\begin{align}
\label{Bsub} 
    B(r) &=  \tfrac 13 \B^2 \, \ln r
	\left[ r^{-4} - 4 \lambda \, r^{-5}
	    + 10 \lambda^2 \, r^{-6} \right] 
	+ r^{-3} B_s(r)\,.
\end{align}
This removes leading logarithmic terms and introduces a convenient
rescaling.
Henceforth, $B_s$ will be referred to as
the subtracted anisotropy function.

In the above subtractions, the series in $1/r$ multiplying each logarithm
are just truncated expansions of $(r{+}\lambda)^{-k}$ for $k = 2$, 3 or 4.
The choice to truncate these binomial series was arbitrary,
but we found that our numerics were sufficiently well-behaved
with the above subtractions.
At higher orders in $1/r$, additional terms appear which 
involve the asymptotic coefficient $b_4$
and its (a-priori unknown) time derivatives,
as well as higher powers of logarithms.

\subsection{Initial data}\label{sec:initial}

As indicated above, one must select the value of the
energy density (or asymptotic coefficient $a_4$)
and the initial value of the radial profile of the
anisotropy function, $B(v_0,r)$.
And one must choose the value of the magnetic field $\B$
or charge density $\rho$.
For the charged case, with vanishing magnetic field,
physics can only depend on the
dimensionless combination of charge and energy densities
$\rho/\varepsilon^{3/4}$, so a (positive) value of $\varepsilon$
may be chosen arbitrarily without loss of generality.
Given a choice of $\varepsilon$, possible values of the
charge density $\rho$ are limited by the
extremality bound $|\rho| \le \rhomax = \tfrac 43 \, \varepsilon^{3/4}$.

For the initial anisotropy function, 
we chose to focus on Gaussian profiles.
In the $\lambda = 0$ frame,
\begin{equation}
\label{eq:gauss}
    B(v_0,r) = 
    \mathcal{A}\>
    e^{-\frac 12 {(r-\rave)^2}/{\sigma^2}} .
\end{equation}
We investigate the dependence of results on the parameters
of the Gaussian 
(amplitude $\mathcal A$, width $\sigma$, and mean position $\rave$)
in the first part of section~\ref{sec:results}.%
\footnote
    {%
    For results from an exploration of a broader range of initial
    anisotropy profiles,
    in the case of vanishing charge density and magnetic field,
    see ref.~\cite{Heller:2013oxa}.
    }
Motivated by the fact that in our coordinates
lines of varying $r$, at fixed $v$,
are radially infalling geodesics,
we will refer to this initial Gaussian as a ``pulse'' of
initial anisotropy.

For the magnetic case,
as discussed above,
the breaking of scale invariance 
implies the presence of logarithmic terms in 
the asymptotic behavior of the anisotropy function.
We simply add the log terms shown in eq.~\eqref{Bsub}
to the Gaussian (\ref{eq:gauss}).
With vanishing radial shift, $\lambda = 0$, 
our chosen initial anisotropy function takes the form 
\begin{equation}
\label{eq:gaussMAG}
    B(v_0, r) = \mathcal{A}\>
    e^{-\frac 12 {(r-\rave)^2}/{\sigma^2}}
    + \tfrac 13 \B^2 \, r^{-4} \ln r \,.
\end{equation}
Arguably,
a more natural choice for the magnetic case initial data
might be to add a Gaussian to the full
equilibrium solution for the anisotropy function in
the chosen magnetic field.
This could be seen as nicely paralleling the charged case
(in which the equilibrium solution has vanishing anisotropy).
Nevertheless, we will stick with our somewhat arbitrary choice
(\ref{eq:gaussMAG}), which is an adjustable initial pulse
added to the correct asymptotics.
As will be seen, the Gaussian pulse will nearly always be the dominant portion
of the deviation from equilibrium and the driving force of the
resulting anisotropy in the boundary stress.
We doubt that differing choices in the precise form of the
slowly varying function to which the Gaussian pulse is added
would impact, in any significant way,
the characteristic equilibration times or other significant features
of the results presented below. 

After choosing the initial anisotropy function (in the $\lambda = 0$ frame)
the initial value of the radial shift, $\lambda(v_0)$,
is adjusted to fix the location of the apparent horizon,
as discussed in section~\ref{sec:horizon}.

It should be noted that, in all cases (charged, uncharged, magnetized)
it is quite possible to select physically inconsistent initial data.
This happens when the initial anisotropy,
for a given energy density,
is so large that no apparent horizon shields a coordinate singularity
produced by a vanishing scale factor $\Sigma$.
This sets a natural limit on the amplitude of the initial pulse
which is meaningful to study.


\section{Results}\label{sec:results}

\subsection{Neutral plasma}\label{sec:resultsSch}

\begin{figure}
\centering
\suck[width=.52\textwidth]{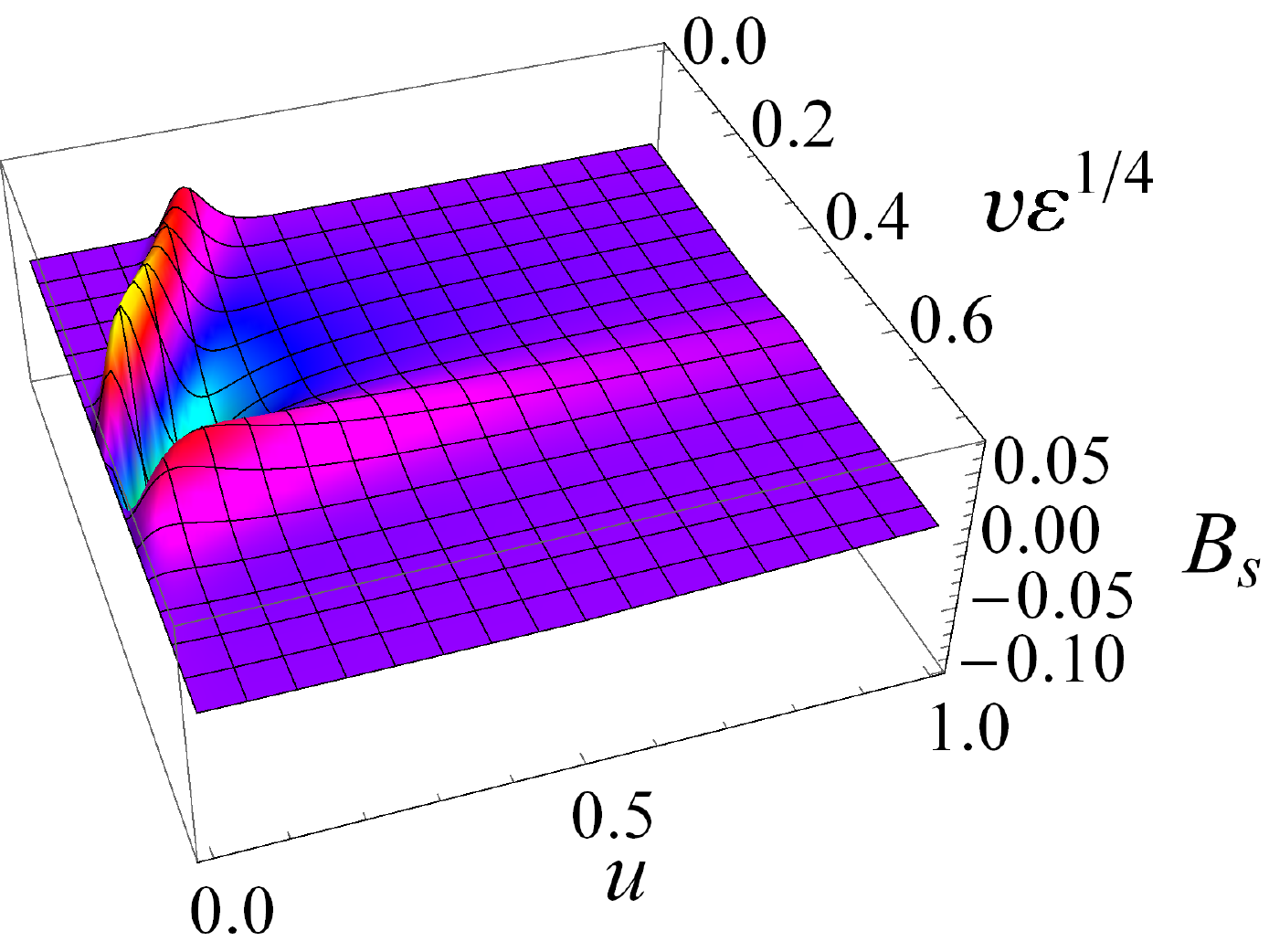}
\hfill
\raisebox{15pt}{\suck[width=.44\textwidth]{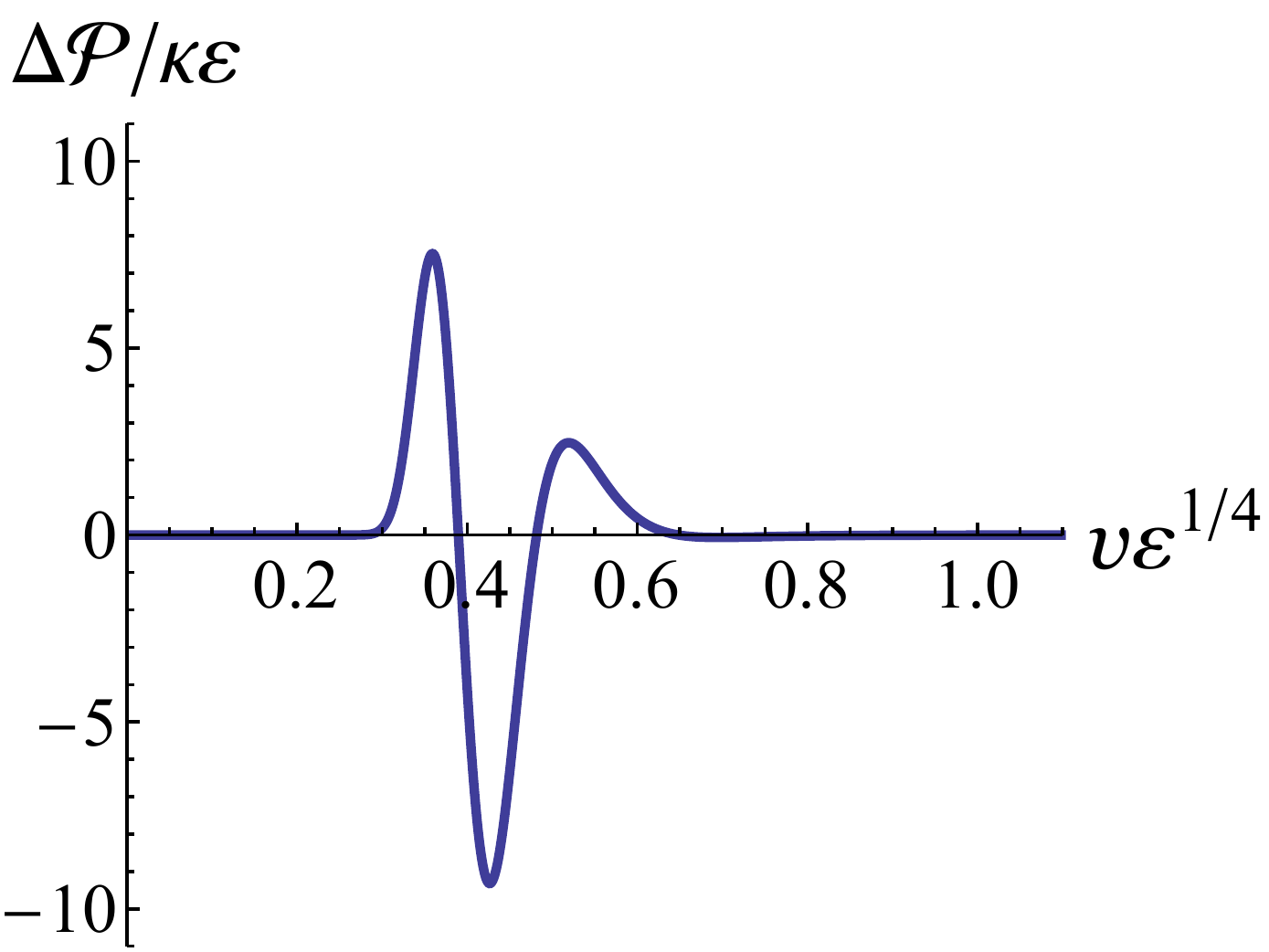}}%
\hspace*{-1em}
\caption
    {%
    Left: Rescaled anisotropy function $B_s = u^{-3} B$, for a typical
    case of equilibration to the Schwarzschild black brane,
    as a function of inverse bulk radius $u$ and time $v$.
    Initial pulse parameters are
    $\mathcal A = 5\times10^{-4}$, $\rave = 4$ and $\sigma = \tfrac 12$,
    with $\varepsilon = \tfrac 34 L^{-4}$.
    The (rescaled) energy density $\varepsilon$
    is used to the set the scale for time.
    One sees that the effect of the initial Gaussian pulse 
    propagates outward, essentially along an outgoing radial
    null geodesic, reflects off the boundary, and then largely
    falls through the horizon (along an ingoing radial null geodesic 
    which is instantaneous in $v$).
    After one ``bounce'', the anisotropy rapidly approaches zero.
    Right:
    The corresponding relative pressure anisotropy,
    $
	\Delta \mathcal P/\kappa \varepsilon \equiv
	\half (T^{11} + T^{11} -2 T^{33})/\kappa \varepsilon
    $,
    induced in the boundary field theory,
    as a function of time.
    Note how the peaks of the pressure anisotropy correspond directly
    to the reflection of the anisotropy pulse off the boundary.
    \label{fig:BevolveSBB}
    }
\end{figure}

Before presenting results for equilibration in
charged or magnetized plasmas,
we first discuss general features of the time evolution in the
uncharged, unmagnetized case and examine the sensitivity of
results to specific features in the initial data.
As noted above, we choose a Gaussian profile (\ref{eq:gauss})
for the initial anisotropy function, with an adjustable
amplitude, width, and mean position.
Typical evolution of our
subtracted/rescaled anisotropy function,
$B_s(v,u) \equiv u^{-3} B(v,u)$, is shown on the left in figure~\ref{fig:BevolveSBB}.
One sees the initial pulse profile on the back side of the plot at $v = 0$.
The figure clearly shows the influence of the pulse propagating outward
and reflecting off the boundary at $u = 0$.
The outgoing portion of the pulse essentially propagates along an outgoing radial null
geodesic which, in our coordinates, near the boundary are $45^\circ$ lines
at constant values of $u + v$.
The influence of the anisotropy pulse, after the reflection,
largely falls through the horizon along an ingoing radial null geodesic
which is instantaneous in our null time coordinate $v$.
The asymptotic coefficient $b_4$,
which equals the slope of $B_s$ at $u = 0$,
controls the anisotropy in the stress tensor,
\begin{equation}
    \Delta \mathcal P \equiv
    \half \langle T^{11} \rangle
    + \half \langle T^{11} \rangle
    - \langle T^{33} \rangle \,,
\label{eq:DeltaP}
\end{equation}
with $\Delta \mathcal P/\kappa = 3 \, b_4$.
Hence, the reflection of the pulse off the boundary
directly produces the pressure anisotropy $\Delta \mathcal{P}$
in the boundary theory.
The time dependence of the relative
pressure anisotropy,%
\footnote
    {%
    Note that, for unmagnetized plasma,
    the energy density is three times the average pressure,
    $
	\varepsilon
	=
	\langle T^{ii} \rangle
	\equiv
	3\overline {\mathcal P}
    $.
    }
defined as $\Delta \mathcal P/(\kappa \epsilon)$,
is plotted on the right in figure~\ref{fig:BevolveSBB}.

As shown in the figure, at late times the anisotropy function approaches zero,
as required for equilibration to the isotropic Schwarzschild black brane solution.
At sufficiently late times, when the departure from equilibrium is small,
the evolution should be well described by a linearized approximation to the
full nonlinear dynamics.
The linearized dynamics of infinitesimal perturbations away from equilibrium
may be represented as a sum of quasinormal modes (QNM), which are
eigenfunctions of the linearized dynamics with complex frequencies,
$\phi(t) = \Re (A \, e^{-i \omega t})$ with $\Im \, \omega < 0$.
The lowest quasinormal mode (for which $-\Im\,\omega$ is minimal)
dominates the late time approach to equilibrium.
For the Schwarzschild black brane, quasinormal mode frequencies have been
previously evaluated by Starinets \cite{Starinets:2002br}.
From the late time behavior of our full nonlinear evolution,
it is straightforward to extract an estimate of the lowest quasinormal
mode frequency.
Comparing with the independent results of ref.~\cite{Starinets:2002br}
provides a useful test of the accuracy of our numerics.
Fitting the late time
($4 \lesssim v \, \varepsilon^{1/4} \lesssim 25$)
portion of our calculated pressure anisotropy
to a decaying, oscillating
exponential, $|A| e^{(\Im \, \omega) v} \, \cos[(\Re \, \omega) v + \phi]$,
yields an estimate of the lowest QNM frequency $\omega$ which agrees
with ref.~\cite{Starinets:2002br} to five digits,
$\omega/(\pi T) \approx 3.11946 - 2.74663 \, i$.

We will see the same vanishing of the anisotropy function at late times
for the case of charged plasmas, whose gravitational duals
equilibrate to an isotropic Reissner-Nordstrom black brane solution.
For the magnetic case, however, at late times 
there is a non-zero profile for the anisotropy function,
reflecting the spatial anisotropy of equilibrium magnetic brane solutions.

We now turn to an examination of the dependence of the pressure
anisotropy on the parameters of the initial Gaussian (\ref{eq:gauss}).
Of particular interest will be the dependence of the response on
the amplitude and position of the initial pulse.
Less interesting is the dependence on the width of the pulse, which
affects the duration of the reflection off the boundary
(and also produces changes more naturally associated with the position of the pulse). 

\begin{figure}
\centering
\suck[width=0.44\textwidth]{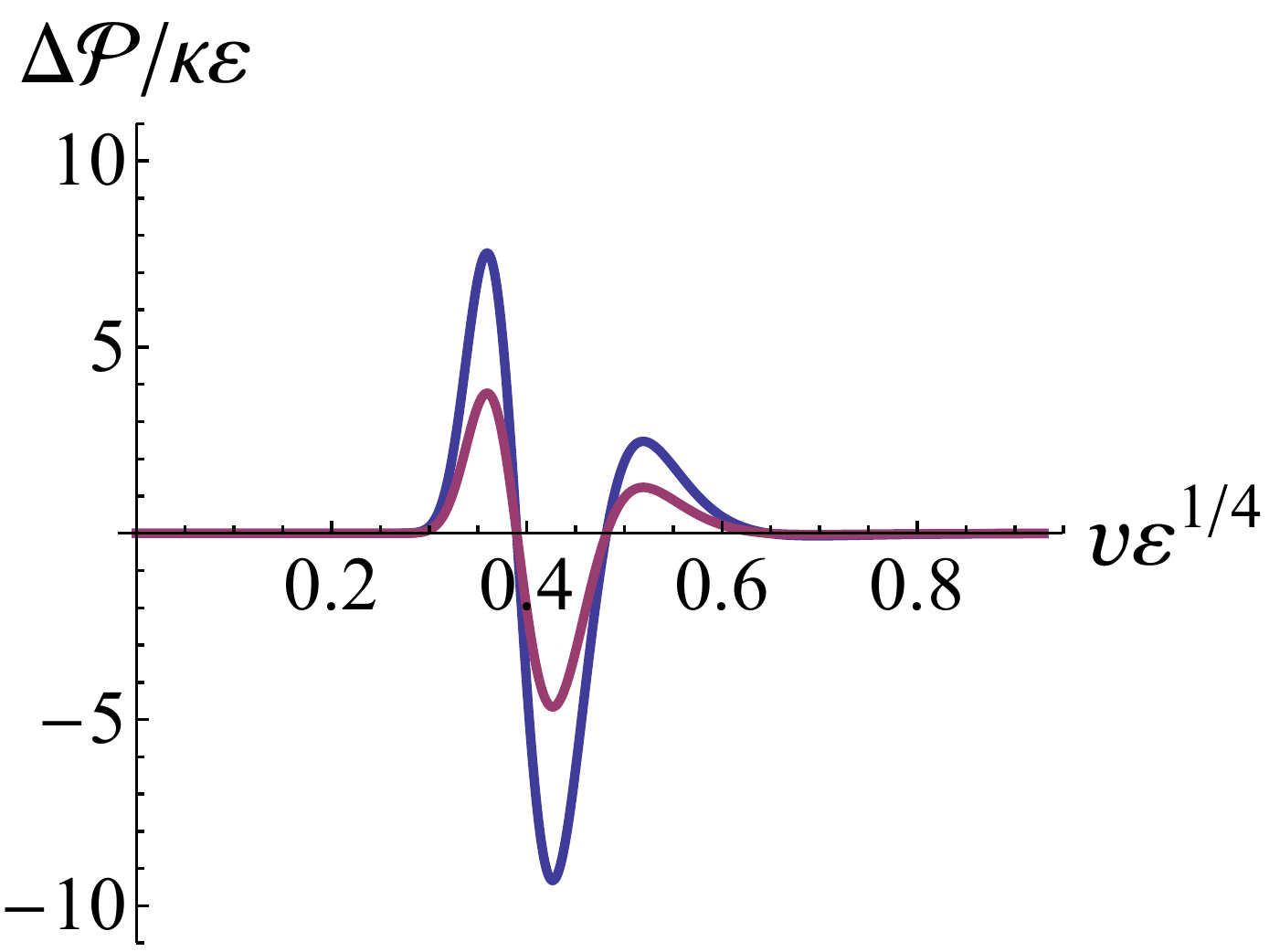}
\quad
\quad
\suck[width=0.44\textwidth]{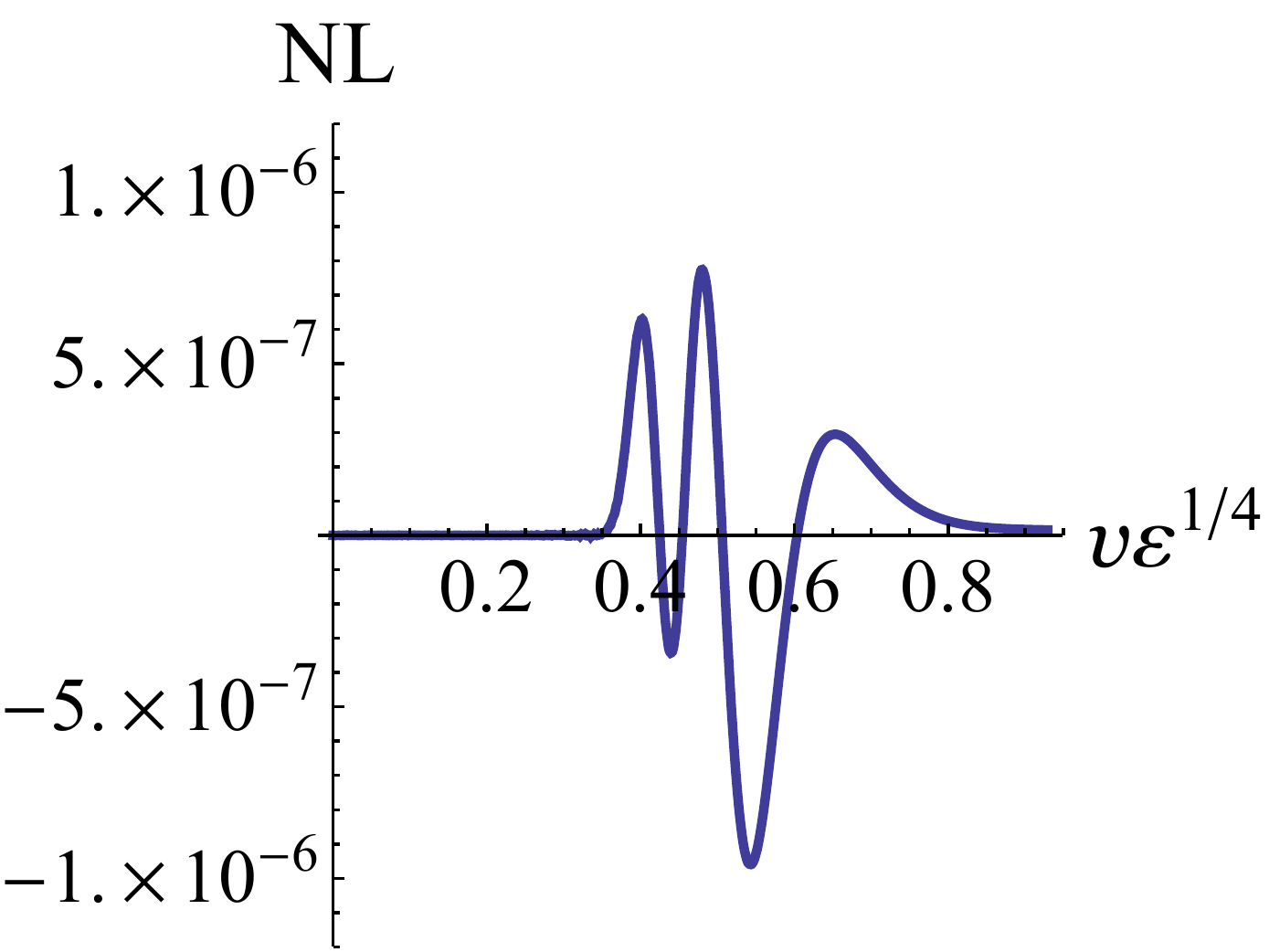}
\caption
    {%
    Left: The pressure anisotropy $\Delta P/\kappa\varepsilon$
    produced by the same initial pulse
    shown in fig.~\ref{fig:BevolveSBB} (blue curve),
    overlaid with the pressure anisotropy produced by an initial pulse
    with half the amplitude (purple curve).
    Halving the initial amplitude roughly halves the induced pressure anisotropy.
    Right: The ``nonlinearity'' (NL), defined as the difference between the
    pressure anisotropy produced by the larger pulse and
    twice the pressure anisotropy produced by the half amplitude pulse.
    \label{fig:pressureDouble}
    }
\end{figure}

In figure~\ref{fig:pressureDouble}, we compare the 
pressure anisotropies created by the pulse shown in
fig.~\ref{fig:BevolveSBB}, and an otherwise identical pulse
with half the amplitude.
Visually, one sees that the smaller amplitude pulse produces
roughly half the pressure anisotropy as does the larger pulse,
but with a virtually identical time course.
The peak pressure anisotropy (divided by energy density),
for both pulses is over 4, significantly larger than unity.
Hence both initial pulses represent large departures from equilibrium.
Given the highly nonlinear nature of the Einstein equations,
one might have expected to see clear signs of nonlinearity in the
dependence of the pressure anisotropy on the initial pulses.
However, even for these pulses producing large departures
from equilibrium, the amplitude of the peaks in the induced pressure
anisotropy are nearly linear in the amplitude of the initial Gaussian pulse.

The right hand panel of figure~\ref{fig:pressureDouble} makes this
comparison quantitative.
This shows the nonlinearity (NL) defined as
the difference between the pressure anisotropy
$\Delta \mathcal P/\kappa\varepsilon$ of the larger initial pulse and
twice the pressure anisotropy produced by the halved initial pulse.
Compared to the pressure anisotropies themselves,
the relative size of the nonlinearity is roughly one part in $10^7$.
This suggests that the dynamics, as probed by these initial pulses,
are surprisingly close to a linear dynamical system.

\begin{figure}
\begin{flushleft}
\hspace{1.05em}\suck[width=0.30\textwidth]{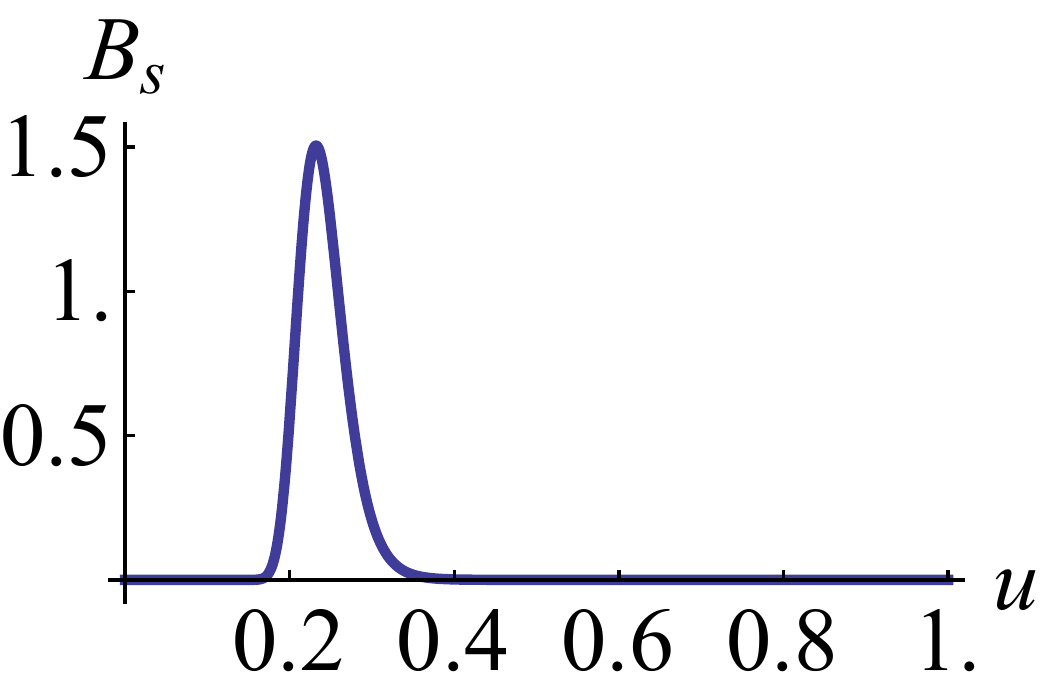}
\hspace{1.1em}\suck[width=0.30\textwidth]{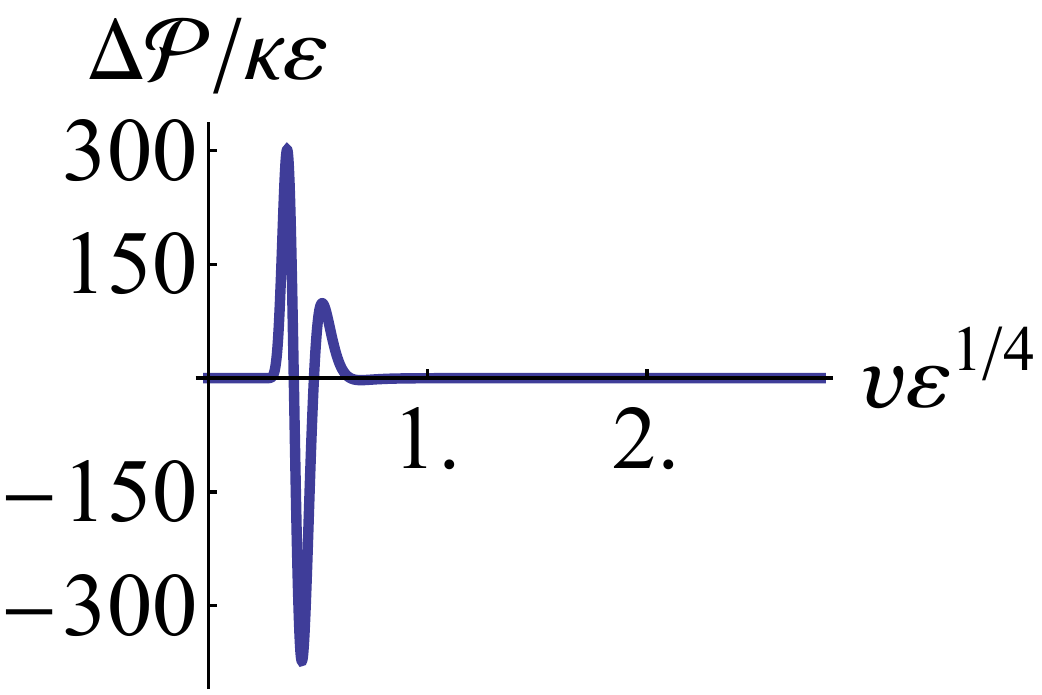}
\hspace{1.6em}\suck[width=0.30\textwidth]{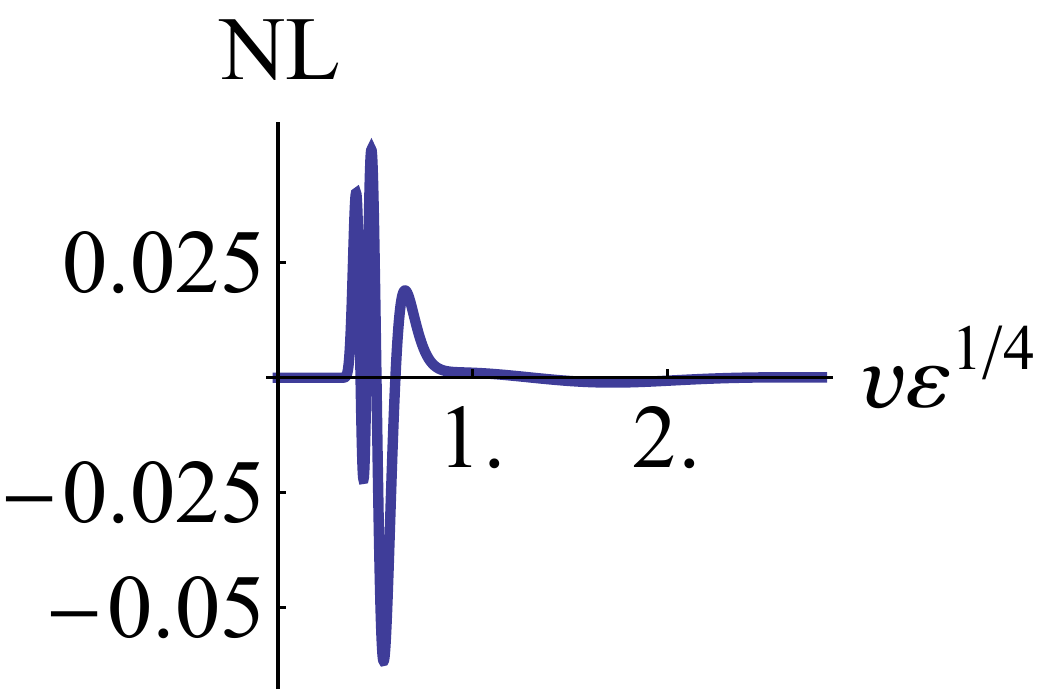}%
\hspace*{-1em}
\\[6pt]
\hspace{0.5em}\suck[width=0.30\textwidth]{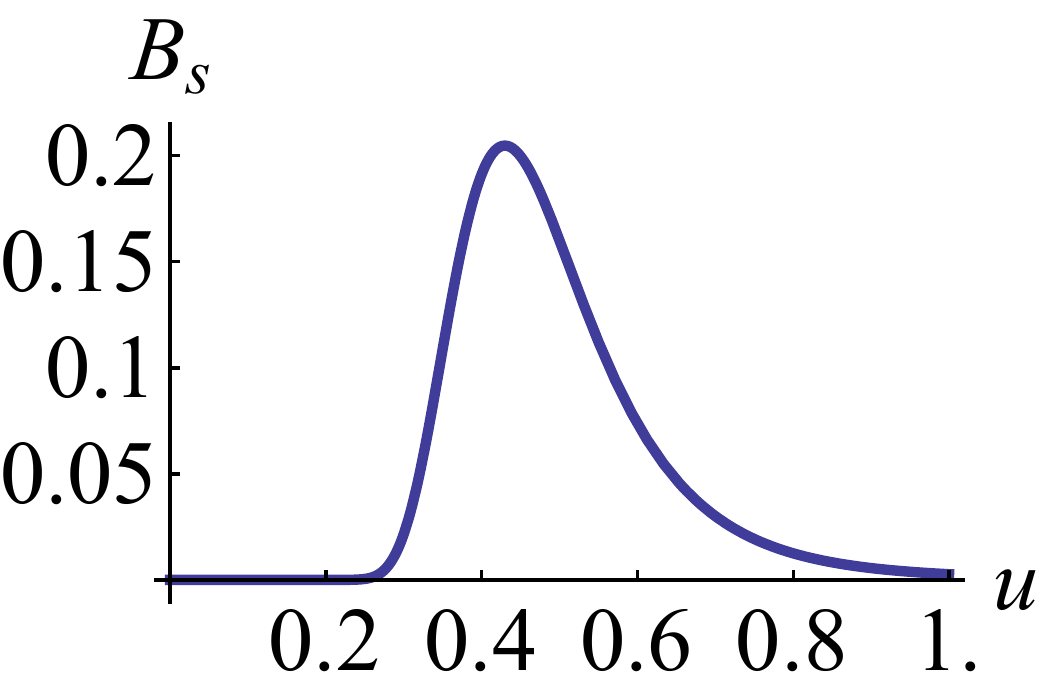}
\hspace{2.65em}\suck[width=0.30\textwidth]{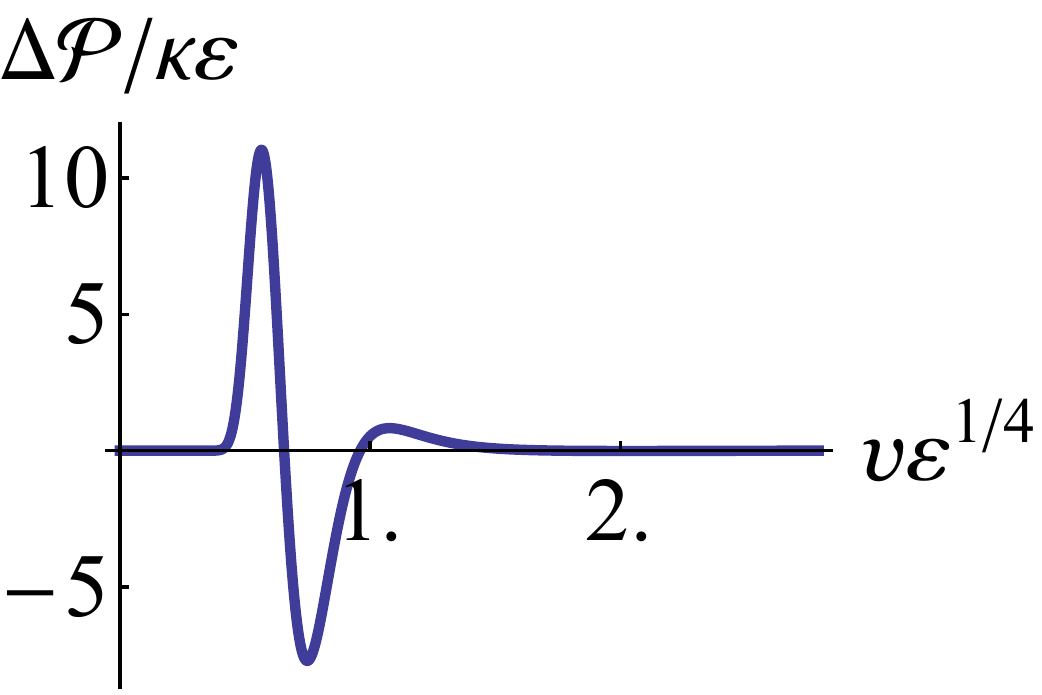}
\hspace{0.8em}\suck[width=0.30\textwidth]{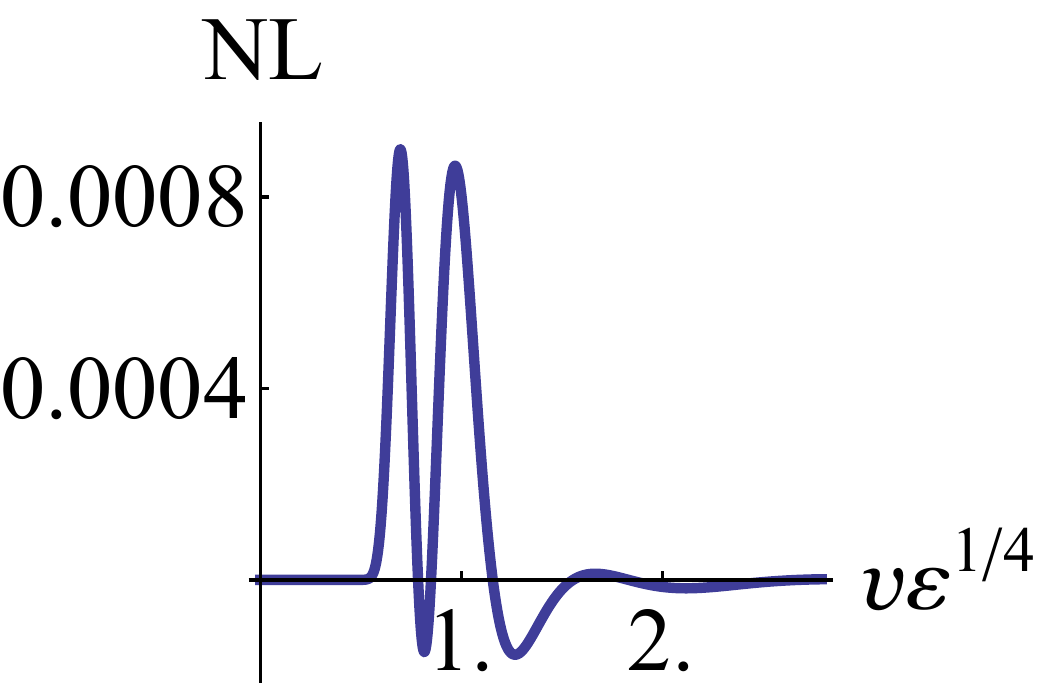}%
\hspace*{-1em}
\\[6pt]
\hspace{0.5em}\suck[width=0.30\textwidth]{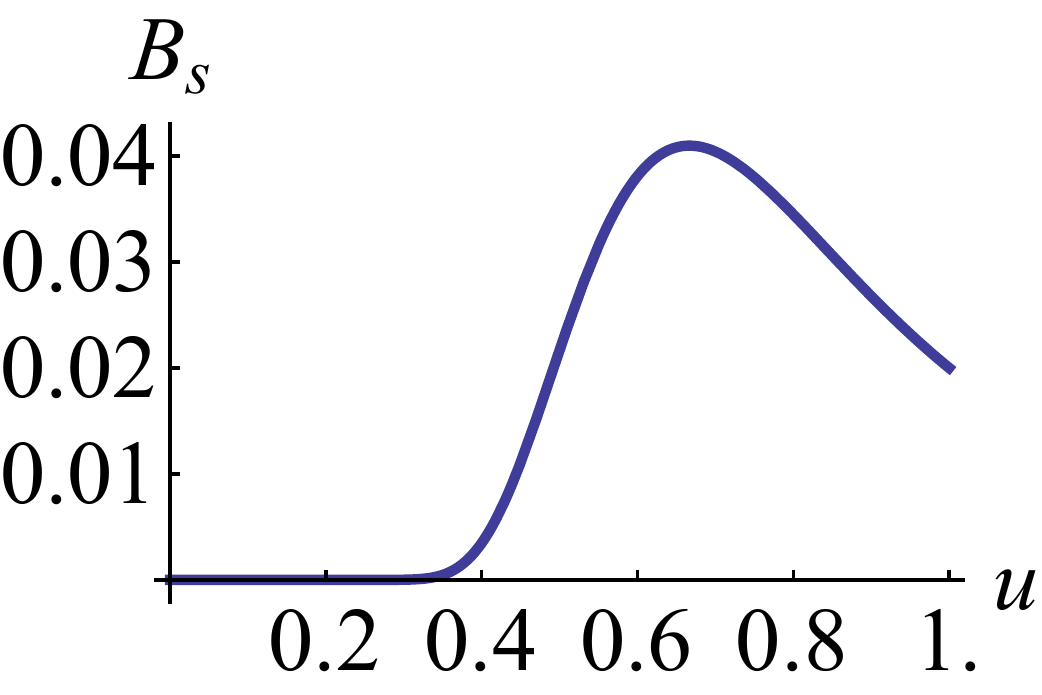}
\hspace{1.4em}\suck[width=0.30\textwidth]{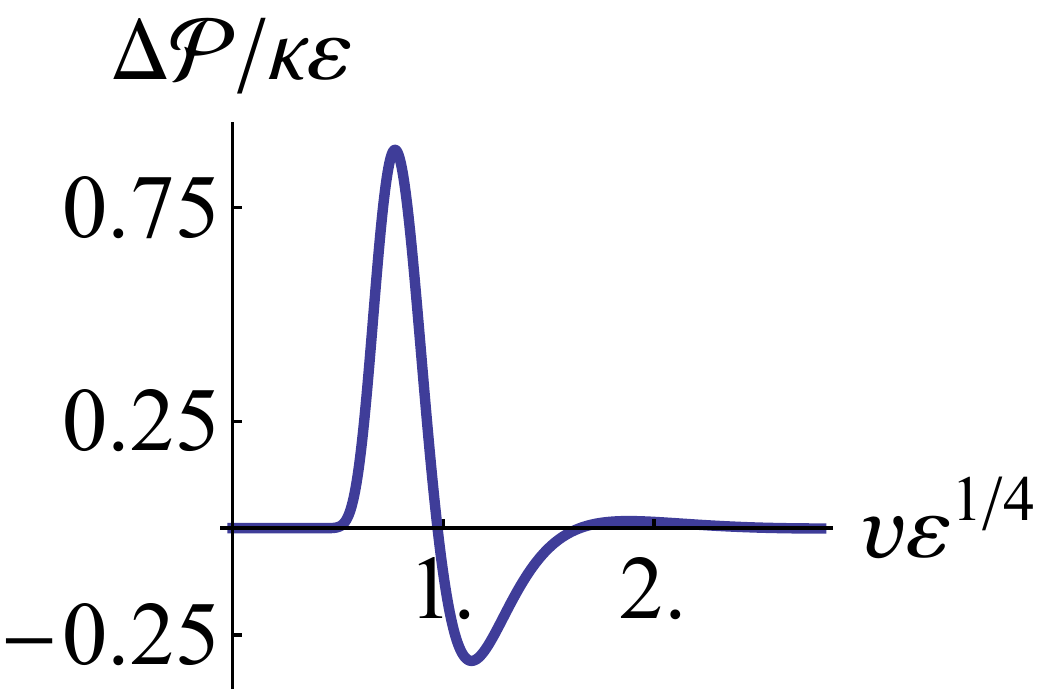}
\hspace{1.55em}\suck[width=0.30\textwidth]{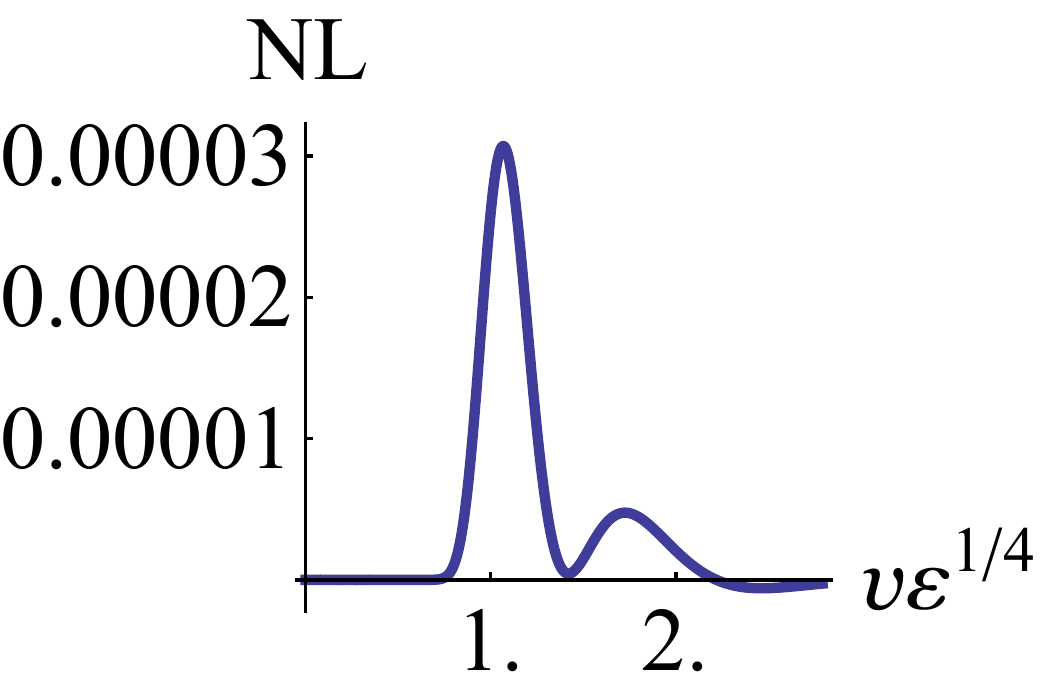}%
\hspace*{-1.0em}
\\[6pt]
\suck[width=0.30\textwidth]{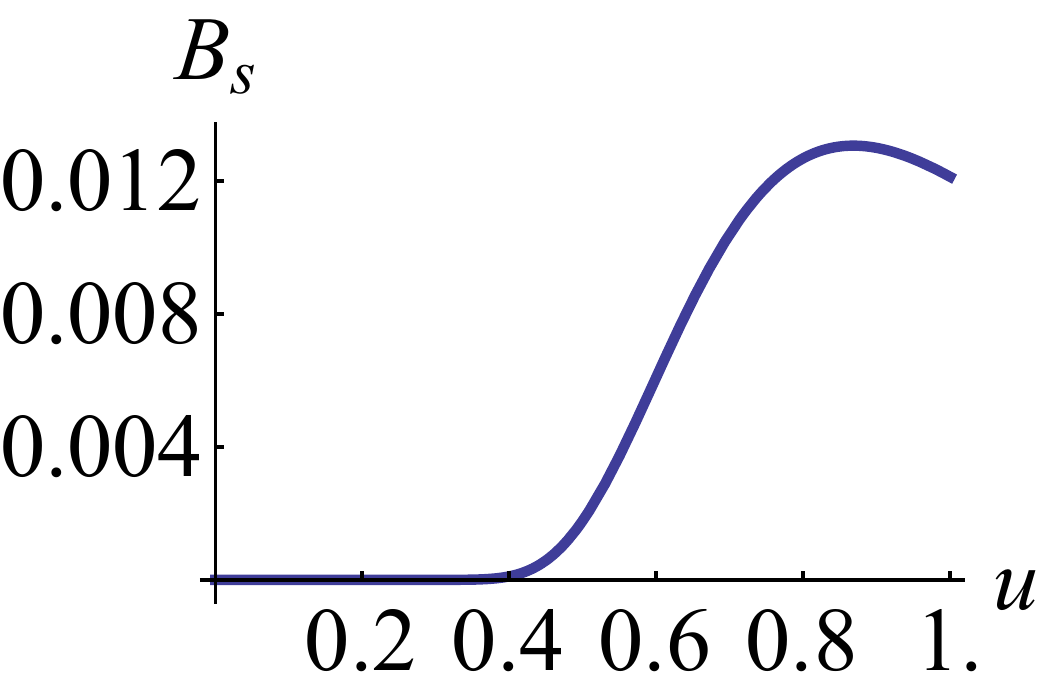}
\hspace{2.55em}\suck[width=0.30\textwidth]{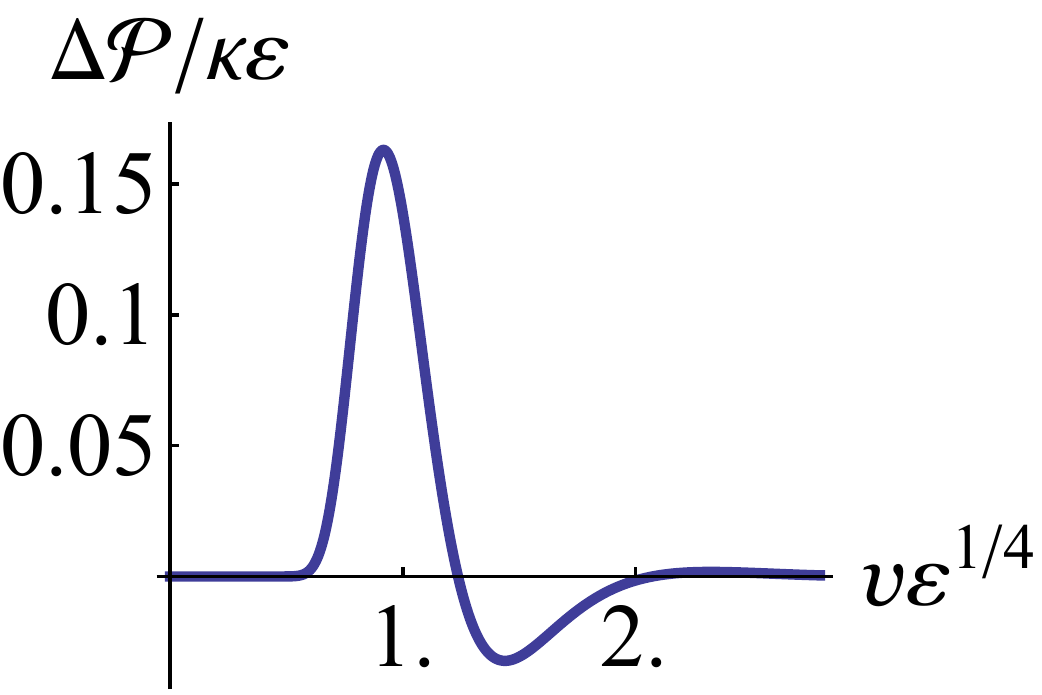}
\hspace{0.10em}\suck[width=0.30\textwidth]{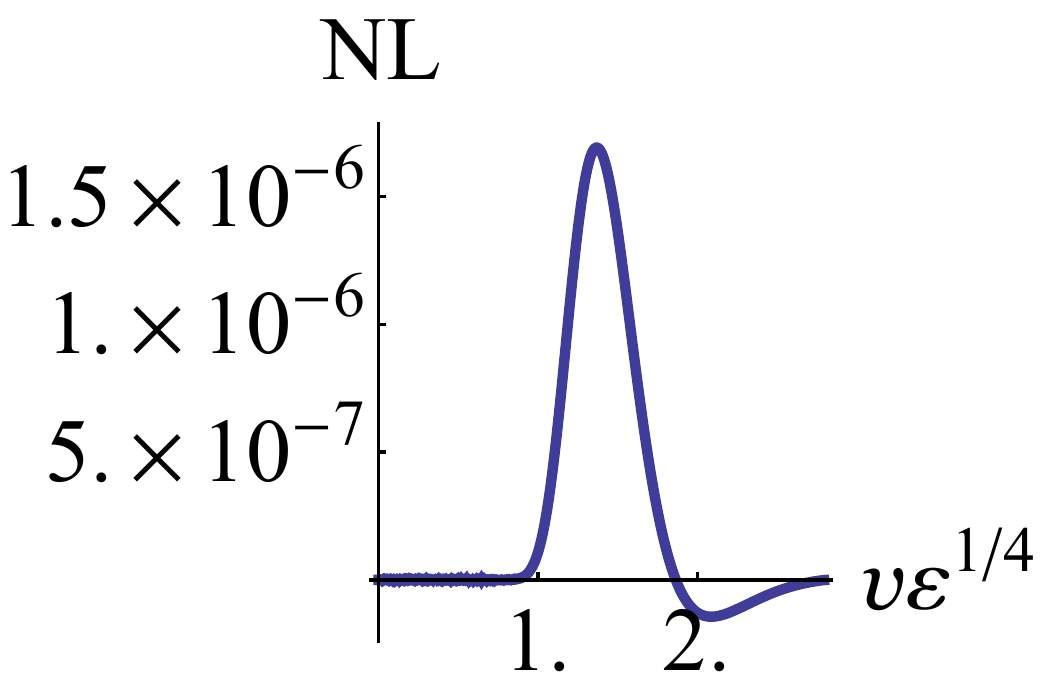}%
\hspace*{-1em}
\\[6pt]
\suck[width=0.30\textwidth]{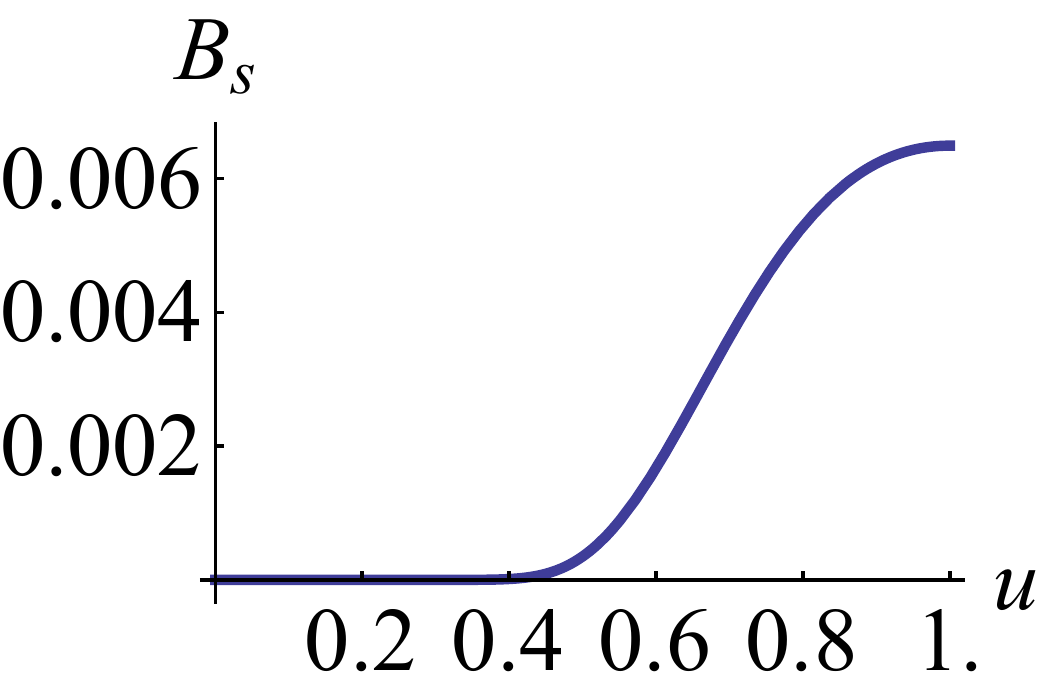}
\hspace{2.55em}\suck[width=0.30\textwidth]{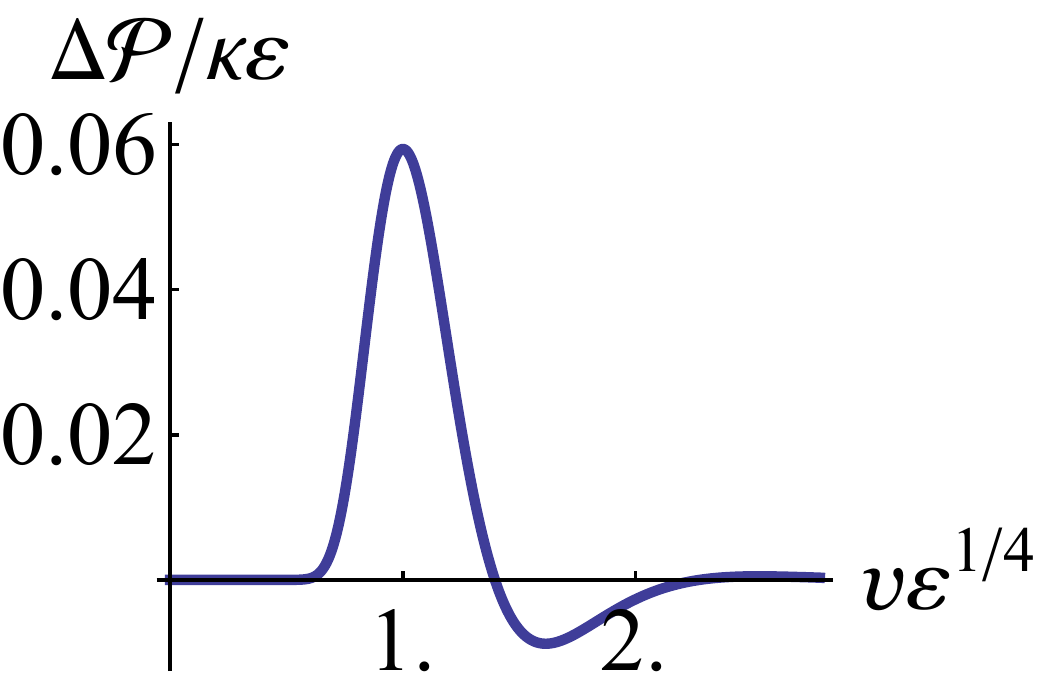}
\hspace{0.58em}\suck[width=0.30\textwidth]{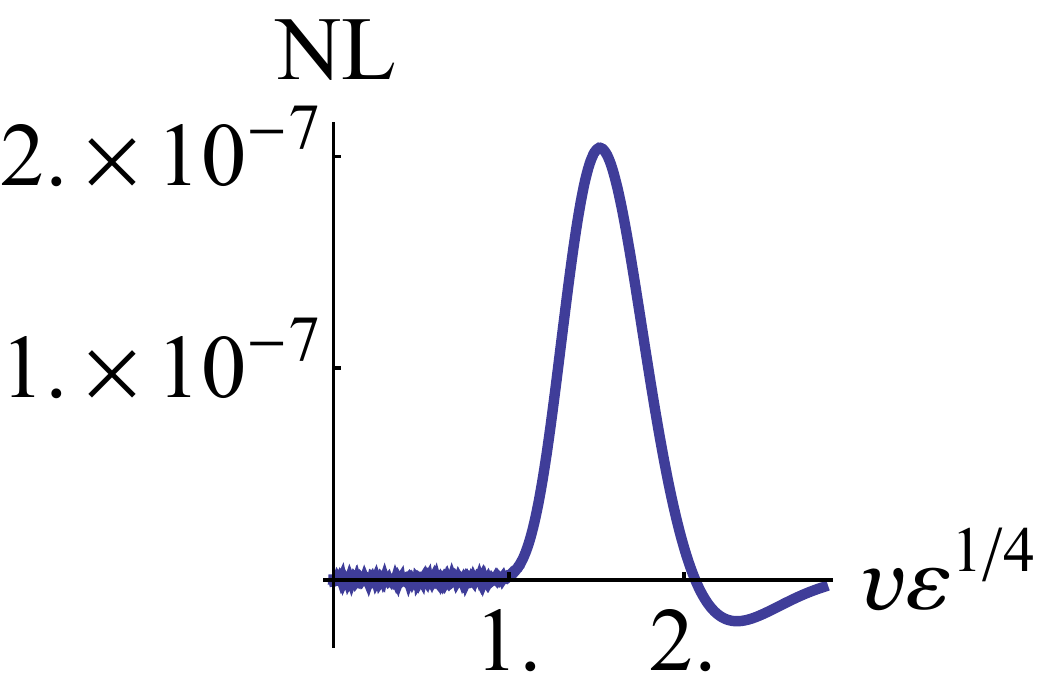}%
\hspace*{-1em}
\end{flushleft}
\vspace*{-4pt}
\caption
    {%
    Comparisons of 
    initial anisotropy functions (left column),
    induced pressure anisotropy (middle column),
    and nonlinearity (right column)
    defined as in fig.~\ref{fig:pressureDouble},
    for a series of five Gaussian initial anisotropy functions
    differing only in their depth in the bulk.
    From top to bottom, the mean position of the initial pulse,
    in the $\lambda = 0$ frame, is $\rave = \{4,2,1,\frac 12, \frac 14\}$.
    In all cases the energy density is $\varepsilon = \tfrac{3}{4} L^{-4}$.
    The plots in the first row come from the same initial data
    as in fig.~\ref{fig:BevolveSBB}, but with the amplitude increased
    by a factor of 40
    ($\mathcal A = 0.02$, $\rave = 4$, $\sigma = \tfrac 12$).
    The left column shows the initial anisotropy
    function scaled by $u^{-3}$ and plotted as a function
    of the inverse radial coordinate $u$, 
    after adjusting the radial shift $\lambda$ to fix the apparent
    horizon at $u = 1$.
    \label{fig:BsubRavepos}   
    }
\end{figure}
\afterpage{\clearpage}

In asymptotically AdS gravitational solutions,
deviations of the geometry from that of pure AdS space
necessarily vanish as one approaches the boundary.
Hence, one might expect nearly linear dynamics to be evident
for initial pulses which are localized sufficiently close to the boundary,
while anticipating much larger nonlinearities 
for initial pulses localized closer to the horizon.
To test this expectation, we used a very large Gaussian --- the same
initial Gaussian profile which generated fig.~\ref{fig:BevolveSBB}
but with the amplitude increased%
\footnote
    {%
    For the chosen values of position and width of the Gaussian,
    plus energy density $\varepsilon = \tfrac{3}{4} L^{-4}$,
    this amplitude is close to the upper limit set by demanding the
    existence of an apparent horizon.
    }
by a factor of 40
--- and then examined the resulting evolution when the mean position of
the initial Gaussian was progressively shifted deeper into the bulk.
Figure~\ref{fig:BsubRavepos} compares the 
evolution for mean positions $\rave = \{4,2,1,\half,\frac 14\}$,
in the frame with radial shift $\lambda = 0$.
In all cases, the energy density was held fixed at a value
($\varepsilon = \tfrac{3}{4} L^{-4}$) which puts the equilibrium horizon position at $r = 1$.
In other words, the only change in the five cases shown in
fig.~\ref{fig:BsubRavepos} is the radial position of the
initial Gaussian anisotropy function (\ref{eq:gauss}),
viewed as a function of $r$.

Each row of fig.~\ref{fig:BsubRavepos} displays results for one of these
five cases.
In each row, the left hand panel shows the initial anisotropy function,
but plotted as a function of the inverse radial coordinate $u$,
after adjusting the radial shift $\lambda$ to fix the apparent
horizon at $u = 1$, as discussed in sec.~\ref{sec:horizon}.
The middle panels show the resulting pressure anisotropy as
a function of time, and
the rightmost panels display the nonlinearity (NL), again defined as the
difference between the pressure anisotropy of the given initial pulse
and twice the pressure anisotropy after halving the initial amplitude.

From the middle column of plots,
one sees that the magnitude of the pressure anisotropy decreases significantly
as the initial pulse is moved deeper into the bulk.
Moreover, both the time it takes for the effect of the pulse to reach
the boundary,
and the width of the resulting peaks in the pressure anisotropy,
grow with increasing depth of the initial pulse.
This reflects the usual holographic mapping between bulk and boundary:
phenomena deeper in the bulk correspond to lower energy or longer time
scales in the boundary field theory.

Turning to the nonlinearity plots in the right hand column,
one sees that the magnitude of the nonlinearity also decreases
as the pulse moves deeper into the bulk.
Dividing the peak nonlinearity by the peak pressure
anisotropy gives a relative measure of nonlinearity.
This is about
$1 \times 10^{-4}$ for the top row,
$3 \times 10^{-5}$ for the middle row, and
$3 \times 10^{-6}$ for the bottom row.
So in this comparison, as the initial pulse 
is pushed deeper into the bulk,
the relative nonlinearity \emph{decreases} systematically.

This comparison does not, however, imply that nonlinearities are
never significant.
The amplitude of the Gaussian pulse in the initial anisotropy
function was kept fixed in fig.~\ref{fig:BsubRavepos}, 
resulting in a decreasing size of the induced pressure anisotropy
as the pulse moves deeper into the bulk.
While the first two rows of the figure show pressure anisotropies
which are large departures from equilibrium,
$\Delta \mathcal P/\kappa \varepsilon \gg 1$,
the final rows with
$\Delta \mathcal P/\kappa \varepsilon \ll 1$
represent small departures from equilibrium.
It is natural to ask what happens if one instead
increases the amplitude as the pulse is moved into the bulk,
so as to keep fixed the size of the induced pressure anisotropy.

\begin{figure}
\centering
\hspace{0.25em}\suck[width=0.36\textwidth]{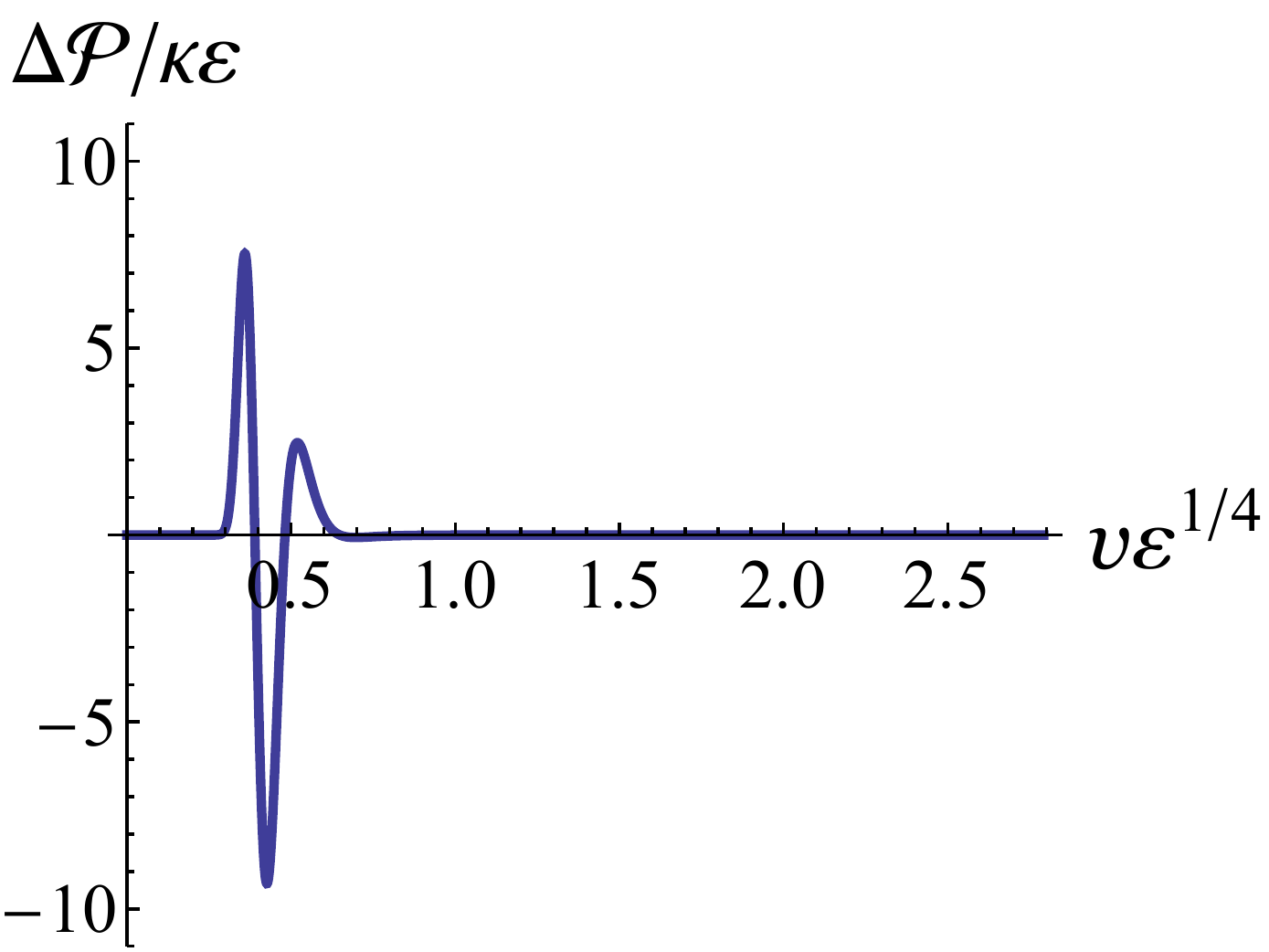}
\hspace{2.35em}\suck[width=0.36\textwidth]{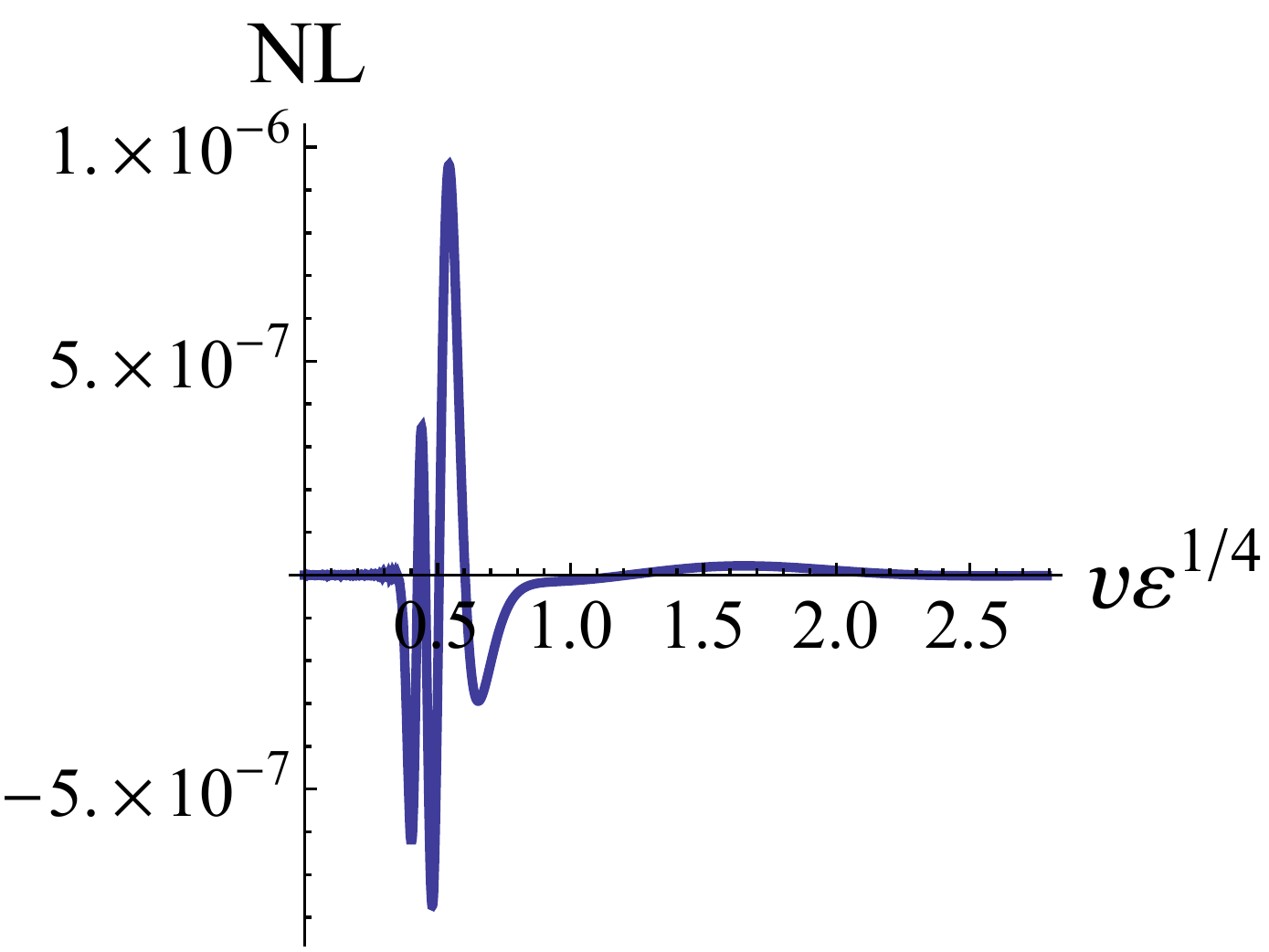}
\\[8pt]
\hspace{0em}\suck[width=0.36\textwidth]{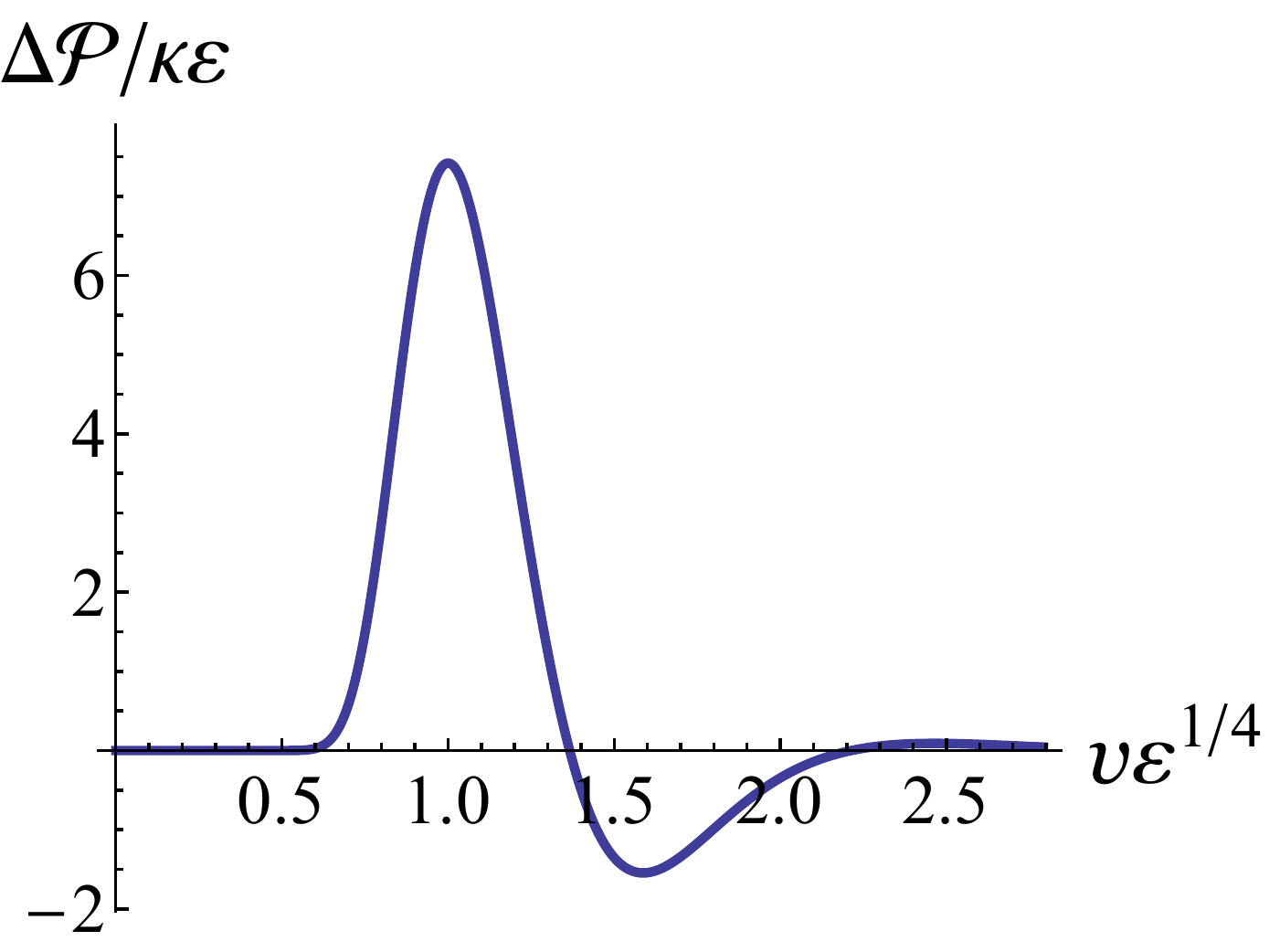}
\hspace{2em}\suck[width=0.36\textwidth]{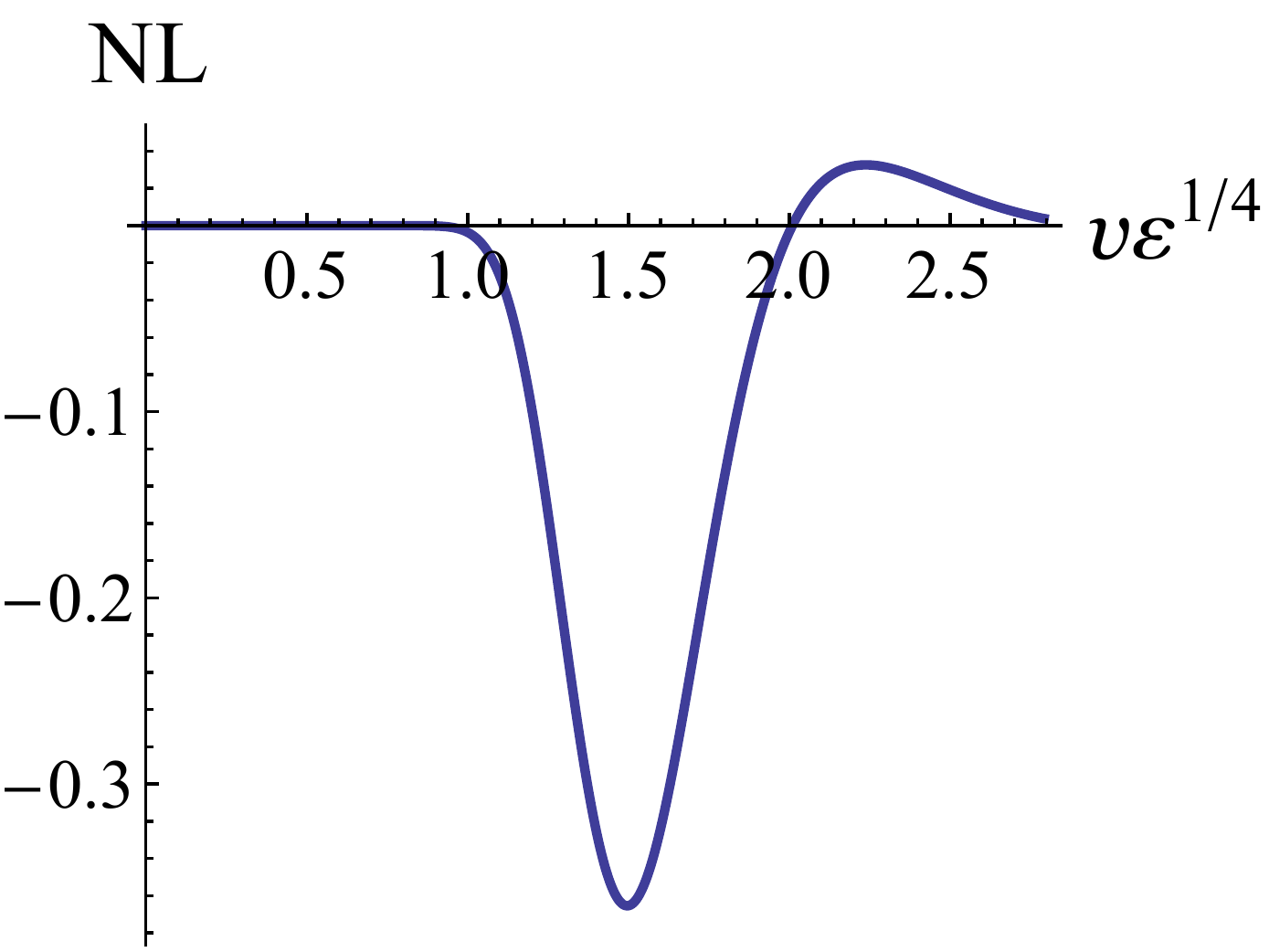}
\caption
    {%
    Top row: pressure anisotropy (left) and nonlinearity (right),
    defined as in fig.~\ref{fig:pressureDouble},
    as a function of time for the initial
    the pulse which created fig.~\ref{fig:BevolveSBB}.
    ($\mathcal A = 5\times10^{-4}$, $\rave = 4$, $\sigma = \tfrac 12$). 
    Bottom row: corresponding plots of pressure anisotropy
    and nonlinearity for a pulse located deeper in the bulk
    ($\mathcal A = 2.5$, $\rave = \tfrac 14$, $\sigma = \tfrac 12$)
    with amplitude adjusted to produce a similar peak
    pressure anisotropy.
    In both cases the energy density is $\varepsilon = \tfrac{3}{4} L^{-4}$.
    Substantial nonlinearity is present for this case,
    where an initial pulse deep in the bulk has
    sufficient amplitude to produce a large departure
    from equilibrium.
    \label{fig:MatchAnisRavepos5}
    }
\end{figure}

Such a comparison is shown in fig.~\ref{fig:MatchAnisRavepos5},
The upper row of the figure shows the time dependence of the
pressure anisotropy and the nonlinearity for the same pulse which generated
figs.~\ref{fig:BevolveSBB} and \ref{fig:pressureDouble}
($\mathcal A = 5\times10^{-4}$, $\rave = 4$, $\sigma = \tfrac 12$),
while the lower row shows the pressure anisotropy and nonlinearity
of a pulse with the same shape as the last row of
fig.~\ref{fig:BsubRavepos}, but with larger amplitude
($\mathcal A = 2.5$, $\rave = \tfrac 14$, $\sigma = \tfrac 12$).
The energy density remains fixed, $\varepsilon = \tfrac{3}{4} L^{-4}$.
The peak pressure anisotropy is similar in the two cases,
and corresponds to a large departure from equilibrium.
For the latter case of a pulse deep in the bulk, large enough
to induce a far from equilibrium pressure anisotropy,
the nonlinearity is significant, much larger than the
previous examples.
However, even for this case, the size of the nonlinearity
relative to the peak pressure anisotropy is only about 5\%,
NL$/(\Delta \mathcal P/\kappa\varepsilon) \approx 0.05$.%
\footnote
    {%
    We have also examined the level of nonlinearity
    by comparing the pressure anisotropy resulting
    from a sum of two different Gaussians to the sum
    of anisotropies induced by the individual Gaussians.
    The results were comparable to those discussed above
    and do not warrant separate discussion. 
    }

The data shown in fig.~\ref{fig:BsubRavepos}
inspire several further questions.
Looking down the middle column of the figure,
one sees that the onset of the response
(i.e., the time of the first peak in the pressure anisotropy)
increases as the initial pulse moves deeper into the bulk
but seems, perhaps, to be approaching a maximum value.
Is this really true, or can one craft initial data for which
the onset of the response is much greater?
As shown on the left panels of the lower rows of the figure, when the average position
$r_0$ of the Gaussian pulse is moved into the bulk, an increasingly large
portion of the Gaussian ends up lying behind the apparent horizon.
And when plotted as a function of our computational coordinate
$u = (\bar r {-} \lambda)^{-1}$
(with $\bar r$ the $\lambda = 0$ frame radial coordinate),
initial pulses with small values of $r_0$ are only moderately localized
near the horizon --- even though these pulses had constant widths
when viewed as functions of $r$.
As may be seen by comparing the left and middle columns
of fig.~\ref{fig:BsubRavepos}, it is the leading edge of the
anisotropy pulse
(the region of near-maximal slope)
which produces the first large response in the boundary anisotropy.
Looking at the last two rows of the figure, 
one may question whether we are doing an adequate job exploring
the response from initial disturbances which are localized close
to the horizon.
Will initial pulses which are more strongly localized near the horizon
show significantly greater nonlinearity?

\begin{figure}
\suck[width=0.32\textwidth]{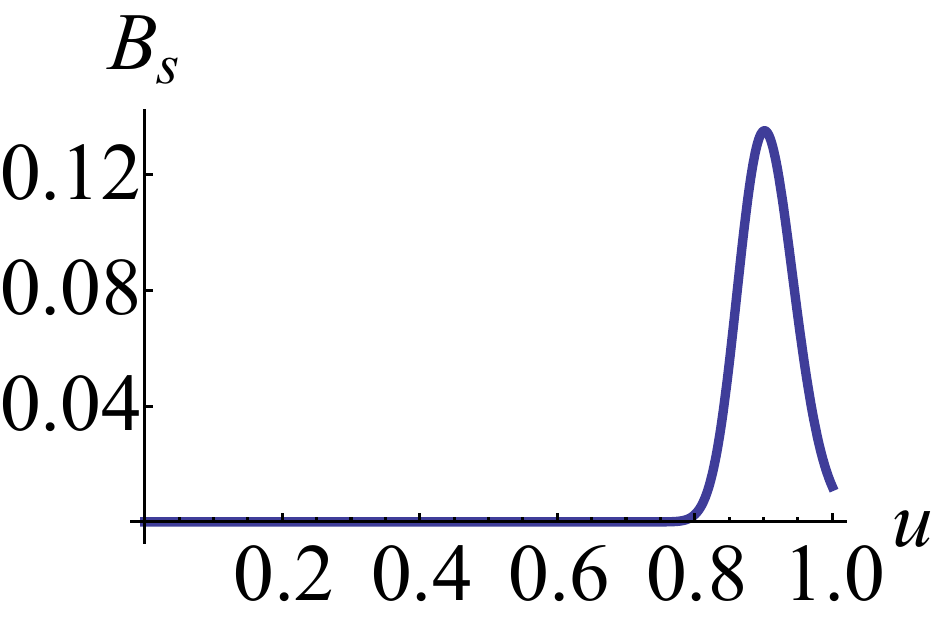}
\hfill
\suck[width=0.32\textwidth]{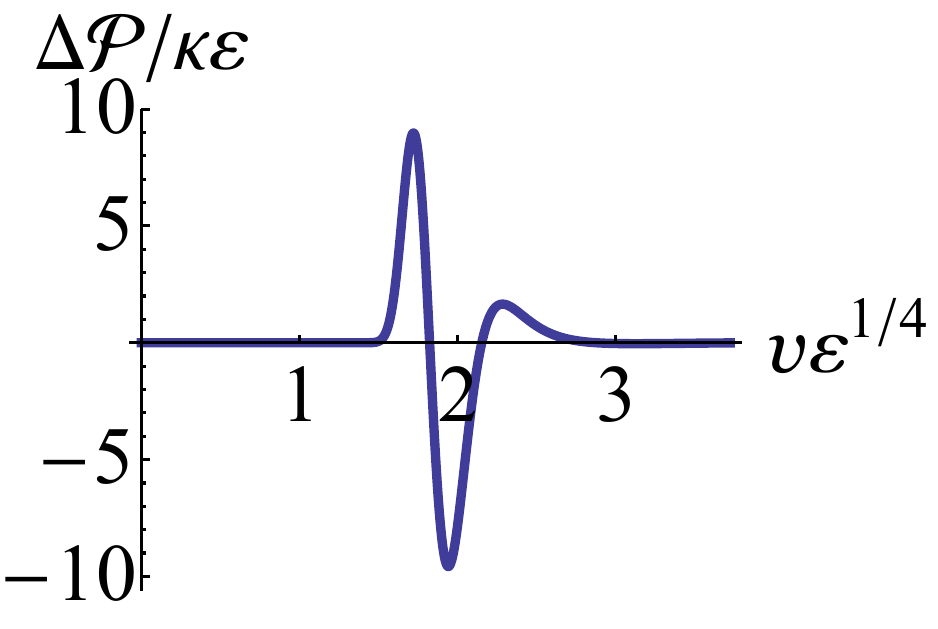}
\hfill
\suck[width=0.32\textwidth]{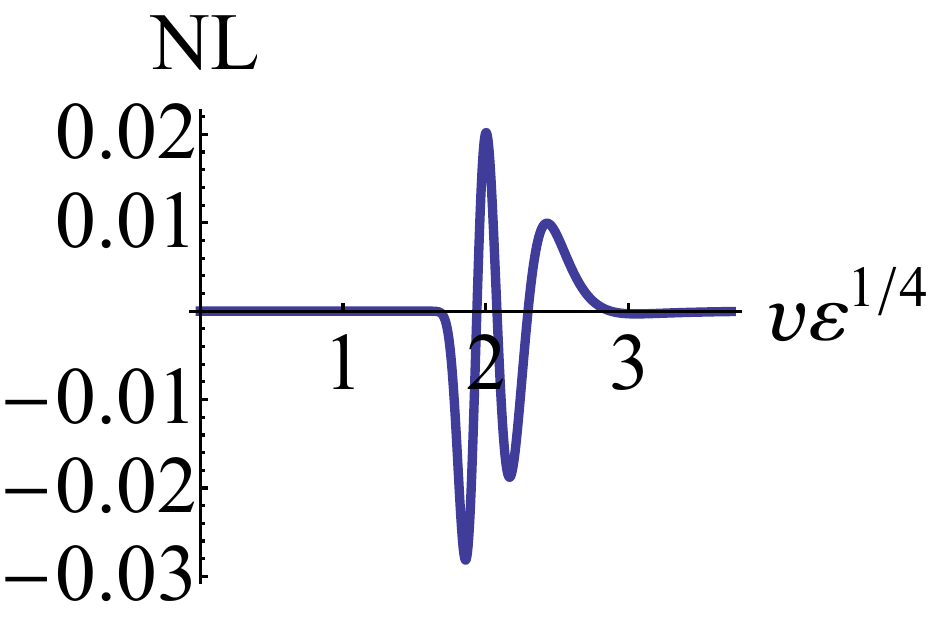}
\caption
    {%
    Initial anisotropy function (left),
    induced pressure anisotropy (middle),
    and nonlinearity (right)
    for a narrow ``deep pulse''
    ($\mathcal{A} = \tfrac{1}{10}$, $\rave = \tfrac {10}{11}$, $\sigma = \tfrac {1}{20}$)
    localized closer to the horizon than the deepest pulses of
    fig.~\ref{fig:BsubRavepos}.
    The energy density remains fixed at the same value, $\varepsilon = \tfrac{3}{4} L^{-4}$.
    Relative to the previous case of fig.\ref{fig:MatchAnisRavepos5},
    the induced pressure response has a delayed onset, but is otherwise very similar.
    \label{fig:DSpulse}
    }
\end{figure}

\begin{figure}
\suck[width=0.35\textwidth]{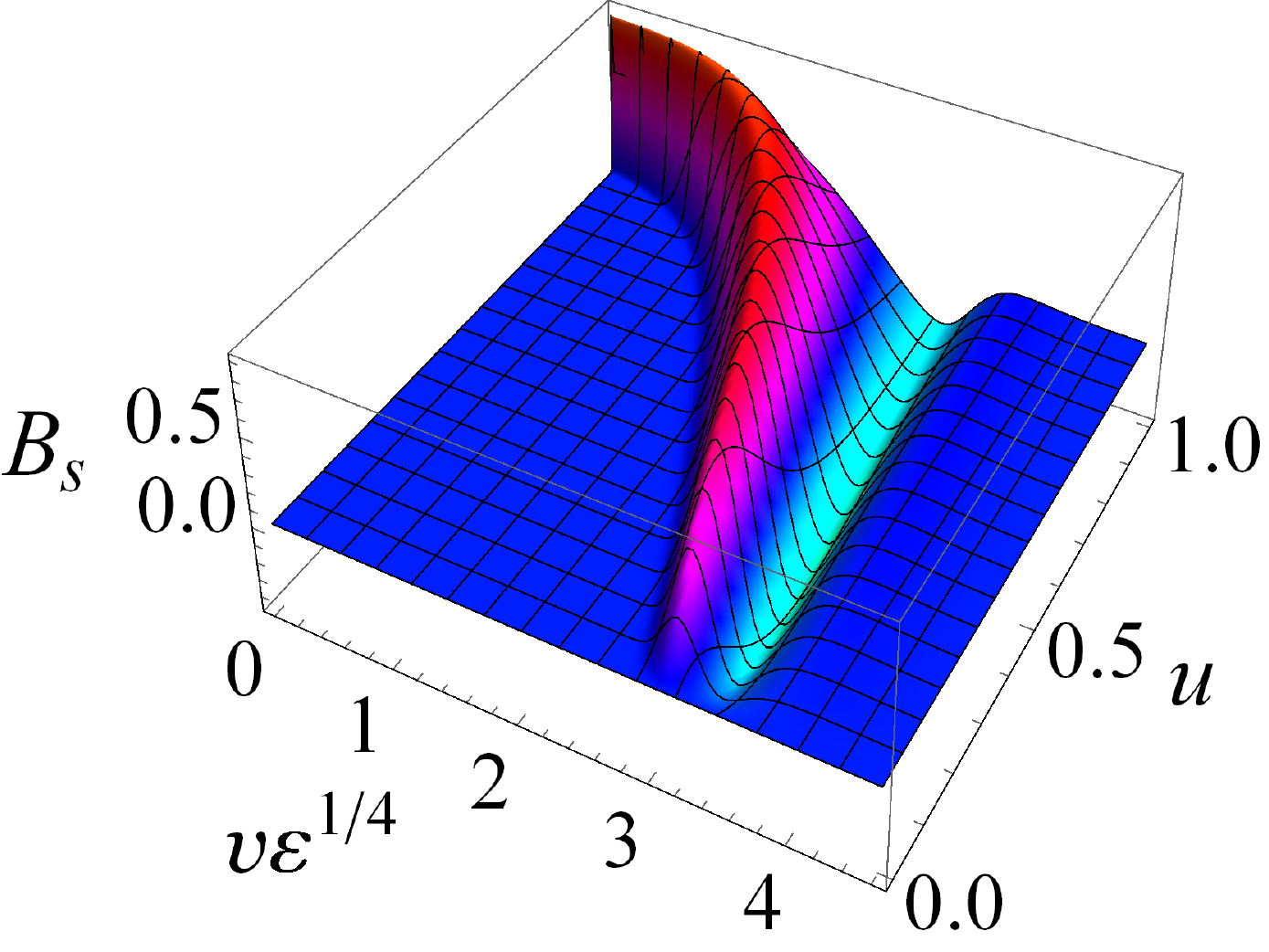}
\hfill
\suck[width=0.31\textwidth]{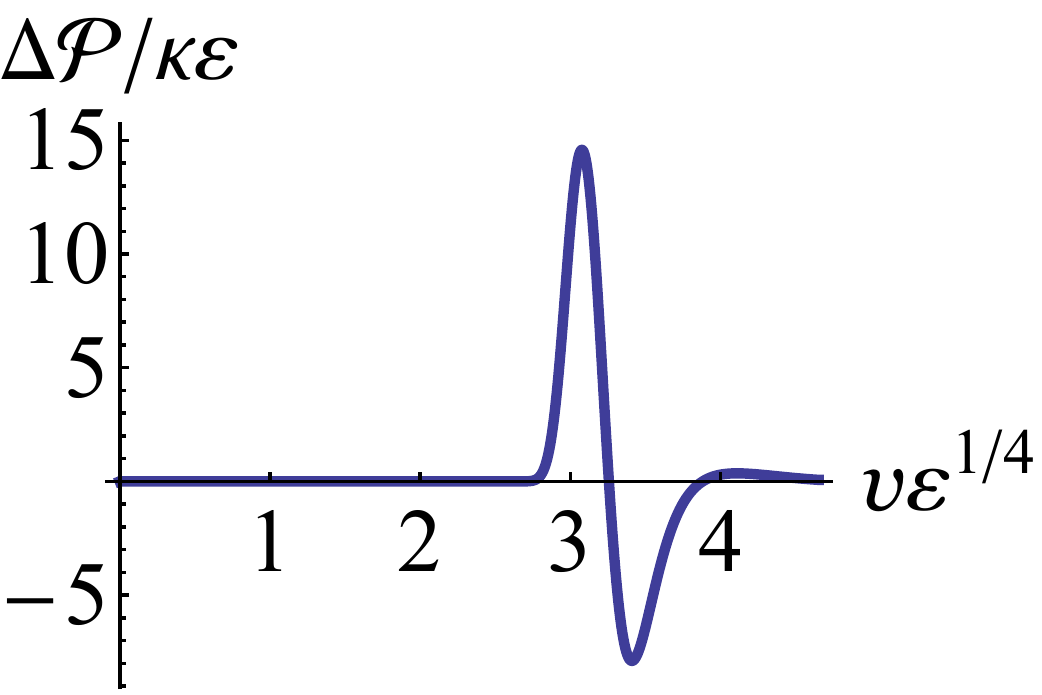}
\hfill
\suck[width=0.3\textwidth]{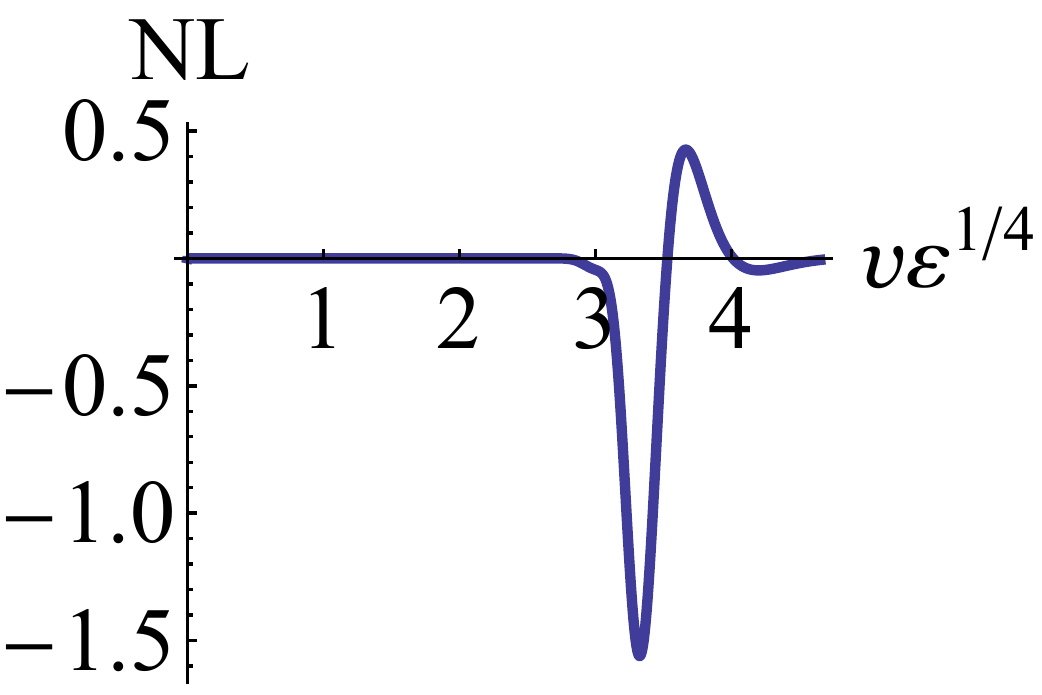}
\hspace*{-1em}
\caption
    {%
    Rescaled anisotropy function $B_s$ (left) as a function of inverted
    radius $u$ and time $v$, for an extremely narrow pulse sitting at
    the horizon
    ($\mathcal A = \tfrac{1}{10}$, $\rave =1$, $\sigma = \tfrac {1}{200}$).
    Resulting pressure anisotropy (middle) and nonlinearity (right)
    as a function of time.
    Energy density $\varepsilon = \tfrac{3}{4} L^{-4}$.
    The outward movement of the pulse toward the boundary clearly resembles
    the behavior of outgoing null geodesics originating very close to an
    event horizon, which can ``hug'' the horizon for extended periods
    before eventually escaping.
    \label{fig:obnoxiouspulse}
    }
\end{figure}

Figures \ref{fig:DSpulse} and \ref{fig:obnoxiouspulse} show results of
an effort to explore these questions.
Fig.~\ref{fig:DSpulse} shows
the initial anisotropy function,
along with the resulting pressure anisotropy and nonlinearity,
for a significantly narrower ``deep pulse'' 
($\mathcal A = \tfrac{1}{10}$, $\rave = \tfrac {10}{11}$, $\sigma = \tfrac {1}{20}$).
And fig.~\ref{fig:obnoxiouspulse} shows a 3D plot of the
time dependent anisotropy function, plus the induced pressure anisotropy,
for an extremely narrow deep pulse
($\mathcal A = \tfrac{1}{10}$, $\rave =1$, $\sigma = \tfrac {1}{200}$).%
\footnote
    {%
    This width is at the limit of what our numerics could do using a 240 point
    spectral grid.
    }
The energy density in both cases remains fixed, $\varepsilon = \tfrac{3}{4} L^{-4}$.
For the narrowest pulse, the amplitude $\mathcal A = \frac 1{10}$
is near the upper limit which can be studied without destabilizing the horizon.
In both figures \ref{fig:DSpulse} and \ref{fig:obnoxiouspulse}, 
the peak pressure anisotropy $\Delta \mathcal P/\kappa\varepsilon$
is large compared to unity, showing that
these pulses are producing far from equilibrium initial states.

As seen in these figures, pulses which are more sharply
localized very near the horizon do lead to a delayed onset
in the resulting pressure anisotropy, occurring at
$v \varepsilon^{1/4} \approx 1.75$ for the case of fig.~\ref{fig:DSpulse} and
$v \varepsilon^{1/4} \approx 3$ for our narrowest pulse in fig.~\ref{fig:obnoxiouspulse}.
But, within the range of pulse widths we have studied,
the onset of the pressure anisotropy response
is only delayed by a factor of 2--3
compared to the case of fig.~\ref{fig:MatchAnisRavepos5}.
From the left panel of fig.~\ref{fig:obnoxiouspulse},
showing the time dependence of the (rescaled) anisotropy function $B_s$,
one sees that the outward movement of the pulse toward the boundary resembles
the behavior of outgoing null geodesics originating very close to an event horizon,
which ``hug'' the horizon for extended periods before eventually escaping.
One may wonder if the onset of the response in the pressure anisotropy
could be delayed indefinitely by going to narrower and narrower initial pulses
localized at the apparent horizon.
We do not have a firm analytic argument, but doubt that this is 
possible if one simultaneously demands that the amplitude of the
response remain bounded away from zero.%
\footnote
    {%
    This expectation reflects the diverging redshift of late emerging
    outgoing geodesics in the geometric optics picture,
    and is consistent with the gapped spectrum of quasinormal mode frequencies
    for translationally invariant perturbations.
    }

The relative nonlinearity for the case of a narrow deep pulse
shown in fig.~\ref{fig:DSpulse} is quite small, about half a percent.
But for the extremely narrow pulse with near maximal amplitude of
fig.~\ref{fig:obnoxiouspulse},
the nonlinearity, relative to the pressure anisotropy, reaches the 10\% level.
Linearization of the dynamics about equilibrium must provide
an accurate approximation to the full nonlinear dynamics
when the deviation of the geometry from the equilibrium
Schwarzschild black brane solution is sufficiently small,
as will be true at sufficiently late times.
For asymptotically anti-de Sitter geometries,
where metric functions have the asymptotic forms (\ref{metricexpansions}),
this will also be the case for initial data involving perturbations
localized sufficiently close to the boundary.
(See refs.~\cite{Heller:2012km,Heller:2013oxa,Bantilan:2012vu}
for related discussion.)
It should be noted that
reasonably good agreement between linearized dynamics and full
nonlinear evolution was previously reported in ref.~\cite{Heller:2012km}
and further explored in ref.~\cite{Heller:2013oxa}.
In these works,
the authors found agreement at a 20\% level between the linearized
and full dynamics for a variety of initial anisotropy profiles.
Our results examining, more systematically,
the dependence of the relative nonlinearity on the parameters of our
initial Gaussian anisotropy function complement and extend this
earlier work.
Overall, despite prior knowledge of refs.~\cite{Heller:2012km,Heller:2013oxa},
we are still surprised by the small, often extremely small,
levels of nonlinearity which we find even at early times when
the induced pressure anisotropy is maximal,
for initial perturbations localized deep in the bulk and producing
large departures from equilibrium.

\subsection{Charged plasma}\label{sec:resultsRN}

We now turn to the equilibration of charged plasmas
(by which we mean SYM plasmas with a non-zero density
of the global $U(1)$ conserved charge --- not a plasma in which
electromagnetic forces are included in the dynamics and
Coulomb repulsion plays a significant role).
As noted earlier in section \ref{sec:eqsolns}, 
the bulk geometry should equilibrate to a non-singular Reissner-Nordstrom
black brane solution provided the charge and energy densities 
satisfy the extremality bound (\ref{eq:rhoext}),
$\rho < \rhomax = \tfrac 43\, \varepsilon^{3/4}$.
However, for values of the charge density near $\rhomax$,
we find that sufficiently large initial metric perturbations
can destroy the apparent horizon.
We expect that such initial data are unphysical, not representing
SYM initial states which could be produced by an operational
procedure such as turning on time dependent external fields
(which correspond to time dependent boundary data in the holographic
description).
In any case, we limit our attention to initial perturbations for which
an apparent horizon is present at all times.
We find that if one suitably decreases the amplitude of the initial
departure from equilibrium while increasing the charge density,
one can approach $\rhomax$ from below while maintaining
the existence of an apparent horizon.

\begin{figure}
\begin{center}
\hspace{0.2em}\suck[width=0.44\textwidth]{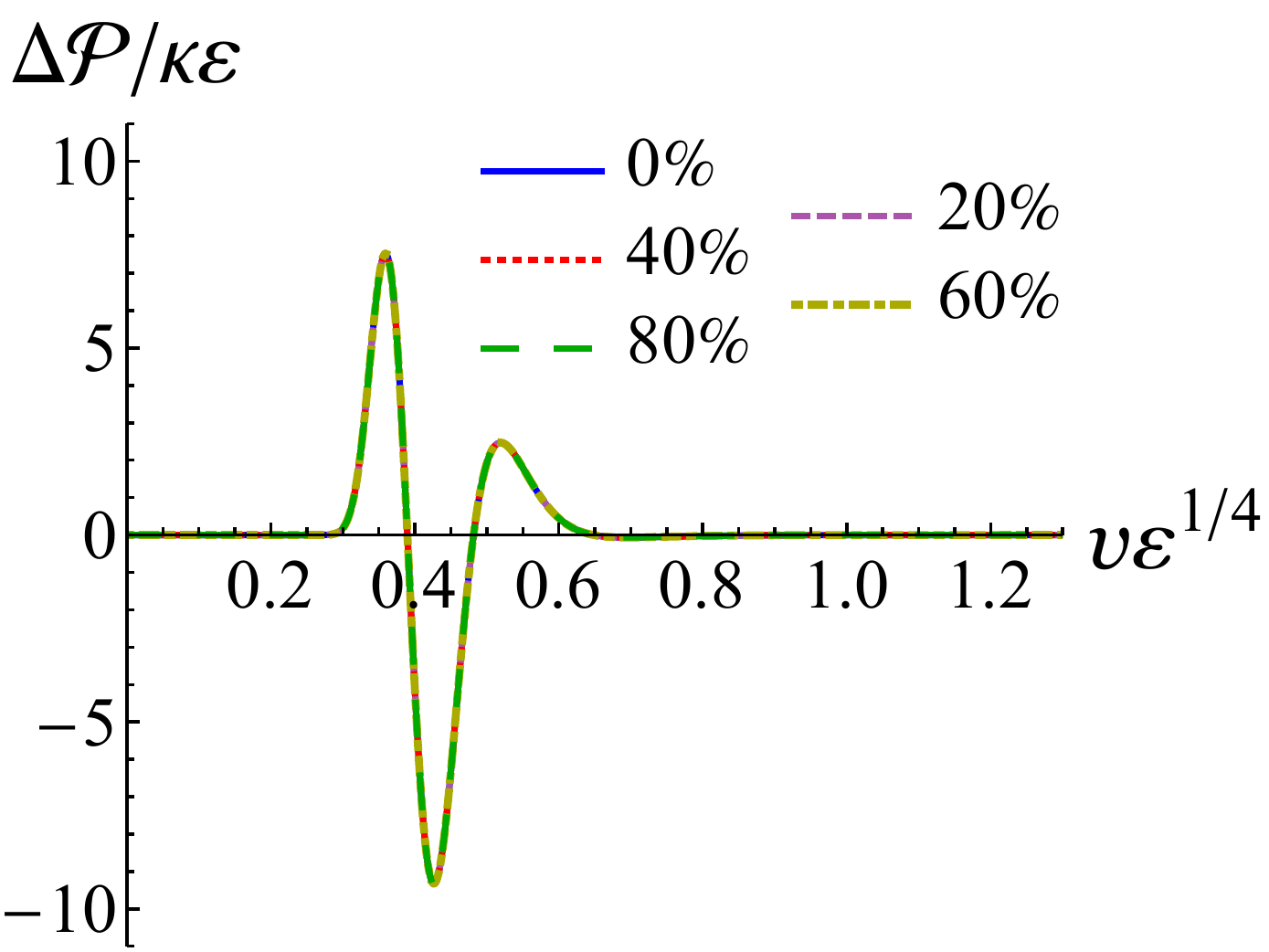}
\label{fig:pressureRNdiff}
\hspace{2em}\suck[width=0.44\textwidth]{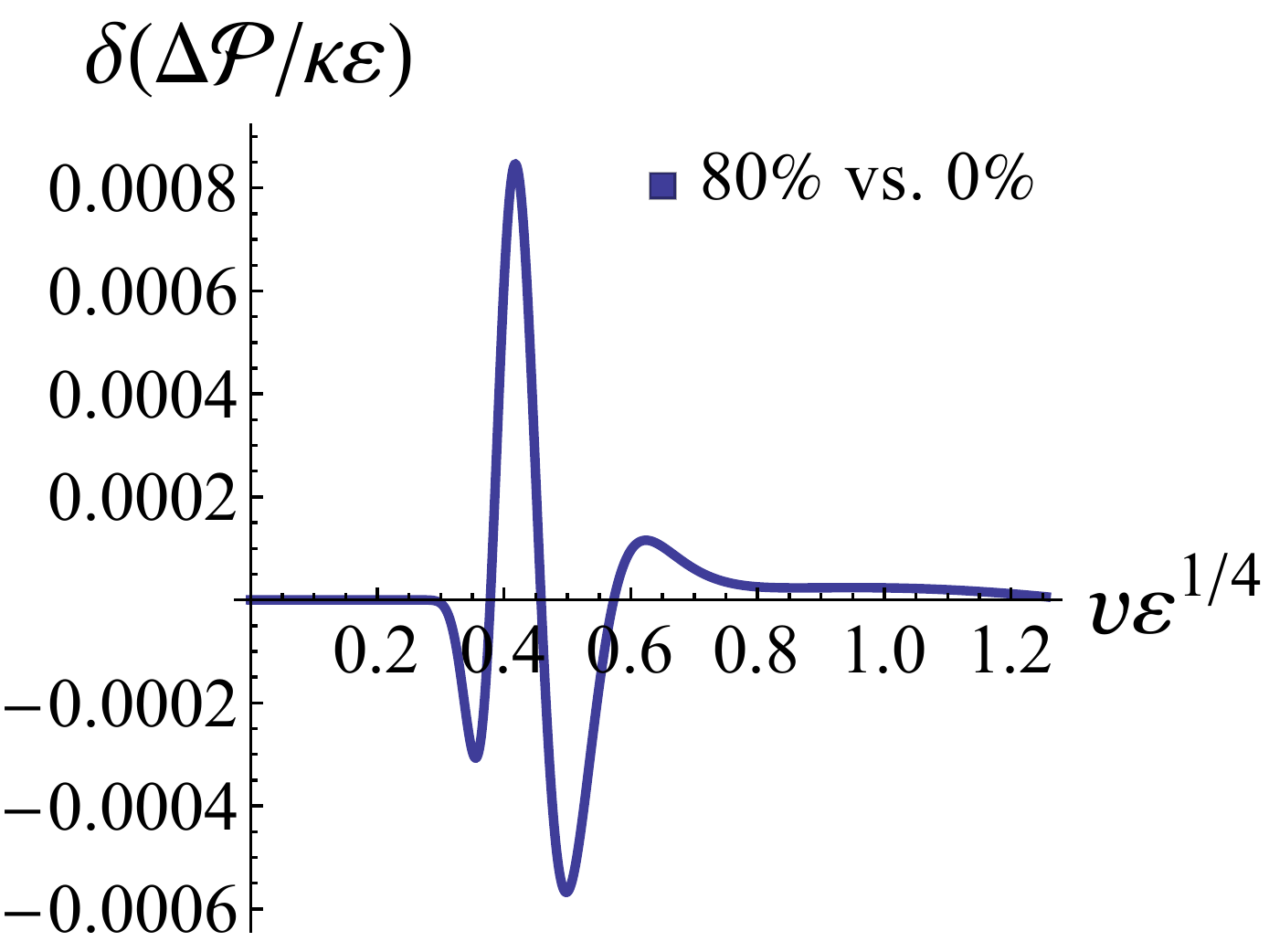}
\end{center}
\caption
    {%
    Left: time dependence of the pressure anisotropy
    (relative to $\kappa\varepsilon$)
    for values of the charge density $\rho$ which are
    0, 20, 40, 60, or 80\% of the extremal density $\rhomax$.
    The different curves are virtually indistinguishable.
    The initial anisotropy function $B(v_0,r)$ and energy density
    $\varepsilon$ are the same as in fig.~\ref{fig:BevolveSBB}.
    Right: difference in $\Delta \mathcal P/\kappa\varepsilon$
    between $\rho = 0.8 \, \rhomax$ and $\rho = 0$.
    \label{fig:pressureRN}
    }
\end{figure}

Figure \ref{fig:pressureRN} (left) compares the time dependence of
the pressure anisotropy
which results from initial data consisting of
precisely the same Gaussian initial anisotropy function $B(v_0,r)$
(in the $\lambda=0$ frame)
and energy density as in fig.~\ref{fig:BevolveSBB}, 
and a charge density $\rho$ equal to 0, 20, 40, 60, or 80\%
of the extremal density $\rhomax$.
The immediately obvious qualitative result is that the
five different curves are so close together than they cannot
be visually distinguished!
Varying the charge density (at fixed energy density and fixed initial
anisotropy function) has stunningly little impact on the
subsequent time evolution.
This is quantified in the right panel of fig.~\ref{fig:pressureRN}
which plots the difference in the pressure anisotropy 
$\Delta\mathcal P/\kappa\varepsilon$ between the cases
of $\rho = 0.8 \, \rhomax$ and $\rho = 0$.
Comparing the scales on the right and left hand plots,
one sees that for this initial anisotropy function
the sensitivity to the charge density is less than one part in $10^4$.

In the plots of fig.~\ref{fig:pressureRN}, we used the fourth root
of the (rescaled) energy density, $\varepsilon^{1/4}$,
to set the scale for time.
Since the energy density was held constant in the comparisons of
fig.~\ref{fig:pressureRN}, this was a simple and convenient choice.
For the five cases shown in the figure, 
$\varepsilon/(\pi T)^4 = 0.75$, $0.76$ , $0.79$, $0.86$, and $1.03$ 
for $\rho = 0$, 20, 40, 60 and 80\% of $\rhomax$, respectively.
And, for comparison, the values of the equilibrium chemical
potentials corresponding to these charge densities are
given by $\mu/T = 0$, 0.34, 0.73, 1.26, and 2.21,
respectively.

\begin{figure}
\begin{center}
\suck[width=0.44\textwidth]{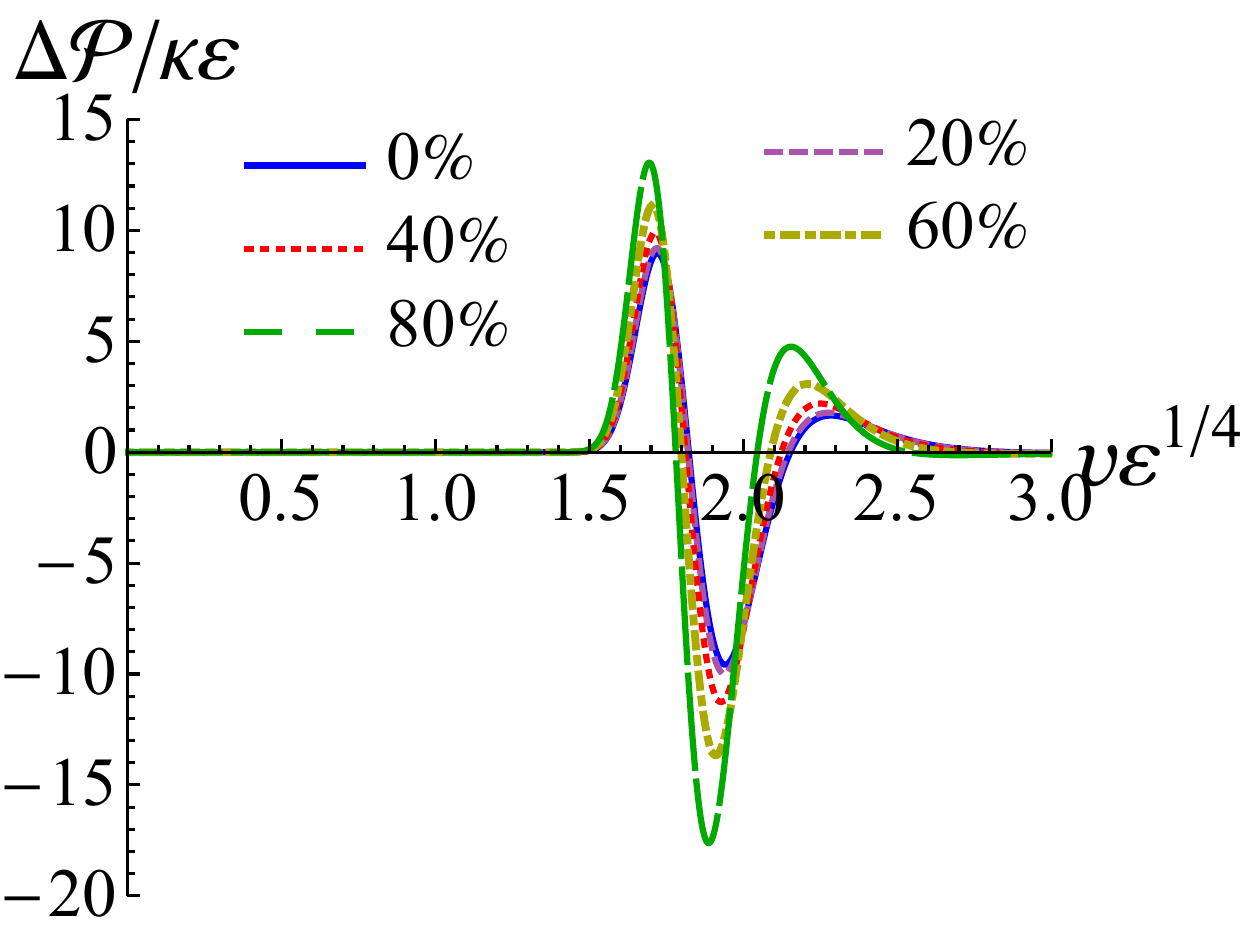}
\hspace{1.95em}\suck[width=0.44\textwidth]{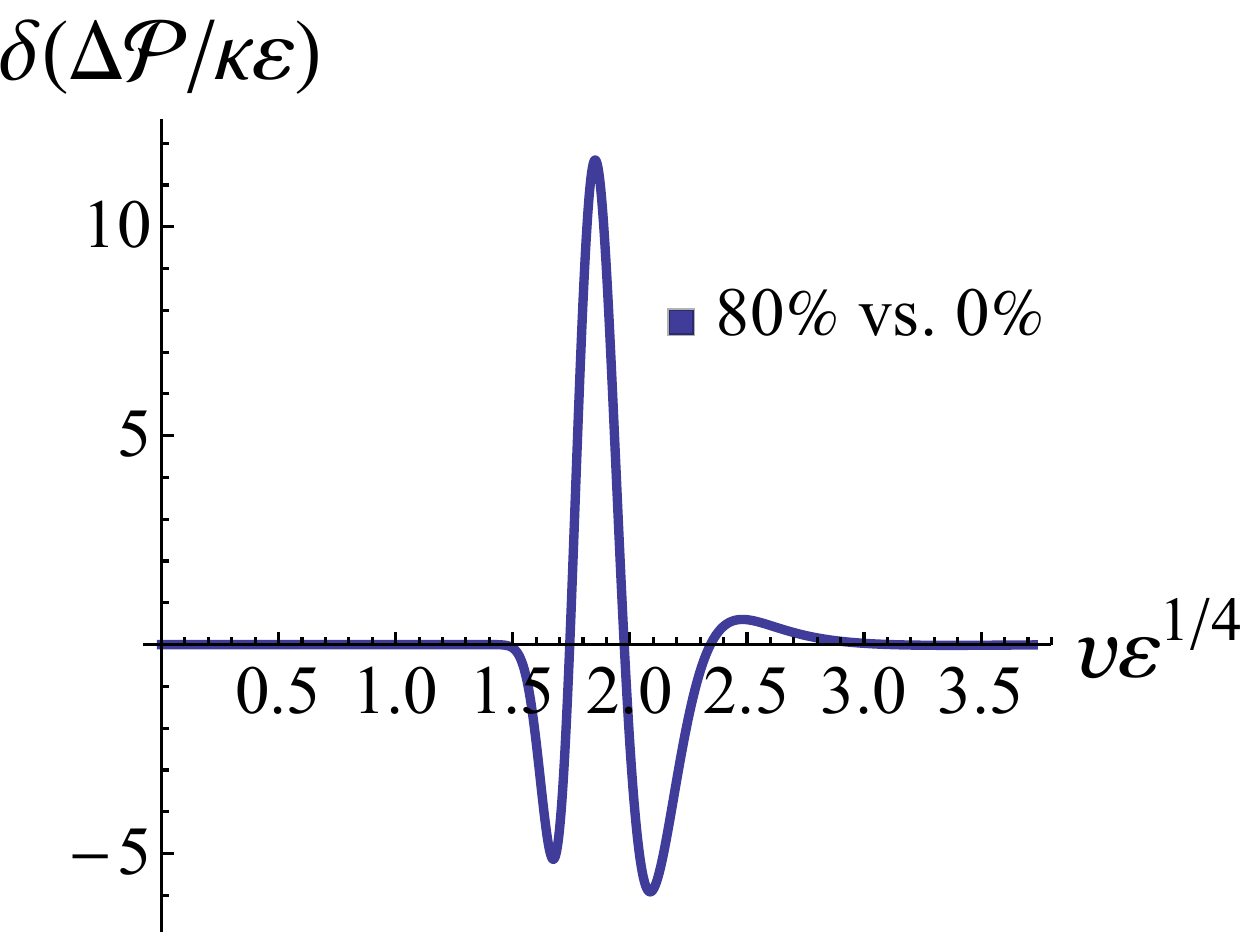}
\end{center}
\caption
    {%
    Left: comparison of pressure anisotropies produced by
    different charge densities, up to 80\% of extremality,
    when the initial anisotropy function is a
    ``deep pulse''
    ($\mathcal{A} = \tfrac{1}{10}$, $\rave = \tfrac {10}{11}$,
    $\sigma = \tfrac {1}{20}$)
    with $\varepsilon = \tfrac 34 \, L^{-4}$.
    Although the amplitude of the response grows, as shown,
    with increasing charge density,
    it is striking how little the time course of the response varies.
    Right: difference in pressure anisotropy
    $\Delta \mathcal P/\kappa\varepsilon$ between
    $\rho = 0.8 \, \rhomax$ and $\rho = 0$.
    \label{fig:DSRN}
    }
\end{figure}

If the initial anisotropy pulse begins deeper in the bulk,
then the sensitivity to the charge density is larger.
Figure~\ref{fig:DSRN} shows a comparison
of pressure anisotropies for differing charge densities,
now using the deep pulse initial anisotropy function
whose radial profile has the shape shown
in fig.~\ref{fig:DSpulse}.%
\footnote
    {%
    One might guess that the pulse used in the bottom row of
    fig.~\ref{fig:MatchAnisRavepos5} would exhibit greater sensitivity
    to charge since it had a larger nonlinearity than the deep pulse
    of fig.~\ref{fig:DSpulse}.
    However, the latter pulse 
    exhibits much greater sensitivity to charge.
    }
As seen on the left panel,
the amplitude of the response increases significantly
as the charge density varies from 0 to 80\% of $\rhomax$.
However, the time course of the equilibration
(e.g., the times of the first or second peaks in the response,
or the zero-crossing between these peaks)
is only modestly affected, with changes of $3\%$ percent or less.

The fact that the sensitivity to the charge density is greatest
for pulses close to the horizon is to be expected.
In the equilibrium geometry (\ref{eq:URN}), one sees that
as the radius $\rtilde$ increases from the horizon,
the influence of the charge density decreases rapidly 
relative to the other terms in the metric.
Only near the horizon, and close to extremality, does the
charge density produce an $O(1)$ effect on the equilibrium geometry.

At the beginning of this work, we expected that
one interesting outcome would be information on the
change in equilibration time produced by varying the
plasma charge density.
By ``equilibration time'', we mean some rough but useful
measure of when the departure from equilibrium is no
longer substantial.
To make this a bit more quantitative we
adopt, somewhat arbitrarily, the criterion
\begin{equation}
    \Delta \mathcal{P}(t)/\kappa\varepsilon \leq 0.1 \,,
\label{eq:teq}
\end{equation}
for all times $t > \teq$, as indicating that
the system is near equilibrium at time $\teq$.

Looking at the left panels of figures \ref{fig:pressureRN}
and \ref{fig:DSRN}, it is clear that the
effect of the charge density on any reasonable measure of
equilibration time can be summarized easily:
there is very little effect! Even in fig.~\ref{fig:DSRN},
where the sensitivity to charge density is
the largest we have found with 
horizon-preserving initial data,%
\footnote
    {%
    Achieving good numerical accuracy becomes increasingly 
    difficult as one pushes toward extremality, where
    the equilibrium solution bifurcates.
    Investigating very near extremal behavior more carefully is
    an interesting topic we leave to future work.
    }
the time $\tpeak$ of the initial response peak and the
approximate equilibration time $\teq$ both vary by less than~5\%.

\begin{figure}
\centering
\suck[width=0.44 \textwidth]{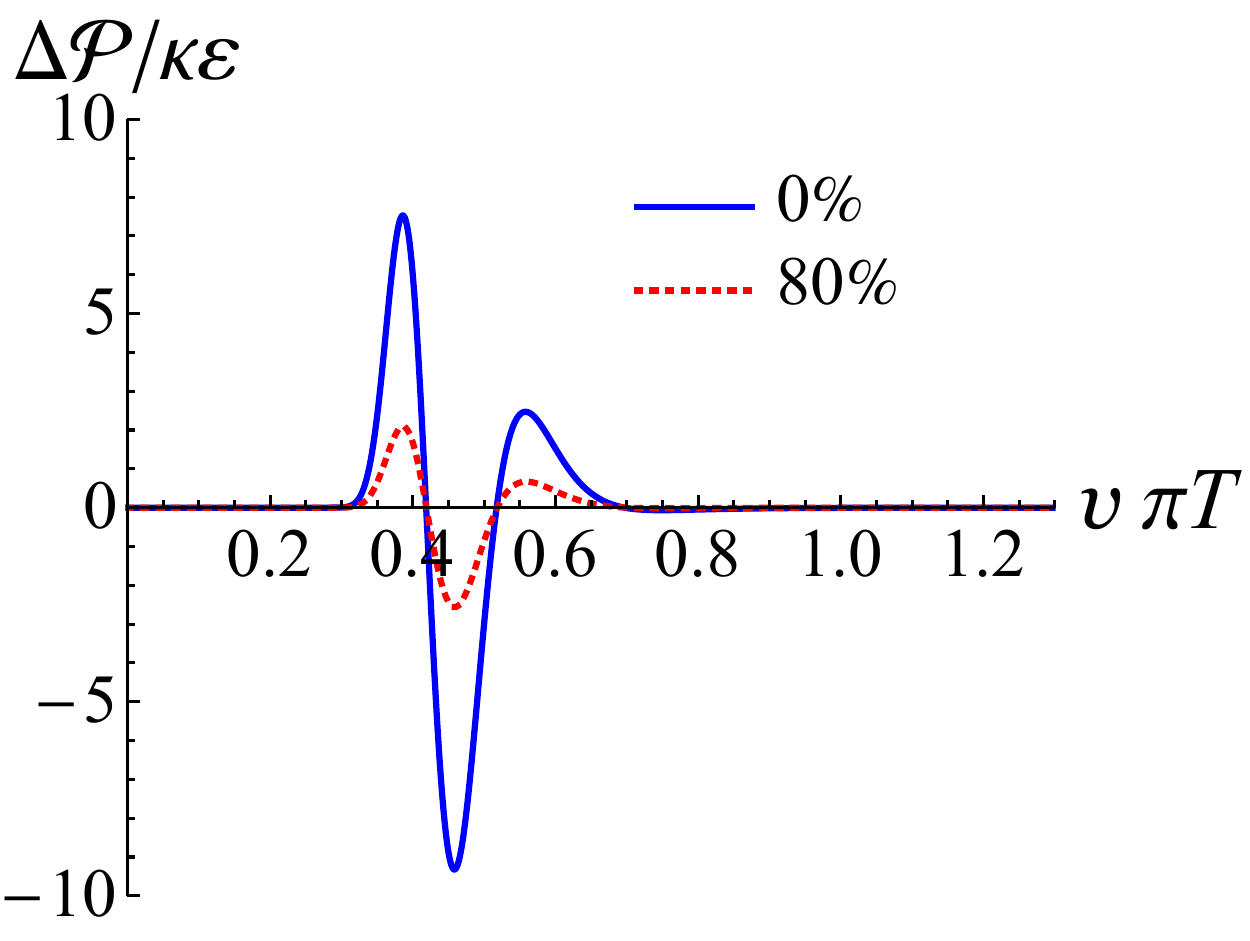}
\caption
    {%
    Comparison of the pressure anisotropies produced by two different
    charge densities, 0\% and 80\% of extremality,
    when holding fixed the equilibrium temperature, $\pi T = 1/L$.
    The energy densities are 0.75 and 2.68, respectively.
    The initial anisotropy function is same as in fig.~\ref{fig:BevolveSBB}
    ($\mathcal A = 5\times10^{-4}$, $\rave = 4$, $\sigma = \tfrac 12$).
    With time is plotted in units of inverse temperature,
    the relaxation time course shows negligible sensitivity to the charge density
    (although the amplitude of the response varies significantly).
    \label{fig:pressureRNconstT}
    }
\end{figure}

Since figures \ref{fig:pressureRN} and \ref{fig:DSRN} plot time in units
set by the energy density,
the high degree of insensitivity of the relaxation time course
to the charge density seen in these figures might lead one to think
that the total energy density is playing a special role
in setting the time scale of relaxation.
But it should be borne in mind that these figures
show comparisons in which, by design,
both the initial anisotropy function (in the $\lambda=0$ frame)
\emph{and} the total energy density have been held fixed.
Because the ratio of energy density to temperature (to the fourth power)
varies significantly with increasing charge density,
it is natural to ask whether the degree of (in)sensitivity
of the relaxation dynamics to the charge density is substantially
different if one holds fixed the equilibrium temperature instead
of the energy density.
Figure~\ref{fig:pressureRNconstT} shows such a comparison.
Plotted are the pressure anisotropies resulting from 
the same initial anisotropy function used in figs.~\ref{fig:BevolveSBB}
and \ref{fig:pressureRN},
and charge densities of either 0 or 80\% of extremality,
but now with the energy density in either case suitably
adjusted to fix the equilibrium temperature, $\pi T = 1/L$.
One again sees a significant change in the amplitude of the
response with increasing charge density, but now with the
temperature held fixed, increasing charge density decreases the
amplitude of the response.
Nevertheless, with time now plotted in units of $(\pi T)^{-1}$,
one again sees negligible ($\approx 0.01\%$) change 
in the time course of the
equilibration dynamics as the charge density varies from
zero and 80\% of $\rhomax$.

Performing the same constant temperature response comparison
using the deep pulse initial anisotropy function
(whose radial profile is shown in fig.~\ref{fig:DSpulse}),
we find a larger --- but still quite small ---
variation in the time course,
approximately $2\%$,
as the charge density varies from zero to 80\% of $\rhomax$.

We have also examined the degree of nonlinearity in the above
examples of equilibrating charged plasmas.
The results for the relative size of the nonlinearity are quite similar
to our earlier results for equilibrating uncharged plasmas.
Because of this, we will refrain from presenting explicit nonlinearity
plots for charged plasmas.

\begin{table}
\begin{center}
\footnotesize
\begin{tabular}{c|ll|c|c}
\toprule
\multicolumn{5}{c}{Charged ($\rho \ne 0$, $\B = 0$)} \\
\midrule
 $~\rho/\rhomax$ 
 & \hspace{2.3em}$\Re\,\lambda /\varepsilon^{1/4}$
 & \hspace{3.0em}$\Im\,\lambda /\varepsilon^{1/4}$
 & $\lambda /\pi T$
 & Linearized $\lambda/\varepsilon^{1/4}$
\\
\midrule
 0.0 & $3.35208 \pm 0.00004$ & $-2.95144 \pm 0.00013$ & $3.11946 -2.74663\,i$ & $3.35207 -2.95150\,i$  \\
 0.1 & $3.34564 \pm 0.00016$ & $-2.95468 \pm 0.00019$ & $3.11948 -2.75763\,i$ & $3.34568 -2.95460\,i$   \\
 0.2 & $3.32624 \pm 0.00020$ & $-2.96429 \pm 0.00019$ & $3.13222 -2.79139\,i$ & $3.32630 -2.96444\,i$  \\
 0.3 & $3.29319 \pm 0.00028$ & $-2.98266 \pm 0.00036$ & $3.14987 -2.85285\,i$ & $3.29327 -2.98287\,i$   \\
 0.4 & $3.24572 \pm 0.00007$ & $-3.01376 \pm 0.00008$ & $3.17857 -2.95141\,i$ & $3.24574 -3.01377\,i$  \\
 0.5 & $3.18366 \pm 0.00016$ & $-3.06498 \pm 0.00007$ & $3.22529 -3.10506\,i$ & $3.18370 -3.06491\,i$  \\
 0.6 & $3.11311 \pm 0.00002$ & $-3.15177 \pm 0.00002$ & $3.31032 -3.35142\,i$ & $3.11311 -3.15176\,i$  \\
 0.7 & $3.07021 \pm 0.00006$ & $-3.29402 \pm 0.00006$ & $3.50617 -3.76176\,i$ & $3.07022 -3.29399\,i$  \\
 0.8 & $3.11863 \pm 0.00265$ & $-3.41376 \pm 0.00295$ & $3.99199 -4.36977\,i$ & $3.11848 -3.42004\,i$  \\
\bottomrule
\end{tabular}
\end{center}
\caption
    {%
    Lowest quasinormal mode frequency
    for charge densities ranging from 0 up to 80\% of extremality.
    The second and third columns show results (with estimated uncertainties)
    for the real and imaginary part of the QNM frequency
    in units of $\varepsilon^{1/4}$,
    obtained from fitting the late time behavior of
    our full nonlinear evolution.
    The fourth column shows these same results
    converted to units of $\pi T$.
    (Fractional uncertainties are the same as in the preceding columns.)
    The rightmost column shows results from an independent analysis
    of the linearized small fluctuation equations by
    Janiszewski and Kaminski \cite{Janiszewski:2015ura}.
    \label{QNM-RN}
    }
\end{table} 

Finally, as noted earlier,
at sufficiently late times the relaxation
must be accurately described by a superposition of quasinormal modes
(eigenfunctions of the linearized dynamics about the equilibrium solution).
Extracting the leading quasinormal mode frequency by fitting the late time
($4 \lesssim v  \varepsilon^{1/4} \lesssim 20 $) 
behavior of our calculated pressure anisotropy to a decaying, oscillating
exponential, as described in the previous section,
yields the results shown in table~\ref{QNM-RN}.
The second and third columns (with uncertainties) show
our estimates for the real and imaginary parts of the leading
QNM frequency in units of $\varepsilon^{1/4}$, while
the fourth column (without uncertainties) shows our estimates
converted to units of $\pi T$.
The rightmost column shows independent results of
Janiszewski and Kaminski \cite{Janiszewski:2015ura} obtained by analyzing the
linearized small fluctuation equations about the RN black brane solution.
The agreement is a satisfying confirmation of our
numerical accuracy.
Interestingly, the imaginary part of the lowest QNM frequency varies by 
over 15\% between $\rho = 0$ and $\rho = 0.8\, \rhomax$.
This is enormously larger than the part in $10^4$ sensitivity
seen in fig.~\ref{fig:pressureRN}, and substantially bigger than the
largest ($\approx 5\%$) sensitivity we found in the evolution
time course with deep pulses,
fig.~\ref{fig:DSRN}.
The implications of these very differing sensitivities will be
discussed further in section~\ref{sec:discussion}.


\subsection{Magnetized plasma}\label{sec:mag}

We now present results of an analogous investigation 
of equilibration in plasmas (with vanishing charge density)
in a homogeneous magnetic field $\B$.
Our discussion will parallel, as much as possible, the previous
treatment of charged plasmas.
But the breaking of scale invariance by the magnetic field,
discussed in section \ref{sec:N=4},
produces some notable differences.
One change, seen in section \ref{sec:asymptotics},
is that the anisotropy function $B(v,r)$ must now contain
logarithmic terms in its near boundary behavior.
We choose our initial anisotropy function to have the form
(\ref{eq:gaussMAG}) in which an adjustable Gaussian is added
to the required leading logarithmic term.
As in the previous discussion of charged plasmas,
we will keep fixed the parameters
of the Gaussian part of the initial anisotropy function $B(v_0,r)$
as we dial up the external magnetic field.
We will also hold fixed the energy density
$\epsL$
defined at a renormalization point $\mu = 1/L$.
As shown by the holographic relation (\ref{eq:T}),
this is the same as holding fixed the asymptotic coefficient $a_4$.
In the following plots, axis labels involving energy density
$\varepsilon$, without any explicit
indication of scale, will denote the energy density evaluated
at the curvature scale, $\varepsilon(1/L) = \epsL$.
Similarly, the pressure anisotropy $\Delta\mathcal P$ should
be understood as $\Delta\mathcal P(1/L)$ unless otherwise
indicated explicitly.

\begin{figure}
\centering
\quad
\hspace*{-1em}
\suck[width=0.52\textwidth]{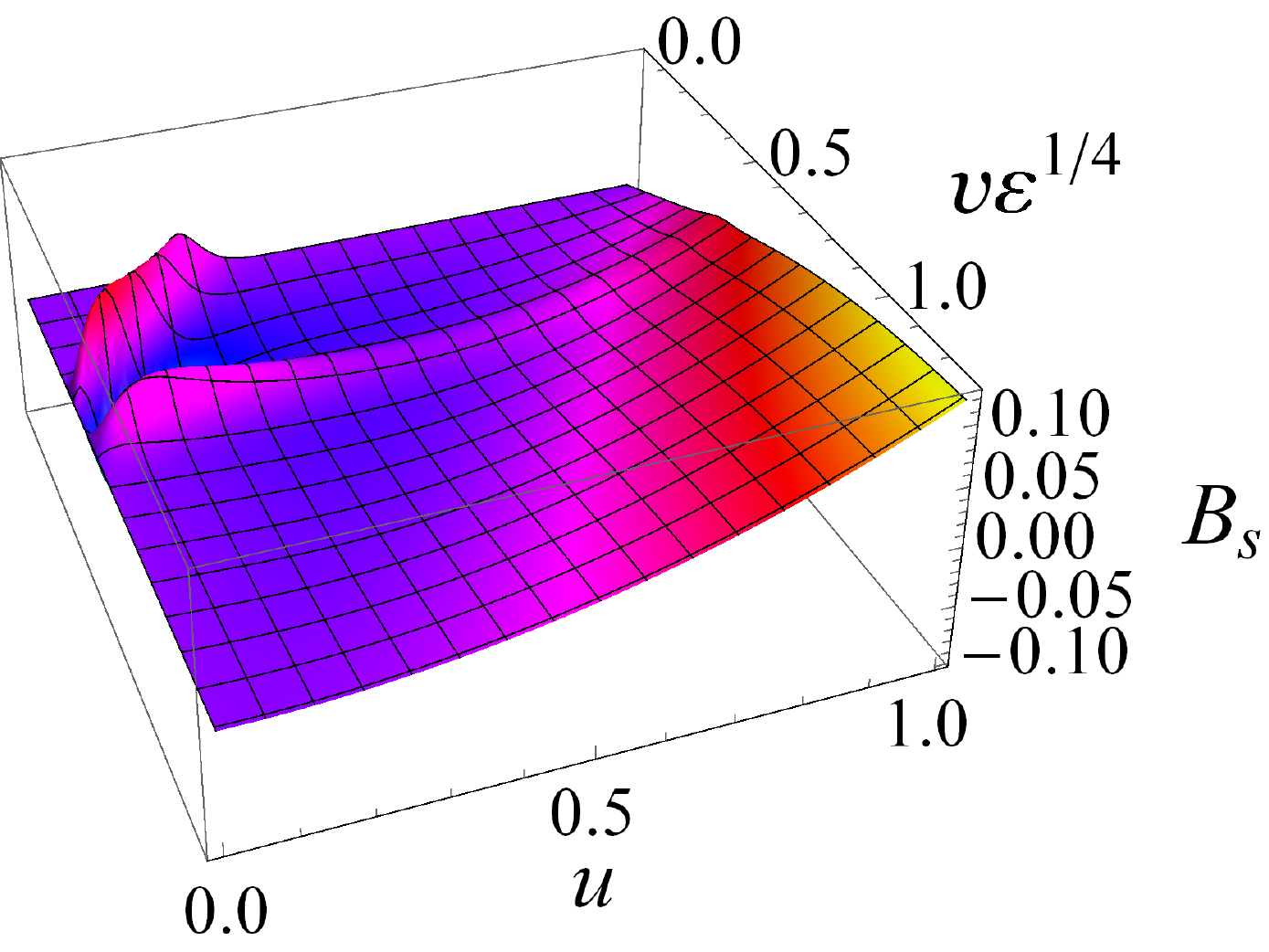}
\quad
\raisebox{20pt}{\suck[width=0.44\textwidth]{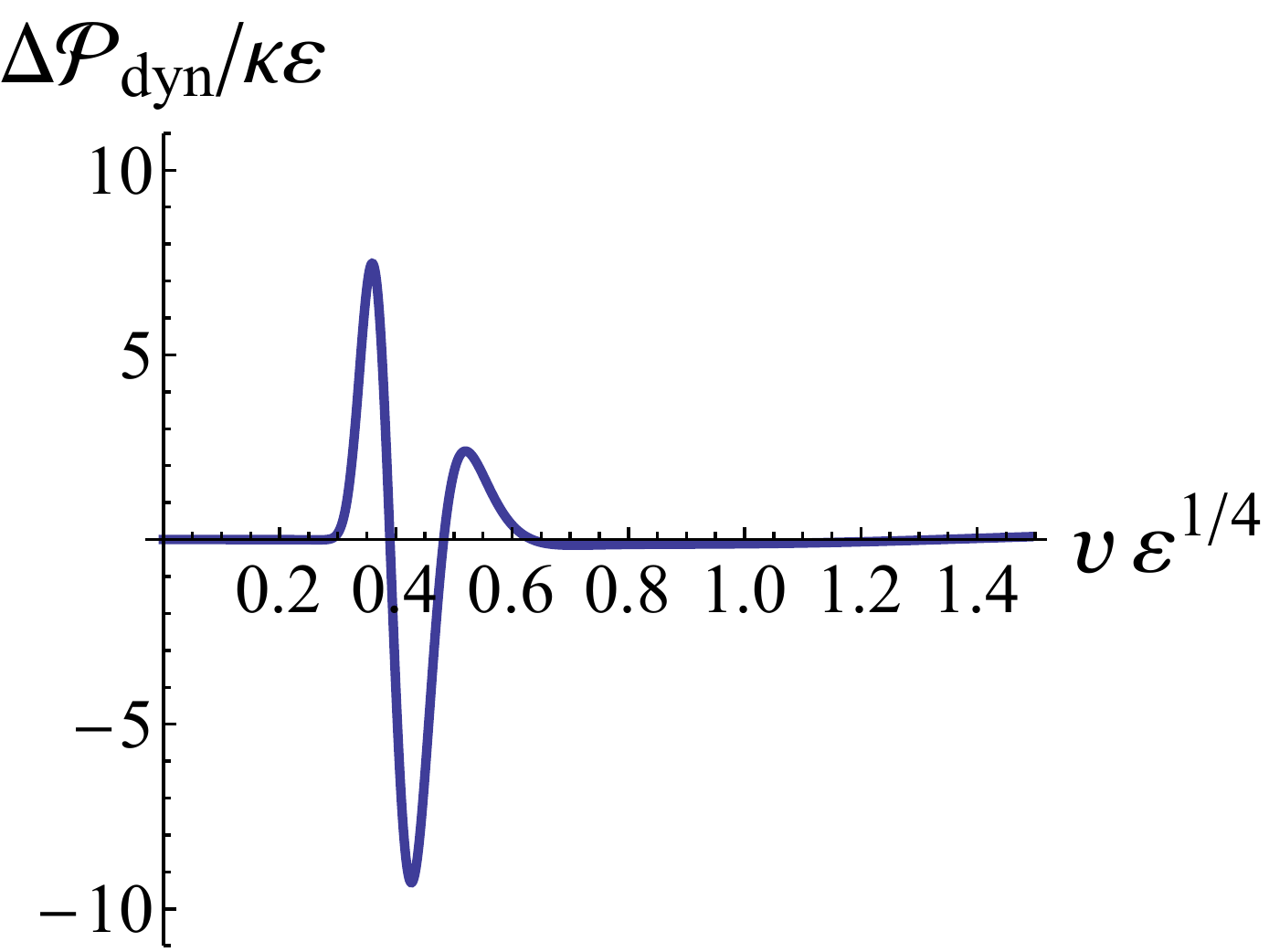}}%
\hspace*{-1em}
\caption
    {%
    Left: The subtracted/rescaled anisotropy function, $B_s(v,u)$,
    as a function of time $v$ (in units of $\epsL^{-1/4}$)
    and inverse radial depth $u$,
    for the case of $\B L^2 = 1.0$, with initial Gaussian 
    parameters chosen to match the initial pulse which
    generated fig.~\ref{fig:BevolveSBB} 
    ($\mathcal A = 5\times10^{-4}$, $\rave = 4$, $\sigma = \tfrac 12$),
    and energy density $\epsL = \tfrac 34 \, L^{-4}$.
    At late times the anisotropy function approaches the non-trivial profile
    of the equilibrium magnetic brane solution discussed
    in section~\ref{sec:eqsolns}.
    Right:
    Corresponding evolution of the dynamical contribution to the
    relative pressure anisotropy ,
    $\Delta\mathcal P_{\rm dyn}/\kappa \epsL$,
    with both $\Delta \mathcal P$ and $\epsL$ evaluated
    at the scale $1/L$.
    The late time limit of the pressure anisotropy is non-zero,
    but too small to be easily visible,
    $
	\lim_{v\to\infty}
	\Delta\mathcal P_{\rm dyn}(1/L)/\kappa\epsL
	= 0.22
    $.
    \label{fig:BevolveMAG}
    }
\end{figure}

Instead of keeping $\epsL$ fixed as 
the magnetic field $\B$ is varied, one could choose to hold fixed
the energy density $\epsB$ defined at the scale set by the magnetic field,
$\mu = |\B|^{1/2}$.
Since the AdS curvature radius $L$ is not a physical scale
present in the dual QFT,
fixing $\epsB$ instead of
$\epsL$ is arguably more natural.
However, for computational reasons it is easier
to hold fixed $\epsL$ as the magnetic field is increased.
The issue is that accurate numerical calculations become
progressively more difficult the deeper one penetrates into
the high-field/low-temperature regime, $T^2/|\B| \ll 1$.
(This is analogous to the difficulty of approaching
extremality in the charged case, where the horizon
temperature also vanishes.)
By holding fixed $\epsL$ instead of
$\epsB$, we are able to perform scans
in $\B$ which avoid dipping too deeply into the
very low temperature region.

Another difference concerns the definition
of pressure (or stress) anisotropy.
With our choice (\ref{eq:C}) for fixing
the scheme dependent ambiguity in the stress-energy tensor,
the resulting holographic relation (\ref{eq:T}),
when evaluated at $\mu = 1/L$, puts the trace anomaly
solely in the transverse components of the stress,
$T^{11}$ and $T^{22}$.
So the pressure anisotropy (\ref{eq:DeltaP}),
defined as the difference between transverse and longitudinal stress,
when evaluated at $\mu = 1/L$
has a ``kinematic'' contribution of $-\tfrac 14 \kappa \B^2$
plus a ``dynamic'' contribution of $3 \kappa \, b_4(v)$.
In presenting results below, we will omit the uninteresting
static kinematic contribution, and just plot the dynamic contribution
\begin{equation}
    \Delta \mathcal P_{\rm dyn}
    \equiv
    \half (T^{11} + T^{22}) - T^{33} + \tfrac 14 \kappa \, \B^2 \,,
\end{equation}
(relative to the energy density),
evaluated at the renormalization point $\mu = 1/L$.%
\footnote
    {%
    Examining the holographic relation (\ref{eq:T}), one sees
    that simply shifting the renormalization point to
    $\mu = e^{1/4}/L$ would accomplish the same removal
    of this uninteresting kinematic contribution to the pressure anisotropy.
    }

A further difference comes from the fact that equilibrium
magnetic brane solutions are intrinsically anisotropic.
The anisotropy function $B$ does not vanish at late times,
but rather settles down to the profile of the equilibrium magnetic
brane solution discussed in section \ref{sec:eqsolns}.
This is illustrated in figure~\ref{fig:BevolveMAG},
which shows the (subtracted/rescaled) anisotropy function $B_s$ 
as a function of time $v$ and inverse radial depth $u$.
One sees similar features as in fig.~\ref{fig:BevolveSBB}:
the initial pulse propagates outward, reflects off the boundary,
and largely disappears into the horizon.
But in addition one also sees that the anisotropy function is
approaching the non-trivial profile of a static magnetic
brane solution.

Correspondingly, the pressure anisotropy in the dual field theory
asymptotically approaches a non-zero constant.
To examine equilibration, it is the difference between the
pressure anisotropy and its asymptotic value which is of interest.
As a measure of (near) equilibration, the condition (\ref{eq:teq})
is naturally replaced by
\begin{equation}
    \left[ \Delta \mathcal{P}(t) - \Delta \mathcal{P}(\infty) \right] /
    (\kappa\epsL)
    \leq 0.1 \,,
\end{equation}
for all times $t > \teq$.

\begin{figure}
\centering
\suck[width=0.75\textwidth]{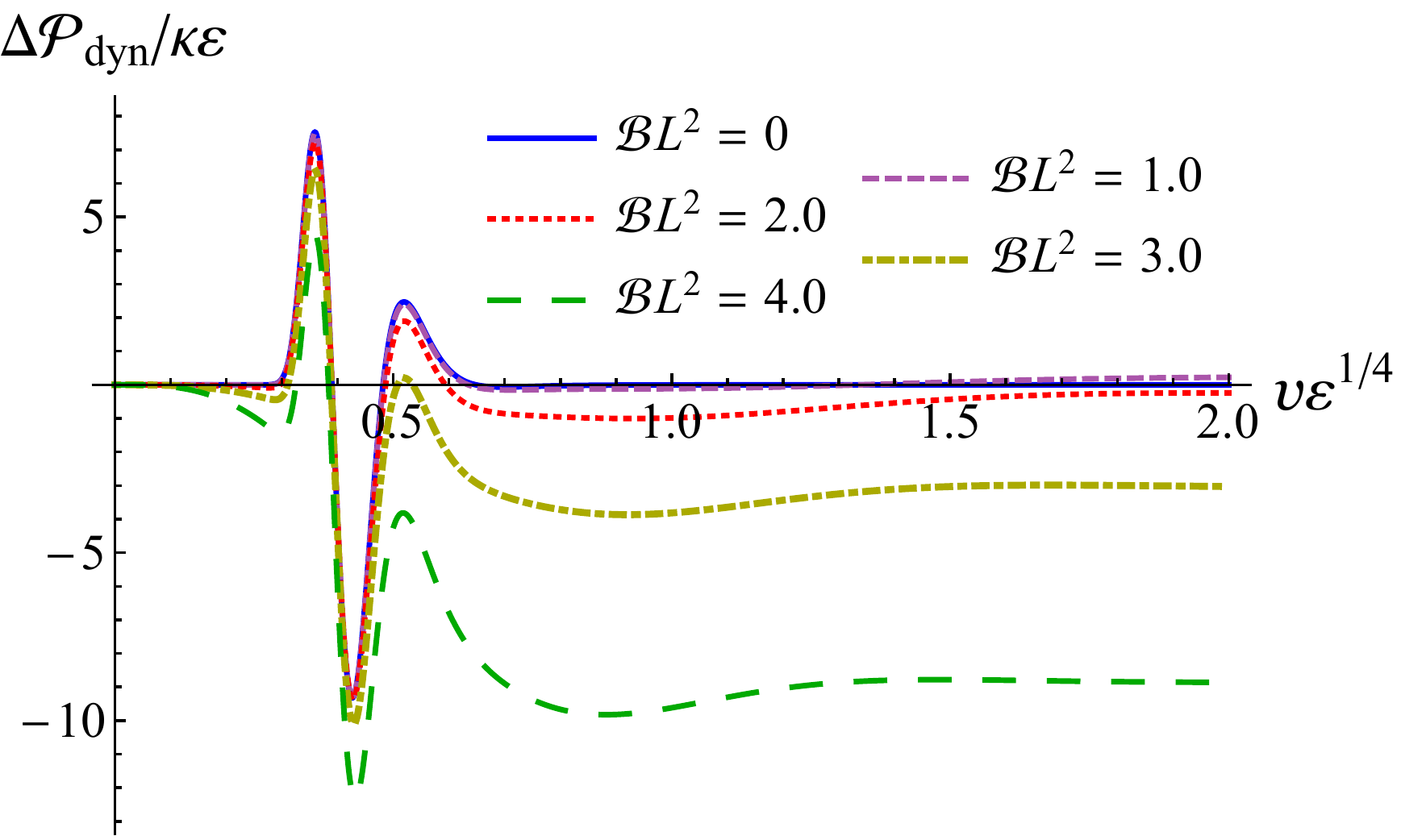}
\caption
    {%
    Time dependence (in units of $\epsL^{-1/4}$)
    of the dynamical pressure anisotropy, at the scale of $1/L$,
    for values of background magnetic field ranging from 0 to $4/L^2$.
    The energy density (at the curvature scale) is held fixed,
    $\epsL = \tfrac 34 L^{-4}$, and the parameters of the
    initial Gaussian pulse in the anisotropy are the
    same as in fig.~\ref{fig:BevolveMAG}.
    \label{fig:pressureMAG}
    }
\end{figure}

The time dependence of the resulting (dynamical contribution to the)
pressure anisotropy is shown in fig.~\ref{fig:pressureMAG}
for a series of magnetic fields, $\B L^2 = 0$, 1, 2, 3, and 4.
The energy density is held fixed at $\epsL = \tfrac 34 L^{-4}$
and the parameters of the Gaussian pulse in the
initial anisotropy function are those of the pulse which generated
fig.~\ref{fig:BevolveSBB} 
($\mathcal A = 5\times10^{-4}$, $\rave = 4$, $\sigma = \tfrac 12$).
For the five cases shown,
the ratios of magnetic field to the equilibrium temperature (squared)
are given by $\B/T^2 = 0$, 13.0, 30.2, 30.5, and 26.3
for $\B L^2 = 0$, 1, 2, 3 and 4, respectively.
Note that, at this fixed value of $\epsL L^4$,
$\B/T^2$ is not monotonic as a function of $\B L^2$.
The energy densities at the intrinsic scale set by the magnetic field
for this series of solutions are given by
$\epsB/\B^2 = \infty$,
$0.75$, $0.36$, $0.36$, and $0.39$,
respectively.

Differences in the late time values of the pressure anisotropy
are obvious in fig.~\ref{fig:pressureMAG}.%
\footnote
    {%
    The tiny positive late time pressure anisotropy barely visible in the $\B L^2 = 1$ curve
    is a consequence of our removal of the static kinematic part of the anisotropy;
    the final value of the total pressure anisotropy, at the scale $1/L$,
    monotonically decreases with increasing $\B L^2$ in this series of solutions.
    }
Small temporal variations are also evident after $v \approx 1.3$.
These are produced by the relaxation of the
initial non-Gaussian background profile of the anisotropy function
to the correct equilibrium form.
These small variations at relatively late times would be
absent if we had constructed our initial anisotropy function 
by adding a Gaussian perturbation to the equilibrium solution
(instead of merely adding a Gaussian to the leading asymptotic term).
Given our choice of initial data,
there are two distinct time scales in the equilibration
shown in figs.~\ref{fig:BevolveMAG} and \ref{fig:pressureMAG},
the first associated 
with the boundary reflection and subsequent infall of 
the Gaussian pulse, and the second with the time it takes for
the background anisotropy profile to reach its equilibrium form.
It is the latter which is responsible for the late time variations;
fortunately, there is relatively little ambiguity in
separating the two contributions to the dynamics.

\begin{figure}
\centering
\suck[width=0.75\textwidth]{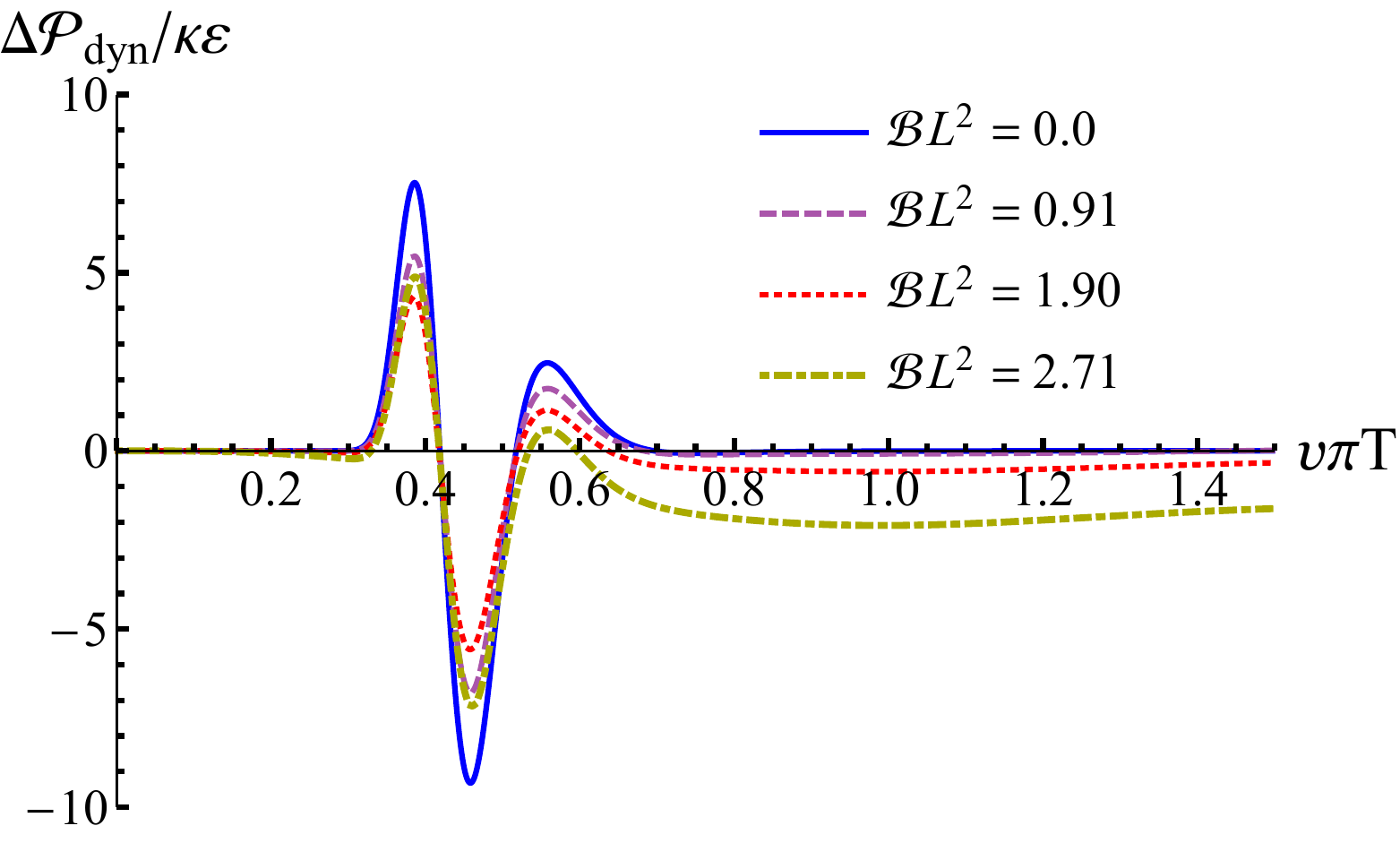}
\caption
    {%
    Time dependence (in units of $(\pi T)^{-1}$)
    of the dynamical pressure anisotropy
    for values of background magnetic field ranging from 0 to $2.71/L^2$.
    The temperature is held fixed,
    $\pi T L = 1$, and the parameters of the
    initial Gaussian pulse in the anisotropy are the
    same as in fig.~\ref{fig:BevolveMAG}.
    The values of $\epsB/\B^2$ for these solutions,
    in order of increasing field,
    are $\infty$, 1.2, 0.51, and 0.39, respectively.
    One sees very little sensitivity to the magnetic field
    in the time course of the response.
    \label{fig:pressureMAGconstT}
    }
\end{figure}

Once again, a notable feature of in the comparison
of fig.~\ref{fig:pressureMAG}
is the similarity in the time dependence of the pressure anisotropy
during the pulse-driven period of significant departure from equilibrium
($0.2 \lesssim v \epsL^{1/4} \lesssim 1.0$),
when the initial perturbation is reflecting from the boundary.
Any reasonably defined measure of equilibration
time $\teq$ clearly does not vary much with magnetic field,
and neither does $\tpeak$ for these relatively near boundary pulses.
This insensitivity result relies, of course, on the constancy of the
initial Gaussian perturbation in the anisotropy function,
and also on our choice to hold fixed the
energy density at the scale $1/L$.

Fig.~\ref{fig:pressureMAGconstT} compares the response,
for different values of magnetic field, when
one holds fixed the equilibrium temperature $T$
(as well as the initial Gaussian perturbation),
instead of fixing the energy density $\epsL$.
With time now plotted in units of $(\pi T)^{-1}$,
one also sees remarkable similarity in the time
dependence of the response, with the times of the
first, second, or third peaks in the pressure anisotropy
varying by less than 0.3\% as $\B/T^2$ varies
from 0 to 26.7.

\begin{figure}
\centering
\suck[width=0.75\textwidth]{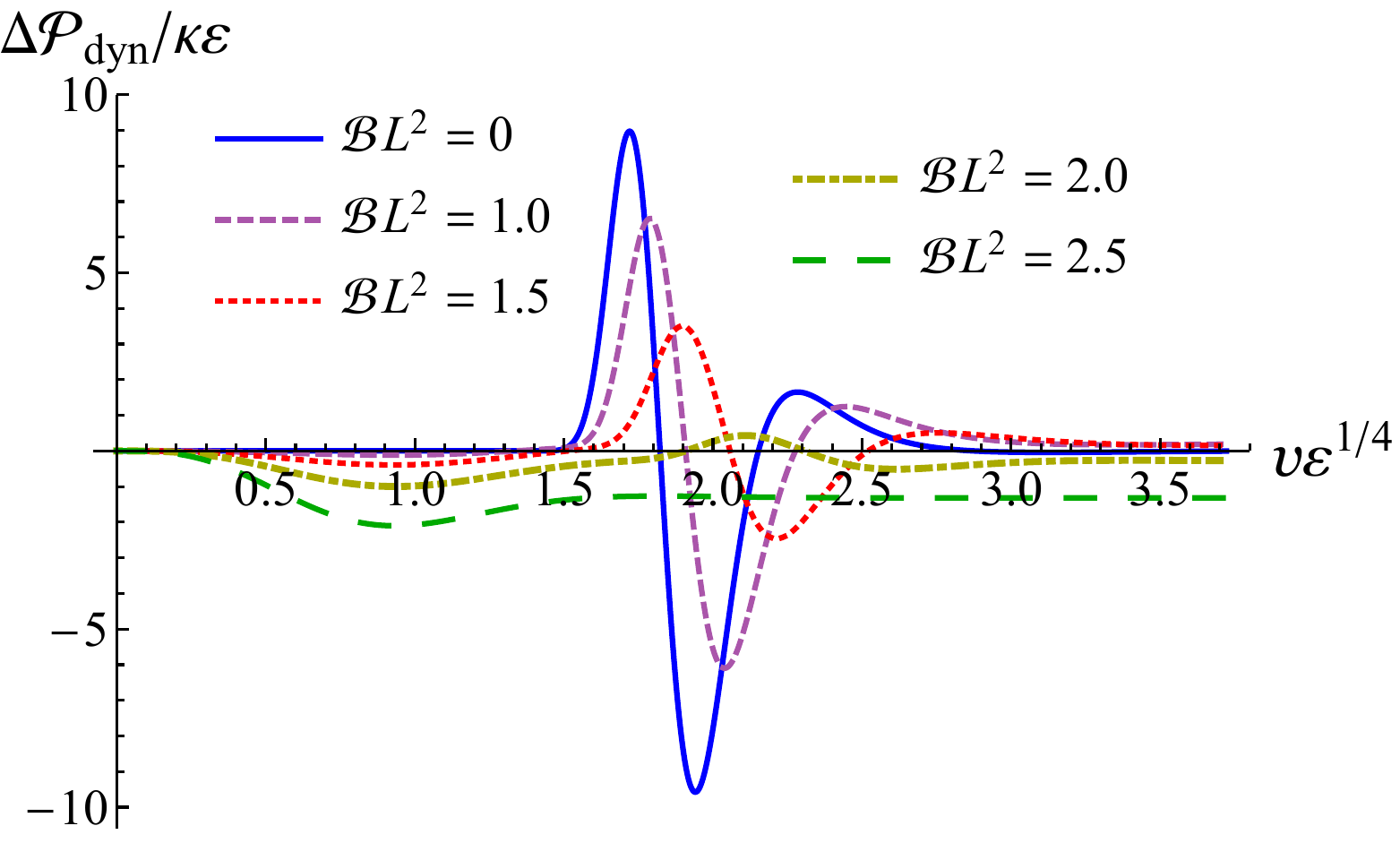}
\hspace{-1.5em}
\caption
    {%
    Time dependence (in units of $\epsL^{-1/4}$)
    of the relative pressure anisotropy
    for values of background magnetic field ranging from 0 to $2.5/L^2$, 
    for the same deep pulse
    ($\mathcal A = \tfrac {1}{10}$, $\rave = \tfrac {10}{11}$,
    $\sigma = \tfrac {1}{20}$)
    used in fig.~\ref{fig:DSpulse} and fig.~\ref{fig:DSRN}.
    The energy density $\epsL = \tfrac 34 L^{-4}$ is held fixed.
    \label{fig:DSMAG}
    }
\end{figure}

Sensitivity to the magnetic field is significantly
larger when
the initial pulse is placed very close to the horizon.
This is shown in fig.~\ref{fig:DSMAG}, which plots
the time dependence of the dynamical pressure anisotropy
for magnetic field values
$\B L^2 = 0$, 1, 1.5, 2, and 2.5,
using the same ``deep pulse'' Gaussian parameters
($\mathcal A = \tfrac {1}{10}$, $\rave = \tfrac {10}{11}$,
$\sigma = \tfrac {1}{20}$),
and fixed energy density $\epsL = \tfrac 34 \, L^{-4}$,
which generated fig.~\ref{fig:DSpulse}.
For this series of
solutions we have $|\B|/T^2 = 0$, 13.0, 22.85, 30.16, and 31.95,
and $\epsB/\B^2 = \infty$, 0.75, 0.43, 0.36, and 0.35,
respectively.

With this deep initial pulse, differences in the time course
of the response are much more pronounced.
Larger magnetic fields greatly suppress the size of the pulse-driven
peaks in the pressure anisotropy
(for a fixed amplitude of the initial Gaussian),
and lead to increasingly large and negative final values
for the anisotropy.
For the lowest curve with $\B L^2 = 2.5$, the contribution
of the Gaussian pulse is completely swamped by the contribution from
the relaxation of the background profile of the anisotropy function to the
correct equilibrium form.
Excluding this curve, the time $\tpeak$ of the first peak,
as well the rough equilibration time $\teq$,
characterizing the portion of the evolution arising
from the Gaussian pulse,
vary at most by 20\% as the magnetic field ranges
from 0 to $2/L^2$.
This is the largest difference in the relaxation time course
we have seen in our exploration of magnetized plasmas
with Gaussian initial perturbations.

A constant temperature comparison (not plotted), analogous to
fig.~\ref{fig:pressureMAGconstT}
but using the same deep pulse as in
figs.~\ref{fig:DSpulse} and \ref{fig:DSMAG},
shows variations in the time course of up to 15\%
as $|\B|/T^2$ ranges from 0 to 22 --- beyond which
the response of the pulse cannot be clearly distinguished
from the background evolution.

\begin{table}
\begin{center}
\def\z{\phantom{0}}
\footnotesize
\begin{tabular}{c|r|r|r|cc|c}
\toprule
\multicolumn{7}{c}{Magnetic ($\B \ne 0$, $\rho = 0$)} \\
\midrule
~$\B/T^2$  &
 $\epsB/T^4$ &
 $\overline{\mathcal P}/\kappa T^4$\!&
 $\!\Delta \mathcal P/\kappa T^4$\!\!&
 $\Re\,\lambda/\epsB^{1/4}$ &
 $\Im\,\lambda/\epsB^{1/4}$ & 
 $\lambda/(\pi T)$
\\
\midrule
 0   & 73.06  & $24.35$   & $0$~~&      $3.3521 \pm 0.0001$ & $-2.9514 \pm 0.0001$ & $3.1195 -2.7466\,i$ \\ 
 %
 0.990 & 72.98 & $24.16$   & $-1.13$~~&  $3.357\z \pm 0.001\z$ & $-2.93\z\z \pm 0.06\z\z$ & $3.124\z -2.73\,i\z\z$ \\ 
 5.344 & 80.74  & $22.15$   & $-10.60$~~& $3.372\z \pm 0.002\z$ & $-2.92\z\z \pm 0.06\z\z$ & $3.217\z -2.79\,i\z\z$ \\ 
 12.953 & 125.85 & $13.98$   & $-15.76$~~& $3.264\z \pm 0.007\z$ & $-2.78\z\z \pm 0.01\z\z$ & $3.480\z -2.96\,i\z\z$ \\ 
 17.821 & 170.16 & $3.79$    & $-9.36$~~&  $3.161\z \pm 0.002\z$ & $-2.69\z\z \pm 0.04\z\z$ & $3.634\z -3.09\,i\z\z$ \\ 
 22.836 & 226.69 & $-11.35$  & $5.00$~~&   $3.061\z \pm 0.008\z$ & $-2.60\z\z \pm 0.03\z\z$ & $3.780\z -3.21\,i\z\z$ \\ 
 30.161 & 328.21 & $ -42.21$ & $39.97$~~&  $2.94\z\z \pm 0.01\z\z$ & $-2.49\z\z \pm 0.03\z\z$ & $3.98\z\z -3.38\,i\z\z$ \\ 
\bottomrule
\end{tabular}
\end{center}
\caption
    {%
    Equilibrium energy density, average pressure, and pressure anisotropy,
    plus lowest quasinormal mode frequency,
    for various values of the external magnetic field.
    Reported values of energy densities and pressures are evaluated at a
    renormalization point $\mu = |\B|^{1/2}$,
    not at the (physically irrelevant) curvature scale.
    The pressure anisotropy is the complete value, including
    the $-\frac 14 \B^2$ kinematic contribution which was
    removed in figs.~\ref{fig:pressureMAG}~--~\ref{fig:DSMAG}.
    Results for the lowest quasinormal mode frequency are reported both in units of $\epsB^{1/4}$
    (middle column, with uncertainties),
    and as well as in units of $\pi T$ (final column).
    }
\label{QNM-MAG}
\end{table}

To conclude this section, we report in table~\ref{QNM-MAG}
equilibrium properties,
plus our estimates for the lowest quasinormal mode frequency,
for values of magnetic field which extend from small fields
well into the strong field regime.
The ratio $\B/T^2$ ranges from zero to just over 30.
The equilibrium energy density, pressure, and pressure anisotropy
are given in units of $T^4$, and have been converted from
the $\mu = 1/L$ renormalization point used in our calculations
(and above presentation) to the intrinsic scale $\mu = |\B|^{1/2}$.
Results for the lowest quasinormal mode frequency are given both
in units of $\epsB^{1/4}$ and in units of $\pi T$.
These estimates are the result of fitting
the late time ($4 \lesssim v \,\epsL^{1/4} \lesssim 9$)
behavior of the pressure anisotropy to a decaying, oscillating exponential
plus a constant equilibrium offset.
To our knowledge,
independent results from a linearized
analysis of small fluctuations about the (numerically determined)
magnetic brane solutions are not currently available.
The leading quasinormal mode frequency varies by about 20\%
as $\B/T^2$ ranges from zero up to 30,
monotonically increasing with increasing field when measured in
units of $\pi T$,
but slightly increasing and then decreasing 
when measured in units of $\epsB^{1/4}$.



\section{Discussion}\label{sec:discussion}

The above results show that to a good
(often extremely good) level of accuracy:
\begin{enumerate}\advance\itemsep -6pt
\item
    the pressure anisotropy response is a linear functional
    of the initial anisotropy pulse profile;
\item
    the time course of the response, measured in units set by the energy density,%
    \footnote {Specifically, $\epsL$ in the magnetic case.}
    is insensitive to the charge density or background magnetic field when
    the pulse profile and the energy density are held fixed;
\item
    the time course of the response, measured in units set by the equilibrium temperature,
    is insensitive to the charge density or background magnetic field when
    the pulse profile and equilibrium temperature are held fixed.
\end{enumerate}
How can one synthesize these observations?
Consider some feature in the time course of the response, such as the
time of the first (or second) peak in the pressure anisotropy,
or the approximate equilibration time discussed above.
For simplicity, we will focus on the time $\tpeak$
of the first pressure anisotropy peak.
This time must be some function of the 
equilibrium state parameters
(energy density plus charge density or magnetic field),
as well as the Gaussian pulse parameters (depth, width, and amplitude).

Consider first the charged case.
The response time $\tpeak$ is a function of the
energy density $\varepsilon$, the fraction $x$ of the extremal charge density,
and the Gaussian pulse parameters $r_0$, $\sigma$, and $\mathcal A$.
But since the temperature decreases monotonically with increasing charge density
(for fixed energy density)
one may equally well regard the equilibrium state as labeled by
the energy density $\varepsilon$ and temperature $T$, and write
\begin{equation}
    \tpeak/ L =
    f(\varepsilon L^4, T L, r_0 /L, \sigma/L, \mathcal{A}) \,, 
\end{equation}
for some function $f$ of the indicated arguments.
We have written all quantities in dimensionless form using, in effect,
our computational units.
Our results on the degree of nonlinearity imply that the function $f$
is nearly independent of the last argument, the Gaussian amplitude
$\mathcal A$.
Only for our narrowest pulse, located right at the horizon,
did the relative nonlinearity reach 10\%.
Away from this corner of parameter space, the nonlinearity was substantially
smaller, rapidly falling to much less than a percent as the initial
pulse becomes less localized at the horizon.
So, to a good approximation, one can regard the function $f$ as being
independent of $\mathcal A$.

For narrow pulses, $\sigma \ll r_0$, the dependence of the
response time $\tpeak$ on the pulse width is negligible;
there is a smooth $\sigma \to 0$ limit.
To the extent that linearity is a good approximation, one may 
regard the response from wider pulses as superpositions of the response
from narrower pulses
(suitably arranged so that their sum reconstructs the wider pulse).
As noted in the discussion of fig.~\ref{fig:BsubRavepos},
the first peak in the response is clearly associated with the
propagation of the leading edge of the anisotropy pulse ---
the region of near-maximal slope on the side of the pulse
closest to the boundary.
Therefore, for pulses of non-negligible width $\sigma$
centered at some depth $r_0$,
one should expect that the time of the first peak in the response
will be most similar to the corresponding response time for a
narrower pulse located not at the depth $r_0$, but rather
at a depth of $r_0 + n \sigma$ for some positive $O(1)$ multiplier $n$.
At the same level of accuracy determined by the
degree of nonlinearity, one should be able to merge the dependence
of the response time $\tpeak$ on the depth $r_0$ and width $\sigma$ of
the initial pulse into a single effective depth
$r_{\rm eff}$ given by  $r_0 + n\sigma$.
(The accuracy of this simplification will be discussed below.)
Hence, the above functional dependence for $\tpeak$ can be replaced
by a simpler form,
\begin{equation}
    \tpeak/ L \approx
    g(\varepsilon L^4, T L, r_{\rm eff} /L) \,,
\label{eq:tpeak-g}
\end{equation}
for some function $g$.
Now, the results of section \ref{sec:resultsRN}
(including figs.~\ref{fig:pressureRN} and \ref{fig:DSRN}
and associated discussion)
show that the time course of the pressure anisotropy response 
has remarkably little dependence on the charge density
when comparisons are made holding fixed
the initial pulse parameters and the energy density.
Since varying the charge density at fixed energy density is,
as noted above, equivalent to varying the equilibrium temperature,
this implies that the function $g$ describing the response time (\ref{eq:tpeak-g})
is nearly independent of the second argument, $T L$.
At the same time, the comparisons at fixed temperature also discussed
in section \ref{sec:resultsRN}
(fig.~\ref{fig:pressureRNconstT} and associated discussion)
show that the time course of the pressure anisotropy response 
\emph{also} has remarkably little dependence on the charge density
when the initial pulse parameters and the equilibrium temperature are held fixed.
This implies that
the function $g$ describing the response time (\ref{eq:tpeak-g})
is nearly independent of the first argument, $\varepsilon L^4$.
Hence, at a level of accuracy determined by the minimal level
of nonlinearity, and the minimal dependence on charge density in comparisons
at both constant energy density and constant temperature,
the response time $\tpeak$ must be a function of \emph{only}
the effective depth of the initial pulse,
\begin{equation}
    \tpeak/L \approx h (r_{\rm eff}/L) \,,
\end{equation}
for some function $h$.
Finally, this function must be consistent with the scaling relations
discussed in section~\ref{sec:scaling}, which imply that $L^2/r$
scales in the same fashion as a distance (or time) in the boundary theory.
Consequently, the dependence of the response time on the effective depth
must have the form
\begin{equation}
    \tpeak \approx C L^2/r_{\rm eff} \,,
\end{equation}
for some dimensionless constant $C$.

A similar line of reasoning is applicable to the magnetic case.
Since we used $\epsL \equiv \varepsilon(L)$ and $\B L^2$ as parameters
labeling the equilibrium magnetic brane geometry in our
comparisons of magnetic plasma response, 
it is convenient to view
the response time $\tpeak$ as depending on these parameters
plus the Gaussian pulse parameters,
\begin{equation}
    \tpeak /L = f(\epsL L^4, \B L^2, r_0 /L, \sigma/L, \mathcal{A}) \,, 
\end{equation}
for some function $f$.
Once again, the observed near-linearity of the response
allows us to simplify this to the form
\begin{equation}
    \tpeak /L \approx g(\epsL L^4, \B L^2, r_{\rm eff} /L) \,, 
\end{equation}
for some function $g$.
The near-independence of the time course of the response on the
magnetic field $\B$, for fixed $\epsL$ and a fixed initial pulse,
implies that the function $g$ is nearly independent of its 
second argument.
Because $\epsL$ does not transform homogeneously under the
scaling relations (\ref{eq:scaling1})-(\ref{eq:scaling3})
(due to the use of the curvature scale $L$ instead of a
physical scale in the dual QFT for setting the renormalization point)
consistency with the scaling relations requires that
the function $g$ be independent of its first argument
and depend inversely on the third.
So, just as for the charged case,
\begin{equation}
    \tpeak \approx C L^2 /r_{\rm eff} \,,
\label{eq:model}
\end{equation}
for some dimensionless constant $C$.

\begin{figure}
\centering
\suck[width=0.48\textwidth]{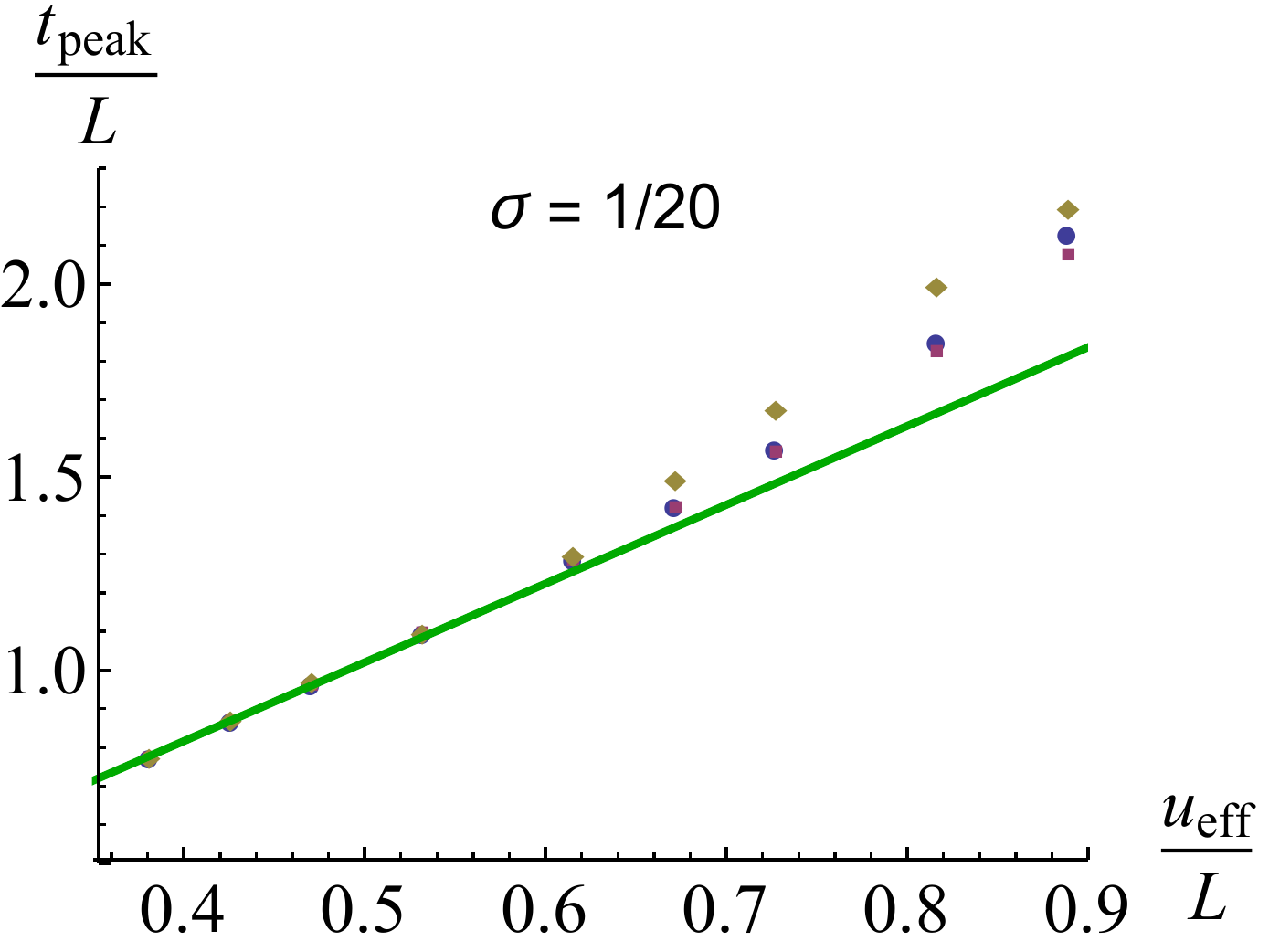}
\suck[width=0.48\textwidth]{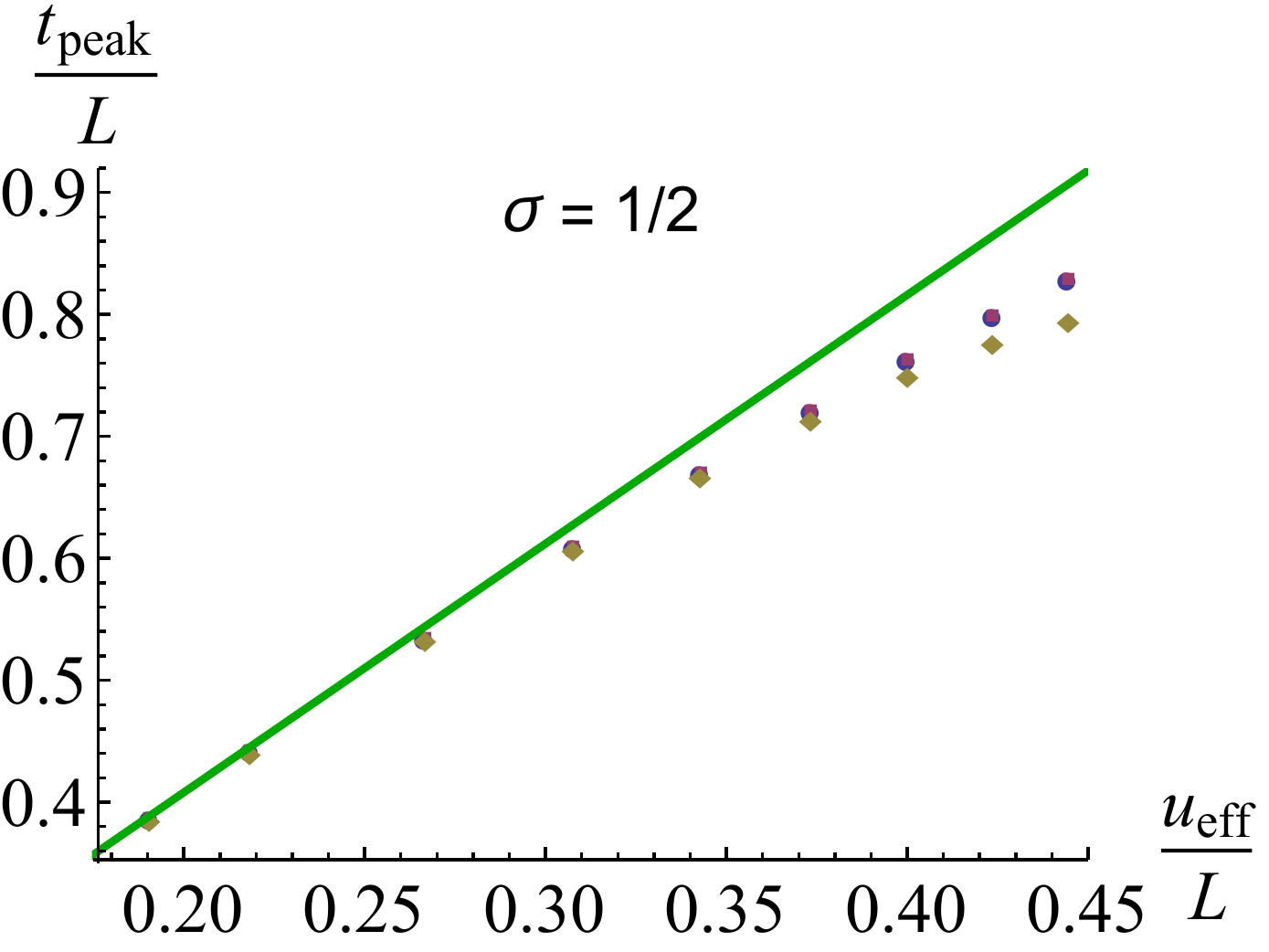}
\vspace*{-0.5em}
\caption
    {%
    Time of the first peak in the pressure
    anisotropy response
    as a function of the inverse effective depth of the pulse,
    $u_{\rm eff} \equiv 1/r_{\rm eff}$.
    The left panel shows results for narrow
    pulses with $\sigma = 1/20$, while the right panel
    shows results from wide pulses with $\sigma = 1/2$.
    Blue circles represent data points from neutral (uncharged,
    unmagnetized) plasmas,
    maroon squares represent data points from charged plasmas
    at 80\% of extremality,
    and gold diamonds represent data points from magnetized plasmas
    with $\B L^2 = 1.5$. 
    The green straight line shows the prediction of our
    simple model (\ref{eq:model}) with $n = 2.5$ and $C = 2.04$.
    \label{fig:model}%
    }
\end{figure}

Figure~\ref{fig:model} compares this simple model with
a sample of our data for neutral, charged, and magnetized plasmas.
The left panel shows data for relatively narrow pulses
of width $\sigma = 1/20$,
while the right panel shows data from rather wide pulses
with width $\sigma = 1/2$.
The abscissa for both panels is the inverse effective depth
$u_{\rm eff} \equiv 1/r_{\rm eff}$.
(Data points at the largest values of $u_{\rm eff}$ shown
in these plots come from pulses centered very near the horizon.)
In both panels,
rather good agreement with the simple model is
found when the multiplier $n$ in the effective
depth $r_{\rm eff} \equiv r_0 + n \sigma$ is chosen to be 2.5,
and the coefficient
\begin{equation}
    C \approx 2.04 \,.
\end{equation}
For both sets of data,
the model is least accurate for pulses located very close to the horizon.
Accuracy for the case of magnetized plasma is a bit worse than
for charged or neutral plasma.
But for all cases, even in the near-horizon regime,
this simple model works at about the 20\% level or better.
As pulses move away from the horizon,
the accuracy rapidly improves.

\section{Conclusion}

In weakly coupled plasmas, adding a conserved charge density to
the system (e.g., flavor charge in a QCD plasma) significantly
changes screening lengths.
When the associated chemical potential is of order $\pi T$,
the relative change in the Debye screening length is $O(1)$
\cite{Kapusta:2006}.
Such changes in screening lengths significantly affect
transport coefficients like the viscosity \cite{Arnold:2003zc}.
Lattice studies \cite{Takahashi:2013mja} of the effect of a
baryon chemical potential in the deconfined phase,
when the system is not asymptotically weakly coupled,
claim to find measurable sensitivity in the Debye screening length,
comparable to perturbative estimates,
but with quite large error bars.
Lattice QCD studies of magnetoresponse
\cite{Bali:2014kia, D'Elia:2010nq, Buividovich:2009wi, Abramczyk:2009gb}
also find substantial
changes in thermodynamics when the magnetic field energy density
becomes large compared to~$T^4$.

Consequently, when this work on strongly coupled \Nfour\ SYM plasma
was initiated, we expected to find significant changes in equilibration
dynamics when a conserved charge density is added to the plasma,
or when the system is placed in a background magnetic field.
The most notable result we have found is that this expectation was wrong.
At least within the range of charge densities we studied,
up to 80\% of extremality,
the equilibration dynamics is remarkably insensitive to the presence
of a conserved charge density.
Additionally, magnetic fields which are well into the strong field region,
$\B/T^2 \gg 1$, induce almost no change in the equilibration time course.

Efforts to use results of holographic calculations in strongly
coupled \Nfour\ SYM as the basis for predictions about real heavy
ion collisions \cite{CasalderreySolana:2011us}
are inevitably hampered by our limited
understanding of the effect on the relevant dynamics of changing
the theory from real QCD to a supersymmetric Yang-Mills model theory.
From this perspective, the insensitivity of the
equilibration dynamics to the charge density is reassuring,
as this provides an example where changes in the plasma constituents
have very little impact on the overall dynamics.

A further notable feature in our results is the remarkably small
degree of nonlinearity in the dynamics governing the pressure anisotropy.
Despite the fact that one is solving the highly nonlinear Einstein equations,
the dependence of the induced pressure anisotropy on the initial anisotropy
function is surprisingly close to linear.
We find deviations from linearity of at most $\approx 30\%$, and this only
for initial disturbances which are crafted to reside very deep in the bulk.
(This is consistent with earlier work in ref.~\cite{Heller:2013oxa}.)
For disturbances localized even modestly above the horizon, the degree of
nonlinearity quickly drops to sub-percent levels.
This near-linearity holds even when the system is far from equilibrium,
with pressure anisotropies which are large compared to the energy density.

In the charged case, the sensitivity of the lowest quasinormal mode frequency
to the charge density is significantly larger than the sensitivity we find
in, and shortly after, the far-from-equilibrium period of the
equilibration dynamics.
The lowest quasinormal mode dominates the equilibration process at sufficiently
late times when all higher modes have decayed away and become negligible
relative to the lowest mode.

The near-linearity which we find in the dynamics implies that the
deviation from equilibrium, even during the far-from-equilibrium portion
of the process, can be represented quite accurately as a sum of quasinormal
modes obeying linearized small fluctuation equations.
The minimal sensitivity of the equilibration dynamics to the charge density,
substantially less than the sensitivity of the lowest quasinormal mode frequency,
suggests that during most of the equilibration process
many quasinormal modes above the lowest one are contributing,
with the sensitivity to the charge density quickly falling with
increasing mode number.
This is something which could be tested directly in a linearized
analysis but, to our knowledge, has not yet been done.

In the magnetic case, we find only modest (few percent) changes induced
by the magnetic field in the time course to equilibration,
and in the lowest quasinormal mode
frequency characterizing very late time behavior.
It would be nice to have independent calculations of quasinormal
mode frequencies for magnetic branes, including studies of the
field dependence of higher mode frequencies.

Although we no longer have reason to expect large effects, it would be
natural to generalize the study of homogeneous equilibration to the case
of plasmas with both non-zero charge density and a background magnetic field.
Extensions to more complicated inhomogeneous settings, such as colliding
shock waves, are also of interest.
It would be desirable to gain a clearer understanding of the
connection between our choices of gravitational initial data
on null slices and boundary observables such as multi-point stress-energy
correlators.
Can more operational procedures,
such as time-dependent background fields \cite{Chesler:2008hg},
produce far-from-equilibrium states which resemble those
produced by our ``deep pulse'' initial data?
We hope future work can shed light on some of these topics.

\acknowledgments
We are grateful to
Han-Chih Chang,
Michal Heller,
Stefan Janiszewski,
Matthias Kaminski,
Andreas Karch, 
Per Kraus, 
Julian Sonner,
and Mikhail Stephanov
for helpful discussions.
This work was supported, in part, by the U.S. Department
of Energy under Grant No.~DE-SC0011637.
We also thank the University of Regensburg
and the Alexander von Humboldt Foundation
for their generous support and hospitality during
a portion of this work.

\bibliographystyle{JHEP}
\bibliography{refs}

\end{document}